# AGE DETERMINATION IN UPPER SCORPIUS WITH ECLIPSING BINARIES

Trevor J. David,[1,2] Lynne A. Hillenbrand,[2] Edward Gillen,[3,∗] Ann Marie Cody,[4] Steve B. Howell,[5] Howard T. Isaacson,[6] and John H. Livingston[7]

[1]*Jet Propulsion Laboratory, California Institute of Technology, 4800 Oak Grove Drive, Pasadena, CA 91109, USA*

[2]*Department of Astronomy, California Institute of Technology, Pasadena, CA 91125, USA*

[3]*Astrophysics Group, Cavendish Laboratory, J.J. Thomson Avenue, Cambridge CB3 0HE, UK*

[4]*NASA Ames Research Center, Moffet Field, CA 94035, USA*

[5]*NASA Ames Research Center, Moffett Field, CA 94035, USA*

[6]*Astronomy Department, University of California, Berkeley, CA 94720, USA*

[7]*Department of Astronomy, University of Tokyo, 7-3-1 Hongo, Bunkyo-ku, Tokyo 113-0033, Japan*

## ABSTRACT

The Upper Scorpius OB association is the nearest region of recent massive star formation and thus an important benchmark for investigations concerning stellar evolution and planet formation timescales. We present nine EBs in Upper Scorpius, three of which are newly reported here and all of which were discovered from *K2* photometry. Joint fitting of the eclipse photometry and radial velocities from newly acquired Keck-I/HIRES spectra yields precise masses and radii for those systems that are spectroscopically double-lined. The binary orbital periods in our sample range from 0.6–100 days, with total masses ranging from 0.2–8 $M_\odot$. At least 33% of the EBs reside in hierarchical multiples, including two triples and one quadruple. We use these EBs to develop an empirical mass-radius relation for pre-main-sequence stars, and to evaluate the predictions of widely-used stellar evolutionary models. We report evidence for an age of 5–7 Myr which is self-consistent in the mass range of 0.3–5 $M_\odot$ and based on the fundamentally-determined masses and radii of eclipsing binaries (EBs). Evolutionary models including the effects of magnetic fields imply an age of 9–10 Myr. Our results are consistent with previous studies that indicate many models systematically underestimate the masses of low-mass stars by 20–60% based on H-R diagram analyses. We also consider the dynamical states of several binaries and compare with expectations from tidal dissipation theories. Finally, we identify RIK 72 b as a long-period transiting brown dwarf ($M = 59.2 \pm 6.8$ $M_{\rm Jup}$, $R = 3.10 \pm 0.31$ $R_{\rm Jup}$, $P \approx 97.8$ days) and an ideal benchmark for brown dwarf cooling models at 5–10 Myr.

## 1. INTRODUCTION

While the basic theory of stellar evolution in the pre-main-sequence (PMS) stage has existed for over 50 years (e.g. Hayashi 1961; Henyey et al. 1965; Iben 1965), direct tests of these model predictions remain infrequent.

Corresponding author: Trevor J. David
trevor.j.david@jpl.nasa.gov

∗ Winton Fellow

Meanwhile, theoretical models have evolved from simple hydrostatic contraction and basic nuclear reaction networks to including the effects of deuterium burning, proto-stellar and circumstellar disk accretion, realistic surface boundary conditions, convection, and magnetic fields or starspots (see D'Antona 2017, for a review). Our best method of evaluating such models is through detailed characterization of benchmark PMS stars allowing dynamical mass determinations from, e.g., circumstellar disk rotation curves or binary orbit determi-nation through either astrometry or spectroscopy (Hillenbrand & White 2004; Mathieu et al. 2007). The most stringent tests are provided by double-lined eclipsing binaries (EBs). For these systems, absolute dimensions and masses can be directly measured in a distance-independent manner with minimal theoretical assumptions and precisions approaching 1% (see Andersen 1991; Torres et al. 2010, for reviews). Thus, PMS EBs allow for direct measurement of the contraction timescales (i.e. PMS lifetimes) of stars.

To date, relatively few PMS EBs have been discovered and characterized. Stassun et al. (2014) provided a review of PMS EBs and a careful assessment of how their parameters compare with predictions from stellar evolution models, clearly indicating systematic inaccuracies in the current generation of models. Since that work, several new PMS EBs have been added and evolution models have been updated.

A sizable chapter of stellar astrophysics is dependent on the calibration of PMS evolution models. For example, the inability of current models to match the observed colors and luminosities of stars less massive than the Sun translates directly into uncertainties in the initial mass function, perhaps the most fundamental relation in stellar astrophysics and our most salient clue towards understanding how stars form (Bastian et al. 2010). Other foundational relationships in stellar astrophysics such as age-activity-rotation relations (Barnes et al. 2005; Mamajek & Hillenbrand 2008; Meibom et al. 2015) are calibrated to clusters and other coeval stellar populations, the ages of which depend on evolutionary models. Likewise, the timescale for protoplanetary disk dispersal and thus giant planet formation is tied to the age scale of young clusters and star-forming regions, which are age-dated using PMS evolution models (Haisch et al. 2001; Hillenbrand 2005; Mamajek 2009). Additionally, it is through the rotational and orbital properties of binaries of different ages that one can empirically constrain tidal circularization and synchronization timescales (Meibom & Mathieu 2005).

Stellar models also underpin much of our knowledge of extrasolar planets, the properties of which are only measured relative to the properties of host stars. Uncertainties in stellar models can thus introduce systematic biases in the masses, radii, and occurrence rates of extrasolar planets (e.g. Mann et al. 2012; Gaidos & Mann 2013), all of which are important for understanding planet formation. The radii of PMS stars are particularly uncertain and quantifying this uncertainty is crucial to ongoing efforts to measure the temporal evolution, if any, in the occurrence rates of close-in planets (Rizzuto et al. 2017). Inaccurate assumptions about the properties of young stars also translate into uncertainties or biases in the masses of planetary companions detected (or missed) via imaging or radial velocity, as well as the locations of condensation fronts in protoplanetary disks.

A double-lined EB in a cluster or other presumably coeval stellar association is particularly valuable, representing a rare benchmark system whose masses, radii, temperatures, luminosities, age, and metallicity can be precisely and accurately determined. EBs also allow for precise distance determinations, and have been successfully used to determine the distances to benchmark clusters such as the Pleiades (e.g. Southworth et al. 2005; David et al. 2016b) and Praesepe (Gillen et al. 2017a). If the age of a cluster or association is somewhat in doubt, as is the case for Upper Sco, then EBs can be used to assess the age of a population in a manner that is independent from traditional H-R diagram analyses.

Here, we present a uniform analysis of EBs discovered with the *K2* mission in the Upper Scorpius OB association, hereafter Upper Sco. Three of these EBs are newly reported, while six have been previously published, though we update the parameters and interpretations of some previously published systems here using additional data. In §2 we describe the observational data considered in this study, including *K2* photometry, follow-up spectroscopy, high-resolution imaging, and supplemental data from the literature. The analysis procedures used to translate these data into stellar masses, radii, temperatures, and luminosities are discussed in §3 and the detailed results for individual EBs are presented in §4. In §5 we discuss the observed properties in the context of stellar multiplicity statistics and the theory of tidal dissipation. In §6 we discuss the implications of our EB measurements for the age of Upper Sco and evaluate the predictions of various stellar evolution models. We summarize our conclusions in §7. Finally, we construct empirical relations in Appendix A and identify two new, non-eclipsing spectroscopic binaries in Upper Scorpius in Appendix B.

## 2. OBSERVATIONS

### 2.1. *K2 Photometry*

The *Kepler* space telescope, during the second campaign of its extended *K2* mission (Howell et al. 2014), continuously observed a 115 deg$^2$ field towards Upper Sco and the $\rho$ Ophiuchi dark cloud from 2014 August 23 to 2014 November 13. The resulting data set constitutes the longest and most sensitive photometric monitoring campaign for a large population of stars in the T Tauri or post T Tauri stages. *K2* observations of Upper Sco and $\rho$ Oph have already enabled a broad range of astrophysical investigations on topics such as the variability of disk-bearing stars (Cody & Hillenbrand 2018; Cody et al. 2017; Hedges et al. 2018; Ansdell et al. 2016), stellar/substellar rotation and angular momentum evolution on the PMS (Rebull et al. 2018; Somers et al. 2017; Scholz et al. 2015), the identification of binaries (Tokovinin & Briceno 2018), stellar magnetospheres (David et al. 2017; Stauffer et al. 2017, 2018), and the first detection of an exoplanet transiting a PMS star (David et al. 2016c).

Prior to the *K2* mission, there were no published eclipsing binaries (EBs) in the Upper Scorpius OB association. We searched the *K2* photometry of known or candidate members of Upper Sco (based on proper motions and color-magnitude diagram positions), and through spectroscopic follow-up observations we confirm nine EBs as members of Upper Sco. To date, six of these systems have been published (Kraus et al. 2015; Alonso et al. 2015; Lodieu et al. 2015; David et al. 2016a; Maxted & Hutcheon 2018). We will refer to some of these publications extensively, and thus abbreviate them hereafter as K15, A15, L15, and D16. Here, we present analyses of the three unpublished systems and updated interpretations of previously published systems in light of new RVs from Keck-I/HIRES. Four of the EBs are double-lined, and one system is triply-eclipsing and triple-lined. Thus, there are eleven stars for which the mass-radius relation in Upper Sco can be mapped over the range 0.1–5 $M_\odot$. The coordinates and photometric properties of the EBs studied here are summarized in Table 1.

Telescope roll angle variations imprint percent-level systematic artifacts in *K2* photometry as a star drifts across the detector, which possesses intra-pixel sensitivity variations. In this study, we use photometry that has been corrected for these systematic effects using one of three detrending algorithms: (1) EVEREST2 (Luger et al. 2016, 2017) based on the pixel-level decorrelation method of Deming et al. (2015), (2) K2SFF (Vanderburg et al. 2016), and (3) K2PHOT (Petigura et al. 2015), which is based on the K2SC algorithm described in Aigrain et al. (2016). In the first two cases, light curves are available from the the Mikulski Archive for Space Telescopes[12] (MAST) and in the third case light curves are available from the ExoFOP page[3] for K2PHOT. Detrended light curves from each algorithm were inspected and the one with the best photometric precision was chosen for analysis, except in the cases of EPIC 204760247 where the EVEREST2 light curve exhibits eclipse depth variations that are certain to be a systematic artifact and EPIC 203476597 where photometry was extracted using a custom aperture with K2PHOT to avoid dilution from a nearby star. For each EB, we summarize the the photometric precision achieved by the different detrending methods and indicate the adopted light curve in Table 2.

## 2.2. *Keck-I/HIRES Spectroscopy*

We obtained high dispersion spectra for the EBs using Keck-I/HIRES (Vogt et al. 1994). From the Keck-I/HIRES spectra we determined radial velocities (RVs), projected rotational velocities, and spectral types. Multiple instrument configurations were used depending on observing conditions and the science goals of different observing programs. The majority of our data were acquired using the B2 or C5 deckers providing spectral resolution of ∼70,000 or ∼36,000, respectively, in the wavelength range ∼4800–9200 Å. In this work, we also include previously published RVs (David et al. 2016a), some of which were derived from HIRES spectra acquired using the setup of the California Planet Search (Howard et al. 2010), covering ∼3600-8000 Å at $R \sim$48,000 with the C2 decker.

## 2.3. *Gemini-S/DSSI Speckle Imaging*

To assess any further multiplicity of binaries in our sample speckle imaging observations were acquired with the DSSI camera (P.I. Steve Howell) at the Gemini South Observatory and the NESSI instrument on the WIYN 3.5-m telescope(Scott et al. 2016, 2018). Both instruments acquired simultaneous observations in two filters, providing colors for any closely projected companions. The DSSI filters are centered at 692 nm and 880 nm, with widths of 40 nm and 50 nm, respectively. The NESSI observations were taken at 562 nm and 832 nm, with widths of 44 nm and 40 nm, respectively. A description of the DSSI data products and processing pipeline is provided in Howell et al. (2011), and the use of the instrument with the Gemini telescope is described in Horch et al. (2012). The NESSI data were also collected and reduced following the methods described in Howell et al. (2011). The *Kepler* bandpass throughput peaks near 600 nm, so the blue filter contrast between a closely projected companion and the source provide an estimate of the amount of third light required in modeling the eclipses. We show the contrast curves from speckle imaging observations for four systems in Figure 2.

## 2.4. *Photometry and Astrometry*

**Table 1.** Coordinates and photometry of Upper Sco EBs[†]

| EPIC | Common name | R.A. (J2000.0) | Dec. (J2000.0) | $G$ | $J$ | $H$ | $K_s$ | Ref. |
|------|-------------|----------------|----------------|-----|-----|-----|-------|------|
| | | hh mm ss | dd mm ss | mag | mag | mag | mag | |
| 204760247 | HR 5934 | 15 57 40.4635 | -20 58 59.081 | 5.7959 ± 0.0008 | 5.764 ± 0.021 | 5.767 ± 0.036 | 5.734 ± 0.033 | D18, MH18 |
| 204506777 | HD 144548 | 16 07 17.7850 | -22 03 36.554 | 8.5091 ± 0.0013 | 7.543 ± 0.027 | 7.146 ± 0.047 | 7.047 ± 0.031 | A15 |
| 203476597 | | 16 25 57.9014 | -26 00 37.672 | 12.0723 ± 0.0013 | 9.575 ± 0.024 | 8.841 ± 0.044 | 8.535 ± 0.021 | D18, D16 |
| 204432860 | USco 48 | 16 02 00.3823 | -22 21 24.200 | 12.5627 ± 0.0036 | 9.824 ± 0.021 | 9.101 ± 0.023 | 8.842 ± 0.022 | D18 |
| 202963882 | | 16 13 18.8960 | -27 44 02.605 | 14.0082 ± 0.0008 | 10.492 ± 0.026 | 9.904 ± 0.025 | 9.623 ± 0.023 | D18 |
| 205207894 | RIK 72 | 16 03 39.2216 | -18 51 29.722 | 14.3511 ± 0.0004 | 11.232 ± 0.021 | 10.466 ± 0.023 | 10.200 ± 0.023 | D18 |
| 205030103 | UScoCTIO 5 | 16 03 59 50.4970 | -19 44 37.683 | 14.5522 ± 0.0009 | 11.172 ± 0.023 | 10.445 ± 0.026 | 10.170 ± 0.021 | D18, D16, K15 |
| 203868608 | | 16 17 18.9697 | -24 37 19.060 | 15.6318 ± 0.0011 | 11.858 ± 0.026 | 11.137 ± 0.024 | 10.760 ± 0.023 | D18, D16 |
| 203710387 | | 16 16 30.6830 | -25 12 20.170 | 16.6459 ± 0.0010 | 12.932 ± 0.023 | 12.277 ± 0.024 | 11.907 ± 0.023 | D18, D16, L15 |

[†] The star EPIC 204165788 ($\rho$ Oph C) has now been classified as an EB twice in the literature (Barros et al. 2016; Rizzuto et al. 2017). However, the K2PHOT, K2SFF, and EVEREST light curves, all of which are contaminated by the brighter nearby star $\rho$ Oph, are inconsistent with such an interpretation.

D18: this work; MH18: Maxted & Hutcheon (2018); A15: Alonso et al. (2015); D16: David et al. (2016a); K15: Kraus et al. (2015); L15: Lodieu et al. (2015).

**Table 2.** K2 light curve properties.

| | Precision (ppm) | | | |
|------|------|------|------|------|
| EPIC | EVEREST2 | K2PHOT | K2SFF | Adopted |
| 204760247 | 196 | 759 | 272 | K2SFF |
| 204506777 | 704 | 783 | 1208 | EVEREST2 |
| 203476597 | 111 | 704 | 1236 | K2PHOT |
| 204432860 | 1807 | 1794 | 2815 | K2PHOT |
| 205207894 | 177 | 534 | 285 | EVEREST2 |
| 202963882 | 2662 | 4003 | 5472 | K2PHOT |
| 205030103 | 213 | 548 | 354 | EVEREST2 |
| 203868608 | 422 | 1220 | 753 | EVEREST2 |
| 203710387 | 888 | 1758 | 1789 | EVEREST2 |

Each of the systems studied here have trigonometric parallaxes and proper motions determined from the *Gaia* mission (Gaia Collaboration et al. 2018), which are summarized in Table 3. Using the trigonometric parallaxes and the high-precision *Gaia* photometry, we constructed color-magnitude diagrams (CMDs) for a sample of high-confidence Upper Sco members. We constructed this sample by cross-referencing the Luhman & Mamajek (2012) and Rizzuto et al. (2015) samples with the *Gaia* DR2 catalog. We then determined the median proper motions ($\langle \mu_\alpha \rangle$ = -10.8 mas yr$^{-1}$, $\langle \mu_\delta \rangle$ = -23.6 mas yr$^{-1}$) from the Luhman & Mamajek (2012) sample and selected stars within a 5 mas yr$^{-1}$ radius of that central value and with distances that satisfied the criterion 100 pc $\leq d \leq$ 205 pc (a range which encompassed more than 99% of the proper motion selected sample). The resulting CMDs, along with the positions of the EBs, are shown in Figure 3. We have not at-tempted to deredden the photometry, and note there is variable extinction in the region.

All of the multiple systems discussed here clearly lie on or above the binary sequence of the association, with two exceptions. The systems which do not appear overluminous for their colors are EPIC 205207894 (RIK 72) and EPIC 203710387. In the case of RIK 72, this fits with our current understanding of the system, which hosts a brown dwarf companion that contributes little flux at optical wavelengths. In the case of EPIC 203710387, it is unclear why the system does not seem to rest on the binary sequence. The system does seem to reside at the more distant end of the association ($d \approx$ 167 pc, approximately in the 93rd percentile for the constructed sample), but this does not fully explain its apparent underluminosity in a color-absolute magnitude diagram. Furthermore, as we will discuss in § 6, the components of the EPIC 203710387 system appear too small for the best-fitting mass-radius isochrones in most but not all model sets. We note that EPIC 202963882, a hierarchical triple, appears very red in the $G - G_{\rm RP}$ color while this is not the case for the $G_{\rm BP} - G_{\rm RP}$ color. This source has a BP-RP excess factor (1.985) that is too high for the $G_{\rm BP}$ and $G_{\rm RP}$ magnitudes to be considered reliable. This can happen if another source was in the slit for the $G_{\rm BP}$ and $G_{\rm RP}$ observations, while the $G$ magnitude, which is acquired using a smaller aperture, would remain unaffected.

### 2.5. Rotation and variability periods

We gathered photometric rotation and variability periods for each of the systems in our sample from the Rebull et al. (2018) catalog, which is based on the K2 photometry. We compare the variability and orbital periods in Table 4. In most cases the variability periods are interpreted to arise from rotational modulation of starspots. One exception is HR 5934, which exhibits low-amplitude photometric variability at a period characteristic of slowly-pulsating B-type stars. We use the variability periods to assess the degree of spin-orbit synchronization in § 5.2.

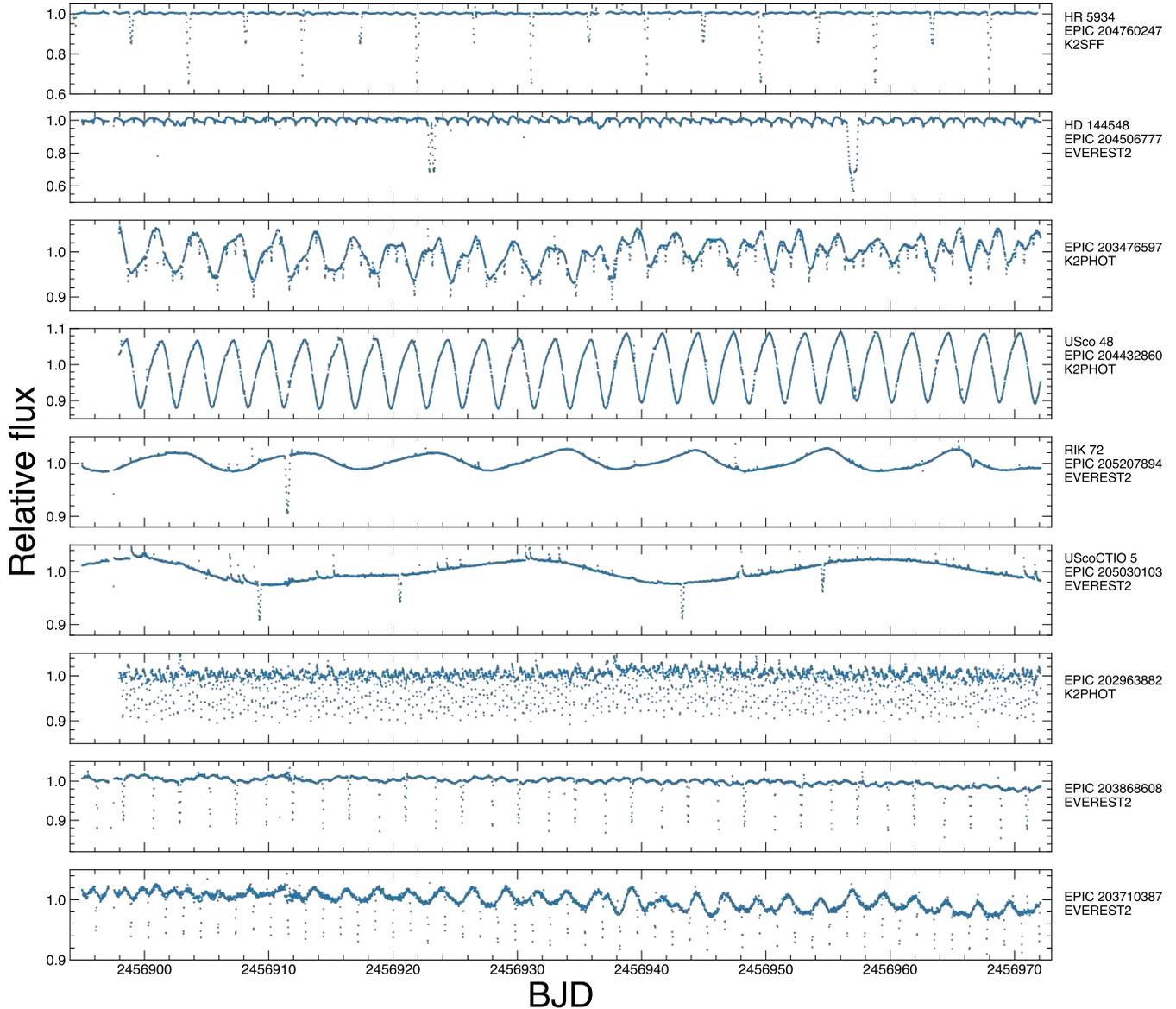

**Figure 1.** Gallery of *K2* light curves for EBs in Upper Sco. At right for each panel, the star's common name, EPIC ID number, and the specific version of the light curve shown. HD 144548 is a triply-eclipsing system, with shallow eclipses at the shorter period and deep irregular eclipses when the close binary eclipses the tertiary. Note, the eclipses of USco 48 are so shallow relative to the stellar variability that they are imperceptible in this figure.

## 3. DATA ANALYSIS PROCEDURES

Here, we describe the common procedures followed for each source in the study to derive physical quantities from the observations.

### 3.1. *Radial velocities and spectroscopic flux ratios*

The IRAF[4] task FXCOR was used to measure relative velocities between program stars and RV standards, with each spectrum first corrected to the heliocentric frame. Spectral orders having sufficient S/N, lacking significant telluric contamination, and with abundant pho-

**Table 3.** *Gaia* DR2 astrometry.

| EPIC | $\varpi$ | $\mu_\alpha$ | $\mu_\delta$ | GOF$_{AL}$ | $\chi^2_{AL}$ | $\epsilon$ | $\sigma_\epsilon$ |
|---|---|---|---|---|---|---|---|
| | (mas) | (mas yr$^{-1}$) | (mas yr$^{-1}$) | | | (mas) | |
| 204760247 | $6.3947 \pm 0.1045$ | $-10.125 \pm 0.188$ | $-21.752 \pm 0.124$ | 44.9130 | 3320.57 | 0.355 | 60.0 |
| 204506777 | $6.8719 \pm 0.0849$ | $-10.954 \pm 0.147$ | $-24.404 \pm 0.103$ | 13.4492 | 533.04 | 0.103 | 3.44 |
| 203476597 | $6.3162 \pm 0.0442$ | $-14.840 \pm 0.094$ | $-23.779 \pm 0.070$ | 10.2677 | 550.12 | 0.000 | 0.00 |
| 204432860 | $6.9181 \pm 0.1181$ | $-11.745 \pm 0.134$ | $-23.822 \pm 0.069$ | 25.0390 | 1409.40 | 0.134 | 9.55 |
| 205207894 | $6.6470 \pm 0.0839$ | $-10.462 \pm 0.182$ | $-21.209 \pm 0.094$ | 21.9507 | 1172.43 | 0.405 | 36.6 |
| 202963882 | $7.0033 \pm 0.0911$ | $-8.931 \pm 0.204$ | $-23.457 \pm 0.108$ | 31.8318 | 2165.70 | 0.442 | 55.8 |
| 205030103 | $6.1391 \pm 0.1188$ | $-11.031 \pm 0.197$ | $-21.005 \pm 0.119$ | 30.8126 | 1785.22 | 0.557 | 61.0 |
| 203868608 | $6.5373 \pm 0.2983$ | $-10.668 \pm 0.649$ | $-21.376 \pm 0.408$ | 93.6149 | 13295.78 | 1.830 | 363.0 |
| 203710387 | $5.9952 \pm 0.1395$ | $-12.159 \pm 0.309$ | $-21.279 \pm 0.189$ | 11.7504 | 666.48 | 0.555 | 14.0 |

GOF$_{AL}$: Goodness-of-fit statistic of astrometric model with respect to along-scan observations.

$\chi^2_{AL}$: Astrometric goodness-of-fit ($\chi^2$) in the along-scan direction.

$\epsilon$: Astrometric excess noise of the source.

$\sigma_\epsilon$: Significance of astrometric excess noise.

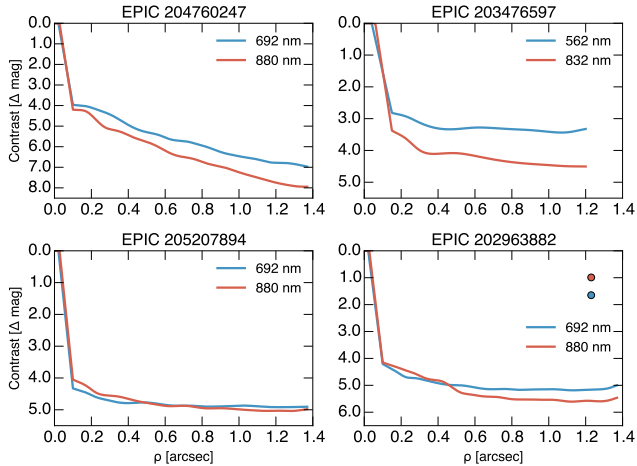

**Figure 2.** Speckle imaging results for four of the EBs discussed here. The data for EPIC 203476597 originate from the NESSI instrument on the WIYN telescope. For the rest, the speckle data were acquired with the DSSI instrument at Gemini South Observatory. EPIC 202963882 has a companion at $1.23''$, and its contrast at 692 nm and 880 nm are represented as the red and blue points respectively. We determined from Keck-I/HIRES spectroscopy that this companion is the EB.

tospheric features were chosen. FXCOR implements the [Tonry & Davis](#) (1979) method of cross correlation peak finding; a Gaussian (or sometimes parabolic) profile was used to interactively fit for the velocity shift for individual components of each binary at each epoch. The measured relative velocities were calibrated to the known RV standard stars. The final velocities at each epoch are derived as error-weighted means from among the individual orders. To establish the epoch of each observation we converted the UTC dates at mid-exposure of each spec-

**Table 4.** Variability and orbital periods.

| EPIC | $P_{var,1}$ | $P_{var,2}$ | $P_{orb,1}$ | $P_{orb,2}$ |
|---|---|---|---|---|
| | (days) | (days) | (days) | (days) |
| 204760247 | $0.9070^\dagger$ | $\cdots$ | 9.1997 | $\cdots$ |
| 204506777 | 1.5325 | 0.8130 | 1.6278 | 33.945 |
| 203476597 | 3.2126 | 1.4403 | 1.4408 | $\cdots$ |
| 204432860 | 2.8752 | $\cdots$ | 2.8745 | $\cdots$ |
| 205207894 | 10.5026 | $\cdots$ | $\cdots$ | $\cdots$ |
| 205030103 | 30.7496 | $\cdots$ | 34.0003 | $\cdots$ |
| 202963882 | $\cdots$ | $\cdots$ | 0.63079 | $\cdots$ |
| 203868608 | 5.6382 | 1.1066 | 4.5417 | 17.9420 |
| 203710387 | 2.5441 | $\cdots$ | 2.8089 | $\cdots$ |

Variability periods originate from [Rebull et al.](#) (2018). Subscripts are not meant to indicate attribution to the primary or secondary, but simply the existence of multiple periods detected in the light curve.

$^\dagger$ This period is believed to be due to pulsations in the primary.

troscopic observation to BJD using the UTC2BJD tool ([Eastman et al.](#) 2010) and the sky coordinates of each target as well as the ground coordinates of Keck Observatory. The optical flux ratio between binary components at each epoch can be approximated from the relative heights of the cross correlation peaks for each of the two components of a double-lined binary system. We note that the flux ratio should formally be calculated from the relative areas of the cross-correlation peaks, but may be approximated from the peak heights when $v \sin i$ is similar for each component. In the present case, we have utilized the peak heights since these are confidently determined from our data while the widths are more uncertain. Final flux ratio values were computed as means among the measured orders. Our new RV measurements are reported in Table 7, and the results from joint fitting of the RVs and eclipse photometry are discussed in §4 where individual EB systems are presented. For some of the spectra of RIK 72 acquired with the California Planet Search set-up, RVs were determined using the telluric A and B bands as a wavelength reference (details presented in Chubak et al. 2012).

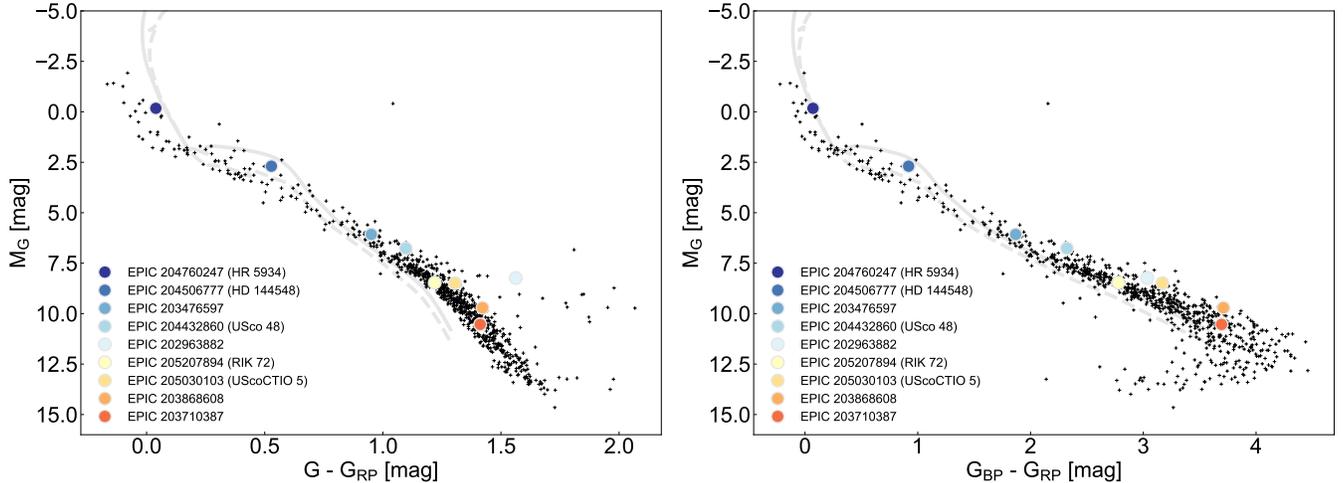

**Figure 3.** Color-absolute magnitude diagrams of high-confidence Upper Sco members (black crosses) and the EB systems studied here (shaded circles). At left, the $G - G_{\mathrm{RP}}$ color is shown on the abscissa while the $G_{\mathrm{BP}} - G_{\mathrm{RP}}$ color is used at right. The solid and dashed lines indicate 5 Myr and 10 Myr MIST isochrones, respectively, with an extinction of A(V) = 0.8 mag applied.

### 3.2. Spectral types, extinction, and effective temperatures

We derived spectral types from the Keck-I/HIRES spectra and comparison with spectral standards. Initial estimates of effective temperatures were then derived from the empirical PMS SpT-$T_{\mathrm{eff}}$ relations presented in Pecaut & Mamajek (2013), hereafter PM13, and Herczeg & Hillenbrand (2015), hereafter HH15. The two relations produce temperatures that are consistent within 100 K, with the HH15 scale being hotter at a given spectral type.

For more precise estimates of the primary and secondary effective temperatures ($T_{\mathrm{eff}}$) for these EBs, we modelled their spectral energy distributions (SEDs) using the method described in Gillen et al. (2017a). Briefly, the PHOENIX v2 (Husser et al. 2013) and BT-Settl (Allard et al. 2012b) model atmospheres were convolved with the broadband photometric data reported in Table 6, and interpolated in $T_{\mathrm{eff}}$—$\log g$ space (fixing $Z = 0$, given the cluster [Fe/H]). Parameters of the fit were the temperatures, radii and $\log g$ of both stars, the distance and reddening to the system, and a jitter term on the observed magnitudes ($T_{\mathrm{pri}}$, $T_{\mathrm{sec}}$, $R_{\mathrm{pri}}$, $R_{\mathrm{sec}}$, $\log g_{\mathrm{pri}}$, $\log g_{\mathrm{sec}}$, $d$, $A_V$, and $\ln \sigma$).

Priors from the joint light curve and RV modeling were placed on the individual stellar radii and $\log g$ values,[5], and the *Gaia* DR2 parallax constraint was used as a prior on the distance. The temperatures and reddening had uniform priors, and the jitter term had a Jeffreys prior (Jeffreys 1946). To break the degeneracy between the two stellar temperatures (and also the distance), we placed a prior on the surface brightness ratio between the two stars in the *Kepler* band from our light curve and RV modeling.

We also explored the effect of imposing Gaussian priors on the reddening. The posterior parameter space was explored via affine invariant Markov chain Monte Carlo (emcee Foreman-Mackey et al. 2013) using 10,000 steps and 160 "walkers". The first 5000 steps were discarded as a conservative burn-in, and posterior distributions derived from the remaining chain after thinning based on the autocorrelation lengths of each parameter.

In general, we found the BT-Settl models imply temperatures about 100–150 K higher (as well as larger $A_V$ values) than those favored by the PHOENIX v2 atmospheres. In the end, we adopted the temperatures implied by the BT-Settl models with Gaussian priors on $A_V$. We summarize the results of the SED fits in Table 5.

Given the stellar radii and effective temperatures from the SED fitting, bolometric luminosities were then calculated from the Stefan-Boltzmann law. The luminosities are calculated from the radii favored by the SED fits so as to be self-consistent, and although strong priors are imposed on the radii based on the EB modeling results, the radii are allowed to vary in the fitting procedure. The corresponding variations in luminosity when using these radius values (as opposed to the median values from EB modeling) are $< 1\sigma$, and insignificant compared to the evolution predicted by models over the timescales of interest here.

---

[5] For EPIC 205207894 we used only the radius ratio, as the system is single-lined.

### 3.3. *Modeling of eclipses and radial velocities*

We performed joint fits to the eclipse photometry and RVs with the widely used JKTEBOP software (Southworth 2013, and references therein). The JKTEBOP program models stars as biaxial spheroids for reflection and ellipsoidal effects (which are negligible for nearly all systems discussed here) and as spheres for eclipse shapes. It is based on the EBOP code (Etzel 1981; Popper & Etzel 1981) originally written by Paul Etzel and based on the Nelson & Davis (1972) model. We note that this software is appropriate for detached EBs where tidal distortion is negligible. Prior to modeling the eclipses, we flattened the light curves by iteratively fitting the out-of-eclipse light curves with a cubic basis spline and rejecting outliers upon each iteration. Our flattened light curves will be made available on the *K2* ExoFOP pages for each individual target.[6]

We note that the practice of flattening or rectifying EB light curves is not equivalent to including the effects of starspots in the modeling. Spots can introduce distortions in the eclipse profiles as well as the spectral line profiles from which the radial velocities are determined. Torres & Ribas (2002) and Stassun et al. (2004) investigated the latter effect for the low-mass EB YY Gem and pre-main sequence EB V1174 Ori, respectively. Both studies found the effect of spots on the RVs were <1 km s$^{-1}$, and in the case of YY Gem the effect on the derived masses was $< 0.2\%$, much lower than the systematic uncertainties we find. Any effect on the radii is also unlikely to change the broad conclusions reached here, which concern the overall behavior of stellar evolution models during a phase when radii are evolving rapidly. Accounting for the impact of spots on either the RVs or eclipse profiles is nontrivial and is left for a future work.

In general, our fitting procedure was as follows: we found the orbital period through a Box-fitting Least-Squares periodogram analysis and estimated the time of minimum light by inspection. The surface brightness ratio was estimated from the ratio of the fluxes at the eclipse minima and practical estimates for the sum of the fractional radii could be approximated based on the orbital period and a plausible total mass. Then, using the Levenberg-Marquardt least-squares minimization routine in JKTEBOP we found best-fit values for the following free parameters: the orbital period ($P$), the reference time of primary minimum ($T_0$), the sum of the radii ($\frac{R_1+R_2}{a}$), the ratio of the radii ($k = R_2/R_1$), the inclination ($i$), two parameters describing the eccentricity and longitude of periastron ($e\cos\omega$ and $e\sin\omega$), and the surface brightness ratio ($J$). The final best fit parameters and uncertainties were then determined through Monte Carlo simulations with JKTEBOP. A quadratic limb-darkening law was assumed, with coefficients chosen specifically for individual systems, as explained in the discussion of each EB ($\S$ 4). Gravity darkening and the reflection effect were ignored. In all cases, eclipse models were numerically integrated to match the *Kepler* cadence of 1766 s.

### 3.4. *A note about model degeneracies*

For several of the EBs studied here, the mass ratio is close to unity ($q > 0.9$). In some of these cases, there is significant covariance between the surface brightness ratio, radius ratio, and inclination, which often leads to solutions that appear non-coeval in the mass-radius diagram (MRD).

Spectroscopic flux ratios, measured in approximately the same wavelength range of the *K2* photometry, are meant to alleviate these degeneracies but they persist nonetheless, perhaps due to insufficient precision in the flux ratio measurements. As such, for systems composed of nearly equal mass stars, we explored three different fits where we (1) fixed the radius ratio at unity, (2) fixed the surface brightness ratio at unity, then (3) allowed both parameters to be free. This decision was based on the facts that the sum of the radii are more robustly determined than the ratio of radii in EB modeling, and the mass ratio in each of these three systems is close to unity, so that large differences between component radii are not expected.

While forcing the radii to be equal in a nearly equal-mass EB will result in a slightly non-coeval solution, allowing the ratio of radii to be a free parameter in some cases results in even more non-coeval solutions. In such cases it is not clear whether the parameters of one component are to be trusted over the other, and the EBs thus become less useful in assessing ages from the MRD.[7]

---

[7] We note that apparent reversals in the temperatures, luminosities, and/or radii of equal-mass PMS EBs have been observed before (Stassun et al. 2007; Gómez Maqueo Chew et al. 2009, 2012;

**Table 5.** Effective temperatures and distance values for each EB estimated from SED modeling and empirical relations.

| Method[*] | Model[†] | $T_{\rm eff}$[‡] Primary (K) | Secondary (K) | Distance (pc) | $A_V$ (mag) | $A_V$ Prior |
|---|---|---|---|---|---|---|
| ............................... | EPIC 203710387 | ............................... | | | | |
| SED | PHOENIX | $2864^{+63}_{-43}$ | $2861^{+63}_{-43}$ | $160.6^{+3.4}_{-2.8}$ | $0.29^{+0.25}_{-0.18}$ | Gaussian |
| SED | PHOENIX | $2819^{+28}_{-23}$ | $2816^{+29}_{-24}$ | $159.7^{+3.1}_{-2.7}$ | $0.07^{+0.11}_{-0.05}$ | Uniform |
| SED | BT-Settl | $3044^{+71}_{-77}$ | $3040^{+73}_{-79}$ | $166.4^{+3.5}_{-3.1}$ | $0.77^{+0.31}_{-0.35}$ | Gaussian |
| SED | BT-Settl | $2906^{+56}_{-36}$ | $2902^{+53}_{-36}$ | $166.1^{+3.4}_{-3.3}$ | $0.13^{+0.21}_{-0.10}$ | Uniform |
| SED | Combined | $2954^{+161}_{-133}$ | $2950^{+162}_{-133}$ | $163.5^{+6.4}_{-5.7}$ | $0.53^{+0.55}_{-0.43}$ | Gaussian |
| ER | HH15 | $3035 \pm 55$ | | | | |
| ER | PM13 | $2950 \pm 70$ | | | | |
| ............................... | UScoCTIO 5 | ............................... | | | | |
| SED | PHOENIX | $3106^{+58}_{-53}$ | $3101^{+56}_{-53}$ | $158.6^{+2.6}_{-2.5}$ | $0.42^{+0.17}_{-0.17}$ | Gaussian |
| SED | PHOENIX | $3029^{+46}_{-33}$ | $3025^{+47}_{-33}$ | $157.6^{+2.7}_{-2.4}$ | $0.14^{+0.16}_{-0.09}$ | Uniform |
| SED | BT-Settl | $3274^{+74}_{-77}$ | $3266^{+74}_{-79}$ | $163.2^{+2.8}_{-2.7}$ | $0.81^{+0.22}_{-0.23}$ | Gaussian |
| SED | BT-Settl | $3122^{+112}_{-62}$ | $3112^{+115}_{-61}$ | $162.6^{+3.0}_{-2.8}$ | $0.31^{+0.36}_{-0.24}$ | Uniform |
| SED | Combined | $3156^{+205}_{-161}$ | $3149^{+205}_{-161}$ | $163.2^{+3.2}_{-3.2}$ | $0.54^{+0.49}_{-0.41}$ | Gaussian |
| ER | HH15 | $3085 \pm 105$ | | | | |
| ER | PM13 | $3020 \pm 140$ | | | | |
| ............................... | RIK 72 | ............................... | | | | |
| SED | PHOENIX | $3229^{+99}_{-60}$ | $2633^{+49}_{-37}$ | $150.5^{+1.9}_{-1.9}$ | $0.25^{+0.19}_{-0.16}$ | Gaussian |
| SED | PHOENIX | $3216^{+94}_{-57}$ | $2624^{+49}_{-34}$ | $150.6^{+2.0}_{-1.9}$ | $0.22^{+0.20}_{-0.15}$ | Uniform |
| SED | BT-Settl | $3360^{+125}_{-127}$ | $2729^{+85}_{-75}$ | $150.5^{+1.9}_{-1.9}$ | $0.54^{+0.30}_{-0.30}$ | Gaussian |
| SED | BT-Settl | $3349^{+142}_{-127}$ | $2722^{+98}_{-78}$ | $150.5^{+1.9}_{-1.9}$ | $0.51^{+0.36}_{-0.37}$ | Uniform |
| SED | Combined | $3294^{+191}_{-126}$ | $2681^{+133}_{-89}$ | $150.5^{+1.9}_{-1.9}$ | $0.40^{+0.45}_{-0.30}$ | Gaussian |
| ER | HH15 | $3485 \pm 75$ | | | | |
| ER | PM13 | $3425 \pm 65$ | | | | |
| ............................... | USco 48 | ............................... | | | | |
| SED | PHOENIX | $3572^{+60}_{-57}$ | $3567^{+59}_{-59}$ | $143.7^{+2.1}_{-1.9}$ | $0.35^{+0.13}_{-0.13}$ | Gaussian |
| SED | PHOENIX | $3563^{+60}_{-59}$ | $3556^{+60}_{-59}$ | $143.6^{+2.0}_{-1.9}$ | $0.33^{+0.13}_{-0.13}$ | Uniform |
| SED | BT-Settl | $3656^{+87}_{-88}$ | $3650^{+87}_{-88}$ | $145.0^{+2.3}_{-2.2}$ | $0.51^{+0.19}_{-0.21}$ | Gaussian |
| SED | BT-Settl | $3643^{+94}_{-96}$ | $3638^{+90}_{-97}$ | $144.8^{+2.3}_{-2.2}$ | $0.48^{+0.21}_{-0.24}$ | Uniform |
| SED | Combined | $3614^{+128}_{-99}$ | $3609^{+128}_{-100}$ | $144.3^{+3.0}_{-2.6}$ | $0.43^{+0.27}_{-0.21}$ | Gaussian |
| ER | HH15 | $3720 \pm 90$ | | | | |
| ER | PM13 | $3630 \pm 70$ | | | | |

[*] SED = spectral energy distribution and ER = empirical relations.

[†] HH15 = empirical SpT-$T_{\rm eff}$ relation from Herczeg & Hillenbrand (2015); PM13 = empirical SpT-$T_{\rm eff}$ relation from Pecaut & Mamajek (2013).

[‡] For the two sets of empirical relations, the secondary $T_{\rm eff}$ is estimated using the temperature ratio in the *K2* band as a proxy for the $T_{\rm eff}$ ratio.

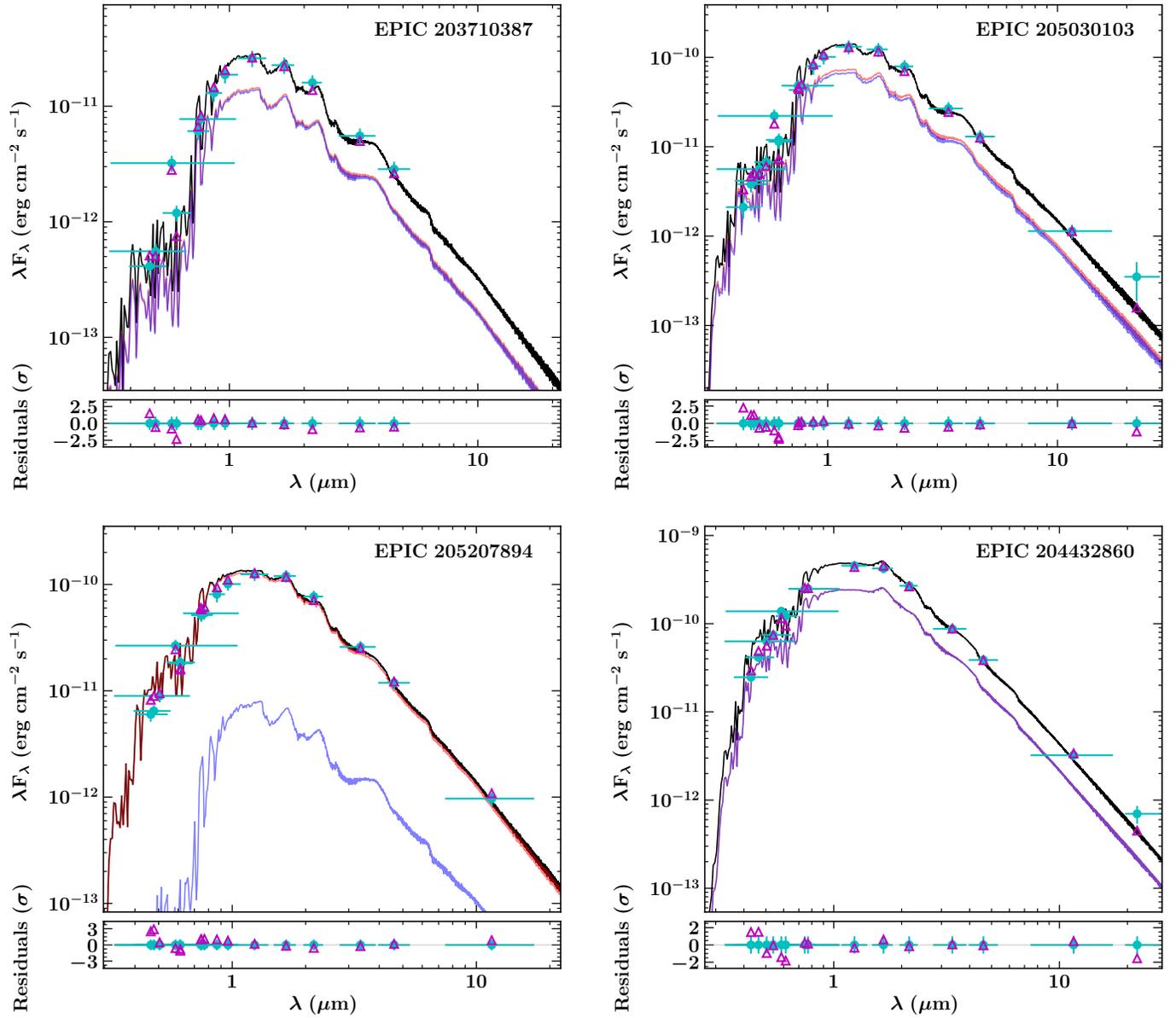

**Figure 4.** SED fits to four of the binaries studied here. Cyan points indicate the observed SED, constructed from broadband magnitudes described in the text, with errorbars representing the spectral coverage in each band. Best-fitting BT-Settl model atmospheres for the primary and secondary are represented by the red and blue curves, respectively. The combined model spectrum is shown in black and the magenta triangles indicate the values obtained by convolving this model with the observed passbands.

We expect, then, that fixing this parameter allows us to determine the average radii and effectively marginalize over the degeneracies that light curve fitting is susceptible to when the ratio of radii are poorly constrained. However, we relax this assumption in separate fits and present both solutions so that the reader may understand the extent to which inherent EB degeneracies and/or apparent non-coevality affects our results. In the following section, we discuss the individual eclipsing systems in detail.

## 4. ECLIPSING BINARY RESULTS

### 4.1. *EPIC 204760247 / HD 142883 / HR 5934*

HR 5934 (also HD 142883 and EPIC 204760247) is a B2.5 type member of Upper Sco that has been studied extensively in the literature.[8] The star was not included in the pioneering work of Blaauw (1946) on the region, but was first proposed as a possible member by Bertiau (1958). The distance to the system from trigonometric parallax and its proper motions, both measured by *Gaia*, are consistent with cluster membership. The systemic RV measured from the RV time series further secures the membership status of these stars.

The system was previously known to be a spectroscopic binary, with RVs measured for both components in Andersen & Nordstrom (1983), though subsequent studies published RVs for only the primary component (Levato et al. 1987; Jilinski et al. 2006). This is simply due to the extreme light ratio between the primary and secondary, which make detection of secondary lines difficult. Prior studies of the primary RVs erroneously assumed an eccentric orbit and a period of 10.5 days but the *K2* light curve shows eclipses and unambiguously defines the true orbital period of 9.2 days with no appreciable eccentricity. This period, and the eclipsing nature of the system, were previously noted by Wraight et al. (2011). We present the literature RVs, which we used in fitting an orbital solution for the system, in Table 8. The literature RVs were published with heliocentric Julian dates, which we converted to barycentric Julian dates using the HJD2BJD tool (Eastman et al. 2010).

Speckle imaging observations of HR 5934 at 692 nm and 880 nm were acquired at Gemini South Observatory with the DSSI instrument (P.I. Steve Howell). We found no evidence for companions brighter than $\Delta m \lesssim 3.97$ mag at 692 nm, or $\Delta m \lesssim 4.2$ mag at 880 nm, in the angular separation range of 0.1–1.37″.

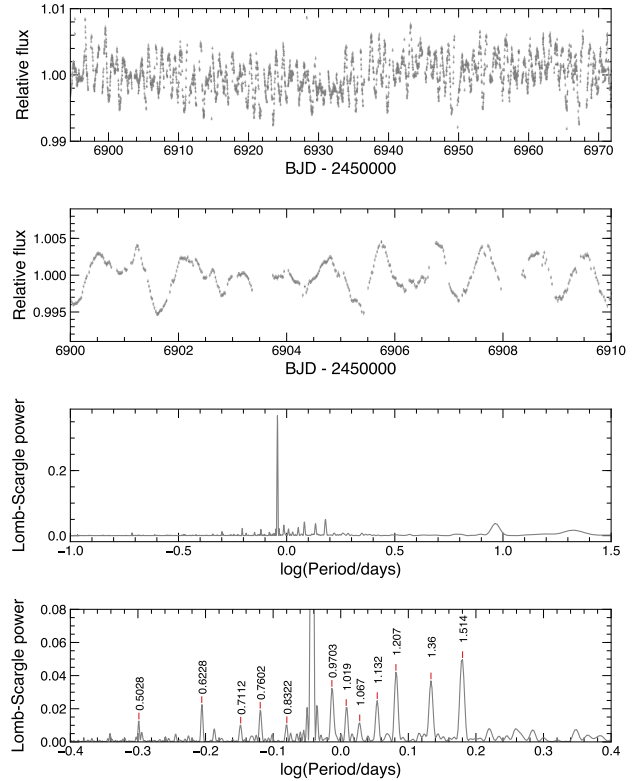

**Figure 5.** Pulsations are visible in the full out-of-eclipse *K2* light curve for HR 5934 (top) and a representative 10-day segment (second panel). A Lomb-Scargle periodogram (third panel) reveals several significant peaks (shown in detail at bottom for clarity), with the most power at $P = 0.907$ d.

Outside of the eclipses, variability of $\lesssim 0.5\%$ amplitude is clearly seen in the *K2* photometry (Figure 5). This variability is almost certainly due to one of the EB components being a slowly-pulsating B-star (SPB). SPBs are typically B2–B9 dwarfs that pulsate with a primary period in the range of 0.5–5 d, a variability amplitude <0.1 mag. If the pulsations are due to the secondary star, the intrinsic variability amplitude would be much larger due to the extreme flux dilution from the primary. We thus find it more likely that the primary is responsible for the pulsations. Unfortunately, the *K2* photometry lack the requisite sensitivity and baseline to perform an asteroseismic analysis for an independent assessment on the primary star's fundamental parameters.

In Table 9, we present parameters of the HR 5934 system resulting from our joint fit of the *K2* light curve, literature RVs, and new RVs determined from Keck-I/HIRES spectra. The best-fitting eclipse and RV models are shown in Figure 6. We find no evidence for eccentricity from the light curve or RVs, and accordingly we jointly fit the data assuming a circular orbit. In fitting the eclipses, we assumed linear limb darkening with coef-ficients of $u_1 = 0.3026$ and $u_2 = 0.4411$ for the primary and secondary, respectively, calculated from interpolation of the Sing (2010) tables for stars of appropriate temperature and surface gravity.

---

[8] The star is misclassified as a Cepheid variable in SIMBAD

From the eclipses, we are able to precisely determine the surface brightness ratio in the *Kepler* bandpass. For systems with a mass ratio close to one, this surface brightness ratio can be used to approximate the temperature ratio of the two components. However, for systems with mass ratios that deviate significantly from one, the surface brightness ratio in the *Kepler* band can differ substantially from the true photospheric temperature ratio. Thus, in order to determine the approximate temperature of the secondary, we calculated theoretical flux ratios by integrating ATLAS9 model atmospheres across the *Kepler* band. The model atmospheres[9] and *Kepler* transmission curve[10] are publicly available online. We found a good match to the observed surface brightness ratio by using model atmospheres of a B3V primary ($T_{\rm eff}$=19000 K) and a B8V secondary ($T_{\rm eff}$=12000 K), both of solar metallicity. Based on an assessment of the literature, we ultimately adopted a primary spectral type of B2.5 $\pm$ 0.5 (c.f. estimates of B2.5Vn and B3 $\pm$ 1; Carpenter et al. 2006; Hernández et al. 2005) and a secondary spectral type of B8 informed by the exercise described above. Using the Pecaut & Mamajek (2013) empirical relations, we ultimately adopted $T_{\rm eff,1} = 18500 \pm 500$ K and $T_{\rm eff,2} = 11500 \pm 500$ K. Our adopted $T_{\rm eff}$ for the primary is in good agreement with previous estimates from the literature, e.g. 18620 K (Hernández et al. 2005), 18700 K (Hohle et al. 2010), though slightly cooler than another estimate of 20350 K (Carpenter et al. 2006).

From joint fitting of the eclipse photometry and RVs, we determined the masses of the HR 5934 binary with 3–4% precision, and the radii with 1% precision. With such highly precise parameters we can test stellar evolutionary models for high mass stars, comparing the predictions of such models in the H-R diagram and the MRD. As we will show in §6 we found that all model sets considered here do an excellent job at predicting the mass of the primary star. For the secondary, however, we note that *only the models that do not include rotation are able to accurately predict the mass*. Models that include rotation (at 40% of the critical velocity) overestimate the mass of the secondary by 20%.

In the late stages of preparing this paper, we became aware of another published solution for HR 5934

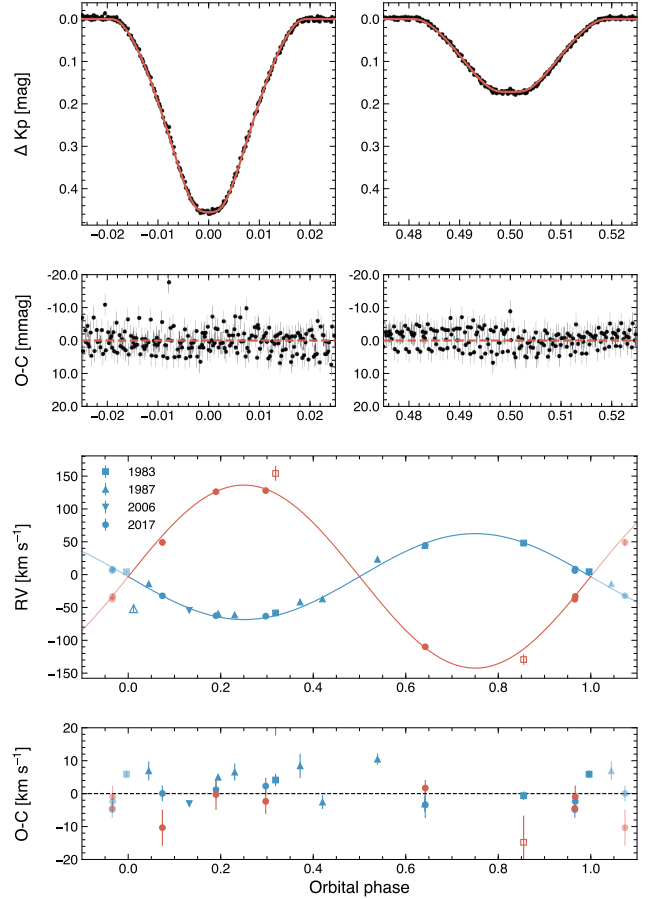

**Figure 6.** Joint fits to the *K2* photometry and radial velocity time series of HR 5934. For this figure and subsequent figures of the same format, the upper left and upper right panels show the primary and secondary eclipses, respectively. Radial velocities from different authors are indicated by different points, with the publication years indicated in the legend. The two most discrepant secondary radial velocities, with large errors, originate from Andersen & Nordstrom (1983). Points with open symbols were not included in our final fit.

in Maxted & Hutcheon (2018). The differences in the masses and radii found in the present study with the values found by those authors are, in some cases, statistically significant. Using our uncertainties, we found our masses to be larger by $2.0\sigma$ (primary) and $3.4\sigma$ (secondary), and the radii to also be larger by $7.6\sigma$ (primary) and $3.0\sigma$ (secondary). The fractional differences are 7–10% in mass, and 3–8% in radius. Notably, those authors made use of five RVs that were not available to us at the time of analysis. A cursory analysis revealed those RVs are in very good agreement with our published solution. Those authors also noted HR 5934 as a triple system, but we do not find any evidence to support this claim.

### 4.2.  EPIC 204506777 / HD 144548

HD 144548 is a triply eclipsing system with masses and radii that have been determined for all three components (Alonso et al. 2015). Modeling this system is a complex task, and we do not duplicate the efforts of those authors here, but adopt the derived masses and radii in this study. We acquired four new epochs of Keck-I/HIRES spectroscopy, for which the data are publicly available through the Keck Observatory Archive[11]. It is worth noting that the most massive component of this system has a location in the MRD that provides powerful constraints on the age of Upper Sco.

Atmospheric parameters for each component of the triple were not published in A15, although those authors did note the use of K5V templates for the secondary and tertiary. Based on this, we adopt spectral types of K5 for both the secondary and tertiary, and effective temperatures of 4210±200 K for each star, based on the HH15 temperature scale. In § 6 we show that these parameters are consistent with the other systems studied here in the H-R diagram. We assign a spectral type of F7.5 to the primary and $T_{\rm eff}$=6210±80, again based on the HH15 relations. These parameters are in good agreement with a previous study (F8V, 6138 K; Pecaut et al. 2012).

### 4.3.  EPIC 203476597

EPIC 203476597 is an apparently single-lined system that was originally published in D16. A star of comparable brightness lies approximately 15 ″ to the southwest. We verified that EPIC 203476597, the brighter star to the northeast, is in fact the source responsible for the eclipses by extracting photometry for each source individually using small circular apertures.

Further, the primary star exhibits no detectable RV variations above the ∼1 km s$^{-1}$ level. However, an Hα emission component is observed to shift by up to ∼70 km s$^{-1}$, exactly in phase with the EB ephemeris. The primary star exhibits Hα in absorption, and depending on the epoch, the emission component is seen either blue-shifted or red-shifted relative to the primary. We believe it is likely that this emission signature is due to a component of the EB, but given that the primary star is apparently stationary (within the limits of our data), it is also possible that the primary star is not a component of the EB.

In this scenario, EPIC 203476597 may be a triple or higher order multiple. The EB may be at a wide separation from the primary star, which dominates the optical spectrum, explaining why no orbital motion is detected in the primary. In either case, the primary RVs exclude the scenario D16 proposed of a stellar mass companion on a close orbit with the primary star, as such a configuration would have led to an easily detectable RV signal. Further evidence in support of this interpretation comes from the rotation period determined from the K2 photometry. The most significant period found from a Lomb-Scargle periodogram analysis of the light curve is 3.2 days, compared with the EB orbital period of 1.4 days. If the primary was in fact a component of the EB, and assuming the rotational modulations in the light curve are due to the primary star, the discrepancy between the two periods would imply that the primary is not tidally synchronized.

To assess further multiplicity, speckle imaging observations of EPIC 203476597 were obtained using the NESSI instrument on the WIYN telescope on UT 2017 May 11. NESSI observes targets at two wavelengths simultaneously, at 562 nm (in a 44 nm wide filter) and 832 nm (in a 40 nm wide filter). The data were processed according to the procedures described in Howell et al. (2011) and Horch et al. (2017). These observations exclude additional companions to the system brighter than $\Delta m \gtrsim 3.0$ mag at 832 nm or $\Delta m \gtrsim 2.6$ mag at 562 nm in the angular separation range of 0.14–1.2″ (or 22–190 AU, given the system's distance). The fact that no additional source could be detected from speckle imaging calls into question the scenario in which EPIC 203476597 is a higher order multiple, e.g. a triple, in which the eclipsing pair is at a wide separation from the primary star. However, the imaging observations do not formally exclude this scenario, as there are regions of parameter space in which a hypothetical binary pair could remain undetected within our data.

In any event, because we do not detect orbital motion of two separate components in the spectra and because we do not know the optical flux ratio between the hypothesized components of the triple system, we are not able to derive masses and radii for this system. We stress that the interpretation of the secondary companion presented in D16 is incorrect. Until spectral lines outside of Hα can be detected for the EB components, fundamental masses and radii can not be determined. Thus we do not present a new analysis of the light curve here, and do not use EPIC 203476597 in assessing the age of Upper Sco or the accuracy of PMS models.

### 4.4.  EPIC 204432860 / RIK-60 / USco 48

USco 48 is newly reported as an EB in this work. The system has a combined light spectral type of M1 and it is located in the Upper Scorpius A region of the asso-ciation. It was first noted for its X-ray emission from *ROSAT* observations, and consequently given the designation USco 48 (Sciortino et al. 1998). It was later included in Preibisch & Zinnecker (1999), a classic reference on the association. From our Keck-I/HIRES spectra we confirm the M1 spectral type and note emission in Hα, Hβ, and the cores of the Ca II triplet, as well as lithium absorption. The system is double-lined, including the emission components.

Several multiplicity studies have targeted USco 48 searching for both close and wide companions, all resulting in null detections (Köhler et al. 2000; Kraus & Hillenbrand 2007; Kraus et al. 2008; Ireland et al. 2011; Lafrenière et al. 2014). Köhler et al. (2000) searched for but did not detect companions to this source with speckle interferometry and direct imaging. Those authors excluded companions with $K$-band flux ratios $F_2/F_1 > 0.12$ ($\Delta K < 2.30$ mag) at $0.13''$ (19 AU) or $F_2/F_1 > 0.05$ ($\Delta K < 3.25$ mag) at $0.5''$ (72 AU). Searches for wide companions in 2MASS (Kraus & Hillenbrand 2007) as well as close companions from Brγ imaging at Keck (Kraus et al. 2008) also resulted in null detections. Given the apparent lack of a relatively bright, closely-projected companion we choose to ignore third light in the eclipse modeling of this system.

The *K2* light curve of USco 48 is characterized by a semi-sinusoidal waveform of approximately 10% in amplitude and period of $P_{rot} = 2.8745$ days, presumably due to rotational modulation of starspots.[12] In phase with this rotation signal are grazing eclipses of ∼1% depth (primary eclipse) and ∼0.5% depth (secondary eclipse). The orbit of the binary is thus inferred to be tidally synchronized (see Figure 7), which is discussed further in § 5. We remove this variability prior to fitting the eclipses. USco 48 was also observed during Campaign 15 of the *K2* mission, the lone EB in this sample to have observations from both campaigns. While we do not make use of the Campaign 15 light curve here, we note that it reveals evolution in the spot pattern, which was stable throughout Campaign 2.

We performed three fits with JKTEBOP to the eclipse photometry and RVs corresponding to the three cases outlined in § 3. In all cases we assumed a quadratic limb darkening law with coefficients of $u_1 = 0.6034$, $u_2 = 0.1506$ for the primary, and $u_1 = 0.5607$, $u_2 = 0.1923$ for the secondary (Claret et al. 2012). The joint fit of the *K2* photometry and HIRES RVs for USco 48 is shown in Figure 8. We note that there is increased scatter in the eclipse centers, which might plausibly be related to eclipses of active regions on both of the stars. Despite the grazing eclipses, we were able to measure the masses and radii of USco 48 AB with 2–3% precision. Solutions for the masses, radii, and orbital elements are presented in Table 10.

---

[12] Kiraga (2012) previously identified this period from ASAS photometry.

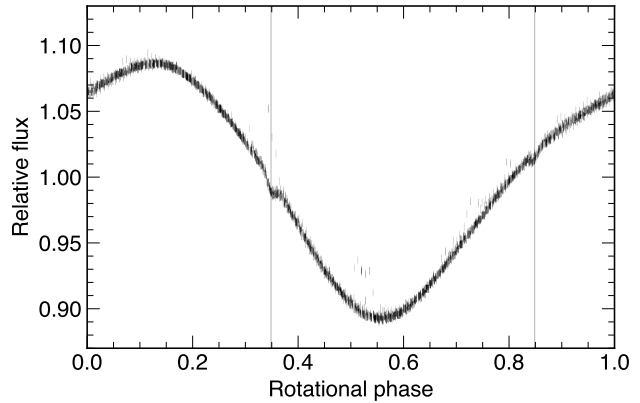

**Figure 7.** *K2* photometry of USco 48 phase-folded on the rotation period, which is commensurate with the binary orbital period. The shallow, grazing eclipses are highlighted by dotted lines. The binary is inferred to be tidally synchronized at a period of 2.8 d and age $\lesssim$10 Myr.

Finally, we note that USco 48 has a modest 24 μm excess at the ∼50% level from *Spitzer*/MIPS observations (Carpenter et al. 2009). There is no observed excess at 8 μm or 16 μm (Carpenter et al. 2006), or in the $W2$, $W3$, or $W4$ bands, leading Luhman & Mamajek (2012) to suggest the system hosts a debris or evolved/transitional disk. Barenfeld et al. (2016) studied the source with ALMA, determining an upper limit to the mass of dust in any putative disk of $M_{dust}/M_\oplus < 0.11$. USco 48 may therefore be a particularly interesting target for future high-contrast imaging programs aiming to study young, self-luminous planets around binary stars, especially in light of the dynamical masses determined here.

### 4.5. *EPIC 205207894 / RIK 72*

RIK 72 is newly reported as an EB here. The primary was first identified and spectroscopically confirmed as a member of Upper Sco in Rizzuto et al. (2015). Those authors assigned a spectral type of M2.5. Our Keck-I/HIRES spectra confirm this spectral type and reveal emission in Hα, Hβ, and the cores of the Ca II triplet, as well as strong lithium absorption. The system is single-lined in our optical spectra. Optical speckle imaging with DSSI did not reveal any additional companions to RIK 72, down to contrasts of Δmag≈4–5 in the angular separation range 0.1–1.37″ (Fig. 2). These constraints effectively rule out most stellar mass companions in the physical separation range of ∼15–200 AU, according to the BHAC15 models (Baraffe et al. 2015). Consequently,

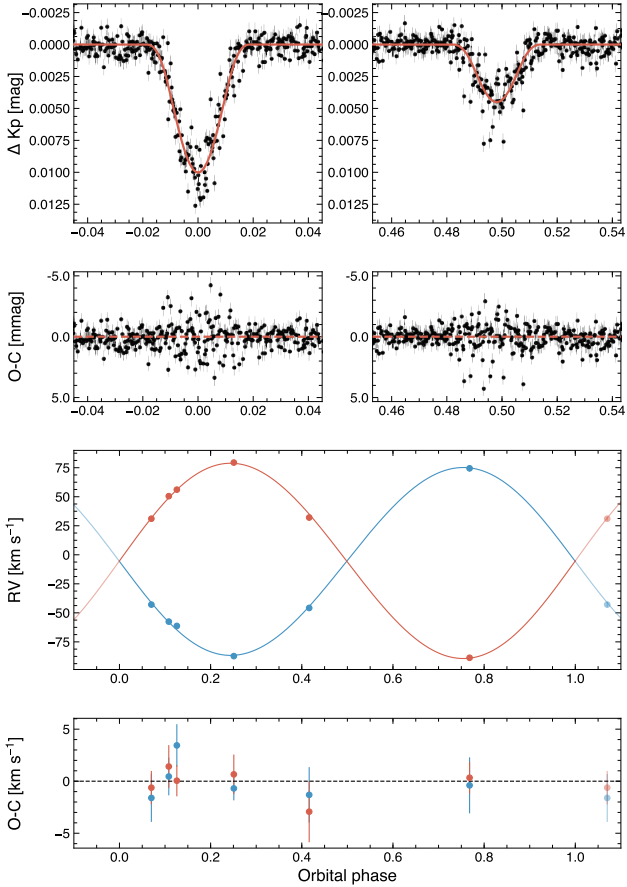

**Figure 8.** Joint fit of *K2* photometry and HIRES RVs for USco 48. In this fit the surface brightness ratio was fixed at unity.

we assumed no third light contributions in our modeling of the eclipses.

The *K2* light curve of RIK 72 is characterized by a semi-sinusoidal waveform of 2% semi-amplitude, presumably due to rotational modulation of starspots on the primary (Figure 1). Assuming this interpretation is correct, the rotation period of the primary is $P_{rot} = 10.47$ d. The waveform evolves throughout the campaign, perhaps indicating differential rotation or evolution of the spot distribution. The light curve shows a single primary eclipse of $\sim 10\%$ depth and a single secondary eclipse of $\sim 2\%$ depth (Fig. 1). As a result, the period is poorly constrained by the *K2* light curve. Furthermore, because only one primary eclipse and secondary eclipse were observed by *K2*, it is possible that these are actually eclipses by two distinct objects on wide orbits. We consider this interpretation unlikely given that the eclipse durations are so similar, and in fact well fit by a circular orbit model. In the present study we will consider only the EB scenario and assume the primary and secondary eclipses are associated with the same companion.

We first performed a fit to the *K2* eclipse photometry assuming a circular orbit. Given an apparent orbital period >75 days this assumption is not justified, but since neither the period nor the eccentricity are strongly constrained by either the light curve or the RVs this exercise allowed us to provide initial estimates of the other parameters of interest, namely the radius ratio and surface brightness ratio. For the JKTEBOP fit we assumed quadratic limb darkening coefficients of $u_1$=0.5187, $u_2$=0.2515 for the primary and $u_1$=0.9001, $u_2$=-0.0735. The results of this fit are presented in Table 11.

Next, we performed a fit to only the RVs using the RADVEL code (Fulton et al. 2018).[13] In this fit, we imposed strong priors on the time of conjunction and the time of secondary eclipse, both of which were measured precisely in the light curve fit described above. We allowed eccentricity and the longitude of periastron as free parameters in this RV fit, and additionally allowed an RV jitter term to account for stellar variability. The parameter space probed by the RVs was sampled using the emcee implementation of the MCMC algorithm (Foreman-Mackey et al. 2013). We used 20 walkers to sample the parameter space and convergence was assessed using the Gelman-Rubin statistic. The results from the RV fit are presented in Table 12.

Finally, we performed a joint fit of the light curve and RVs using GP-EBOP (Gillen et al. 2017a). This method uses Gaussian processes coupled with an EBOP model to simultaneously model the stellar variability and eclipses. In this last fit we allowed the eccentricity and periastron longitude to be free parameters.

We ultimately adopt the parameters from this fit, which are presented in Table 11.

Unlike the systems discussed to this point, RIK 72 does not sit above the single star sequence in a color-magnitude diagram. This observation is congruent with the scenario proposed below in which the eclipsing companion to RIK 72 is substellar in mass, contributing little optical flux. Given that RIK 72 is intermediate to USco 48 and UScoCTIO 5 in spectral type and CMD position, and by extension in mass, we can conservatively assume a range of plausible masses and radii for the primary of $M_1 \sim 0.3$–$0.7$ $M_{\odot}$ and $R_1 \sim 0.85$–$1.15$ $R_{\odot}$. We can characterize the primary better by using empirical relations derived later in §A. Based on the *Gaia* $G$-magnitude and parallax, we estimate the mass and radius of RIK 72 A as $M_* = 0.439 \pm 0.044~M_\odot$ and $R_* = 0.961 \pm 0.096~R_\odot$, respectively. We have assumed *ad hoc* uncertainties of 10% on each parameter. For the primary effective temperature, we adopt the value resulting from the SED fit using BT-Settl models and assuming a uniform prior on $A_V$.

Based on the primary radius, and the distribution of radius ratios resulting from the EB light curve fit, we determined the secondary radius to be $R_2 = 0.318 \pm 0.032~R_\odot$ or $R_2 = 3.10 \pm 0.31~R_{\rm Jup}$. Then, over a dense grid of primary $T_{\rm eff}$ and companion $T_{\rm eff}$, we calculated the expected surface brightness ratio in the *Kepler* band by convolving blackbody curves of the corresponding temperatures with the instrumental response. We then inferred the companion $T_{\rm eff}$ through a 2D linear interpolation within this grid using the observed distribution of surface brightness ratios resulting from the light curve fit and a plausible distribution for the primary $T_{\rm eff}$ as input. The resulting companion temperature is $T_{\rm eff,2} = 2660 \pm 50$. This estimate, based on the assumption of blackbodies for both the primary and secondary, is actually in relatively good agreement with our results from SED fitting ($T_{\rm eff,2} = 2720 \pm 100$ K).

In §6 we compare the derived properties of RIK 72 b with evolutionary models, and show that these models produce a remarkably self-consistent picture of a ~50 $M_{\rm Jup}$ brown dwarf aged between 5–10 Myr. To our knowledge, RIK 72 b constitutes the youngest example of a brown dwarf transiting a stellar host. A younger pair of eclipsing brown dwarfs were discovered in the Orion Nebula (Stassun et al. 2006), but that system does not contain a star. There are relatively few known transiting brown dwarfs, most of which have orbital periods ≲30 days around stars more massive than RIK 72 (Csizmadia 2016). The only known example of a transiting brown dwarf with period ≳100 days is KOI-415 b, which orbits an evolved solar-type star (Moutou et al. 2013). Additionally, RIK 72 b is only the third transiting brown dwarf with a known age through cluster membership, the other two being AD 3116 b in the Praesepe cluster (Gillen et al. 2017a) and EPIC 219388192 b in the Ruprecht 147 cluster (Nowak et al. 2017).

Finally, we note that RV monitoring on four consecutive nights in July 2017 revealed an RV trend with an apparent period in the ~17 day range based on its slope. Given that the star is young and demonstrates relatively large intrinsic stellar variability, we do not present a detailed discussion of the 17 day signal here. Our RV fit effectively models out additional RV variations up to 1 km s$^{-1}$ in amplitude through the jitter term, and so the uncertainty in the mass of RIK 72 b accounts for either stellar activity or the presence of an additional companion of a reasonable mass. Further RV monitoring will determine the nature of that signal and refine the mass and eccentricity of RIK 72 b.

### 4.6. *EPIC 202963882*

EPIC 202963882 is newly reported as an EB in this work. The system has not been previously studied in the literature, and we establish its membership to Upper Sco here. The system's distance and kinematics are highly consistent with membership to the association (Table 3) and our Keck-I/HIRES spectra reveal Hα and Hβ emission as well as strong lithium absorption. The system is observed to be double-lined from multiple spectra and has a combined light spectral type of M4.5.

Keck-I/HIRES guider camera imaging in 0.4″ seeing revealed the system is a visual binary with NE and SW components. For all spectroscopic observations in which the components of this visual binary could be resolved, the slit was rotated to include both components. In good seeing conditions we were able to obtain spatially resolved spectroscopy of the two components in the visual binary. From these observations, we were able to determine that it is the fainter of the pair (the SW component) that is a spectroscopic binary and the EB. The primary component has a systemic RV that is consistent with that of the EB, and with the nominal Upper Sco value. Thus, we define EPIC 202963882 A as the primary star and EPIC 202963882 Ba and EPIC 202963882 Bb as the components of the double-lined EB. Both components of the EB appear to be rapidly rotating, consistent with expectations that the binary is tidally locked. The RVs of this system are more uncertain due both to the rapid rotation of the EB as well as the fact that the primary and the EB were not spatially resolved in all observations. Speckle imaging with DSSI at Gemini observatory resolved the companion at a separation of 1.22" and position angle of 204.1°, with a contrast of $\Delta m = 1.65$ mag at 692 nm, or $\Delta m = 0.99$ mag at 880 nm. The speckle imaging data were acquired at approximately 2016-06-22 3:55:00 UT, corresponding to an orbital phase of ≈0.68 for the EB so that our measurements should accurately reflect the out-of-eclipse contrast between the EB and the primary. Given the trigonometric distance to the system, we calculated the minimum physical separation between the primary and the EB is $175.6 \pm 2.3$ AU.

The *K2* light curve for EPIC 202963882 reveals eclipses with a period of 0.63 d. The EB is semi-detached, as indicated by the out-of-eclipse ellipsoidal modulation in the *K2* light curve. In addition to the el-lipsoidal modulation, the out-of-eclipse brightness max-ima are unequally high. This phenomenon, known as the O'Connell effect, is still unexplained (O'Connell 1951; Wilsey & Beaky 2009). More specifically, the system apparently exhibits the negative O'Connell effect, where the maximum after the secondary minimum is brighter. The ellipsoidal modulations indicate that the stars in the close binary are tidally deformed and that the spherical geometry approximation of JKTEBOP is not well-suited for modeling the data. Nevertheless, we perform a preliminary analysis of this system with the aim of crudely approximating the masses and radii and leave a more detailed analysis of the system for a later work. We note that North & Zahn (2004) investigated the effect of non-sphericity within massive EBs by comparing the EBOP code (upon which JKTEBOP is built) with the Wilson-Devinney code, finding the spherical approximation of EBOP compromises the radii by $\sim$5% at an average fractional radius of $R_*/a \sim 0.3$, a value close to that for the EPIC 202963882 EB. Thus, it is safe to assume that the preliminary masses and radii we report are uncertain to at least the 5% level. Due to the difficulties of modeling this system, we only report a preliminary fit from JKTEBOP and do not report statistical uncertainties on the fit parameters. We additionally introduced three new free parameters in order to achieve a good fit: third light ($l_3$) to account for dilution from the wide companion star, a photometric mass ratio ($q_{\rm phot}$) which does not necessarily reflect the true binary mass ratio but is used to simulate ellipsoidal modulation, and a light scale factor ($s$) which affords flexibility in the median out-of-eclipse light level. In modeling the eclipses, we initialized the third light value to $l_3 = 0.82$ as implied by the 692 nm contrast found in the speckle imaging observations (the final value was $l_3 = 0.81$). We assumed a quadratic limb darkening law with coefficients of $u_1$=0.5607 and $u_2$=0.1923 for each component. The parameters of this preliminary fit are presented in Table 14 and the fit is shown in Figure 10. Surprisingly, despite the short orbital period, the light curve and RVs are best fit by including modest eccentricity ($e$=0.04).

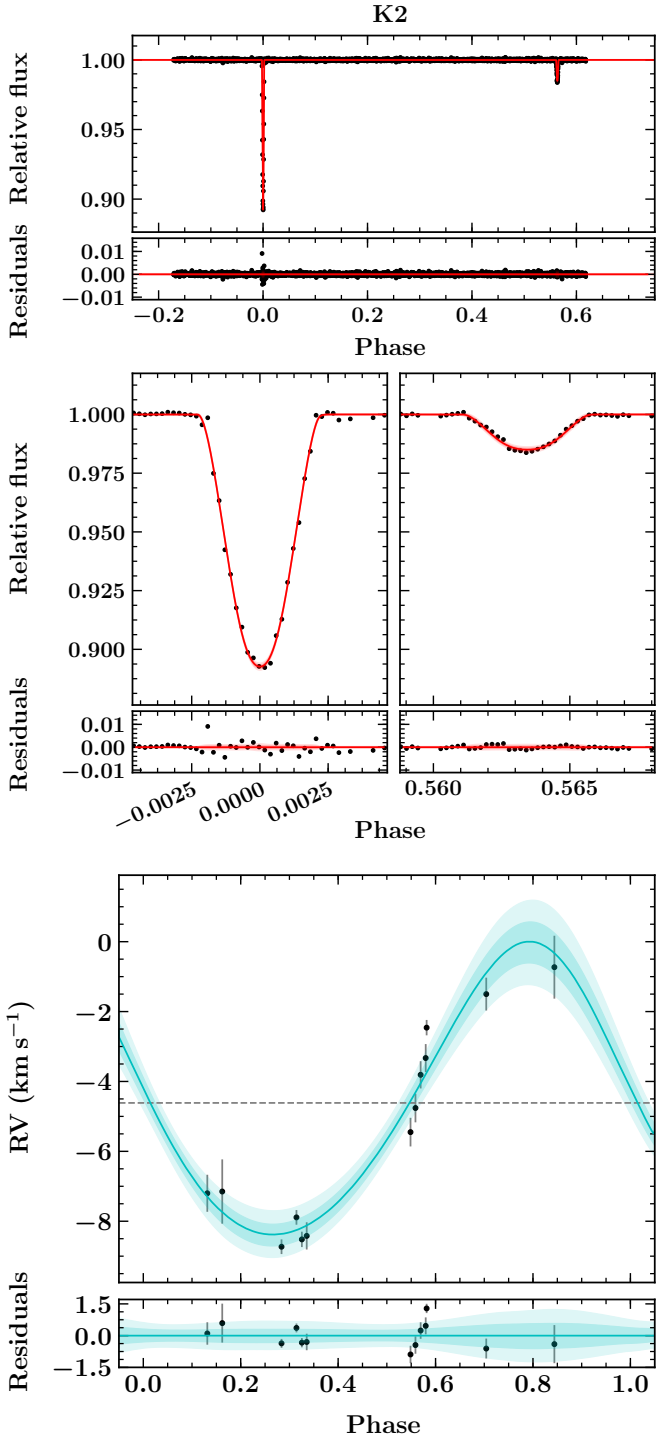

**Figure 9.** Joint GP-EBOP fits to the *K2* photometry and HIRES RVs for RIK 72. The top panel shows the full *K2* light curve after removing stellar variability and the fit residuals. The middle panels show detailed views of the primary and secondary eclipses, along with the fit residuals. The bottom panel shows the phased RVs with the best fit curve in teal, along with $1\sigma$ and $2\sigma$ shaded contours.

### 4.7. *EPIC 205030103 / UScoCTIO 5*

UScoCTIO 5 is comprised of two low-mass ($\sim$0.3 $M_\odot$) stars with nearly equal masses and radii, and a full solution was first published by K15 using archival RVs from Reiners et al. (2005). A parallel analysis was performed by D16 of the system using the same RVs but a different detrending of the *K2* lightcurve. In this analysis, we perform updated fits using the EVEREST 2 light curve, additional RVs at phases that were previously not covered, and assuming equal radii. As in K15 and D16, several photometric eclipse measurements were ex-cluded from our analysis due to a likely a spot-crossing event which induced a change in the eclipse morphology. We adopted quadratic limb-darkening coefficients of $u_1 = 0.4764$ and $u_2 = 0.3137$, based on Claret et al. (2012) calculations for stars with $T_{\mathrm{eff}} = 3400$ K and $\log g = 4.0$ observed in the *Kepler* bandpass. This fit is shown in Figure 11. Compared with the solution presented in D16 we find masses and radii that are consistent within 2% (the new masses are about $3\sigma$ larger but the radii are consistent within $1\sigma$).

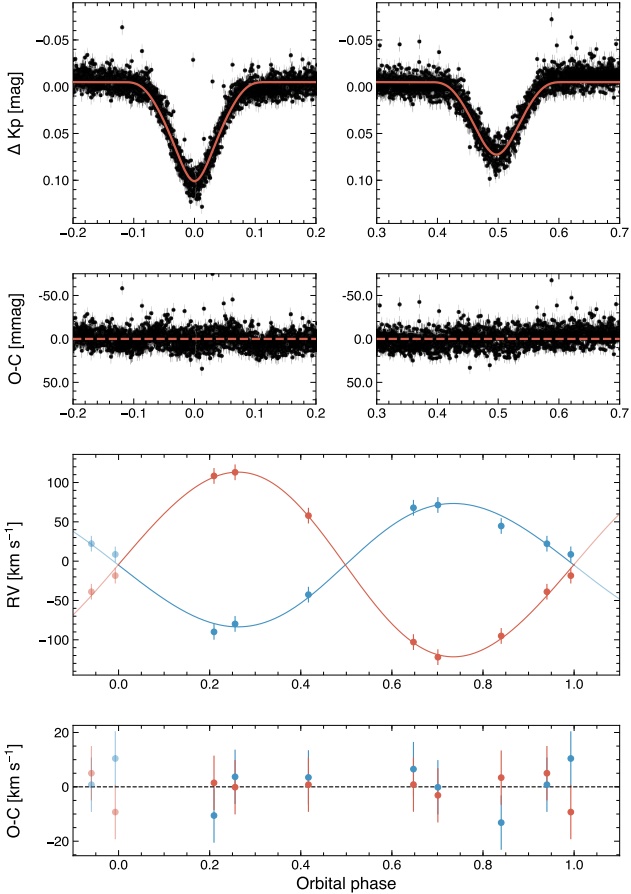

**Figure 10.** Joint fit to the *K2* light curve and Keck-I/HIRES RVs for EPIC 202963882.

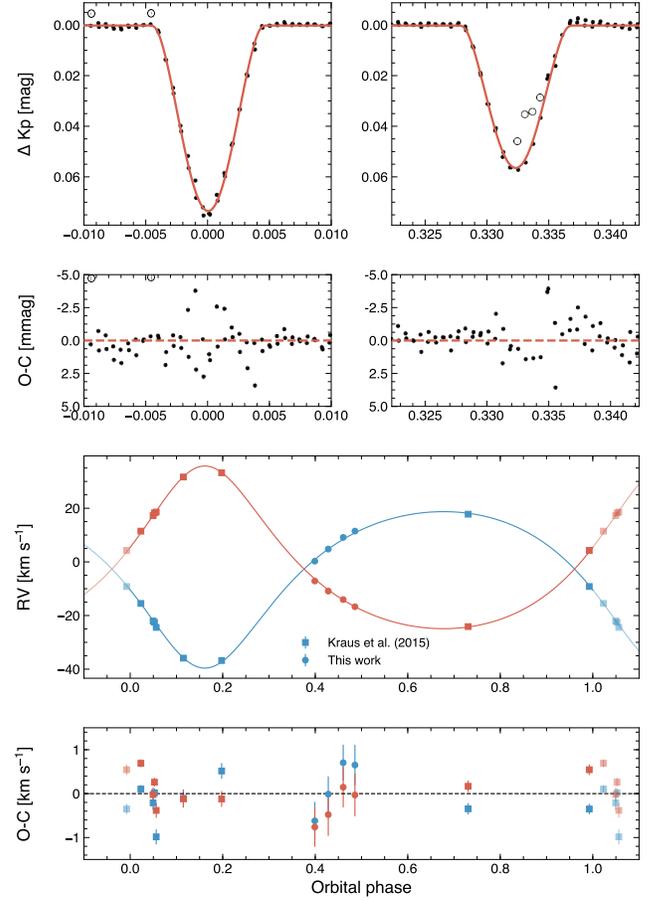

**Figure 11.** Joint fits to the *K2* photometry and radial velocity time series of UScoCTIO 5. In the top panels, open circles indicate eclipse observations excluded from the fitting. These discrepant points are likely due to a starspot crossing.

### 4.8. *EPIC 203868608*

EPIC 203868608 was first published in D16 as a possible eclipsing brown dwarf binary in a hierarchical triple system (a companion was discovered in Keck-II/NIRC2 imaging at a separation of $\rho = 0.12''$). Those authors noted residuals of $\sim$10 km s$^{-1}$ in the RV fits, which led to some trepidation in this interpretation. From sustained RV monitoring with Keck-I/HIRES we are able to confirm that the RVs presented in D16 are in fact due to a spectroscopic binary with an orbital period distinct from that of the eclipsing system. The initial difficulty in characterizing the system is in part due to the fact that a CCF analysis of the HIRES spectra reveals only two obvious peaks, despite the fact that the two components of the visual double have nearly-equal NIR brightnesses. Since the two peaks revealed by the CCF analysis of the HIRES spectra correspond to the SB2 and not the EB, precise masses and radii could not be determined from our data.

With JKTEBOP we performed joint fits to the RV time series alone in order to determine orbital and physical parameters of the SB2. We present these parameters in Table 15, where the uncertainties were determined from 10,000 Monte Carlo simulations. We show fits to the RV time series in Figures 12. We find a minimum system mass of $(M_1 + M_2) \sin^3 i = 0.3685 \pm 0.0050 \, M_\odot$. Assuming the expected value of $\langle \sin^3 i \rangle = 3\pi/16$ implies a total system mass of $(M_1 + M_2) \sim 0.63 \, M_\odot$.

From resolved infrared spectroscopy using NIRSPEC+AO on the Keck II telescope, it was determined that the brighter, eastern component of the visual binary is the SB2 (Wang et al. 2018). The fainter component to the west is thus the EB. The systemic RV of the EB component in our NIRSPEC+AO observations is approximately $\gamma = -5.2 \pm 1.1$ km s$^{-1}$, consistent with the systemic RV of the SB2 and the mean value for Upper Sco members. The agreement between the systemic velocities of the EB and the SB2, combined with the close angular separation on the sky suggests that all four components are physically associated. EPIC 203868608 is thus a hierarchical quadruple system. EPIC 203868608 A is an SB2 with a period of $P_A = 17.9$ days, and EPIC 203868608 B, at a minimum physical separation of 19.3 AU from the SB2, is an EB with period of $P_B = 4.5$ days. The orbital architecture of EPIC 203868608 is similar to that of LkCa 3, another PMS hierarchical quadruple of M-type stars (Torres et al. 2013). Such a 2+2 hierarchy is the most common architecture for quadruple stars in the field (Tokovinin 2008).

The NIRSPEC+AO observations were scheduled to coincide with eclipses to measure the obliquity of the EB. Thus, while both the SB2 and EB are spatially resolved, we were not able to resolve the individual velocities of the EB components. As such, fundamental masses and radii can not be determined from these data. The line profile of the EB pair in the NIRSPEC+AO observations is broadened to ∼20 km s$^{-1}$ in FWHM, indicating at least one of the components is a fast rotator. A Lomb-Scargle periodogram analysis of the *K2* lightcurve reveals two significant periods, $P_1 = 5.64$ days and $P_2 = 1.11$ days. Neither of these periods correspond to the orbital period of the EB, which is $P_{\rm EB} \approx 4.5$. d, nor do they coincide with the orbital period of the SB2, $P_{\rm SB2} \approx 17.9$ d. If the smaller of the two rotation periods corresponds to one or both components of the EB, this might explain the apparently rapid rotation of the EB observed in the NIRSPEC+AO data. However, it is also possible that this period is due to a component of the SB2. Detailed modeling of the NIRSPEC+AO data and an obliquity measurement for EPIC 203868608 B are presented in Wang et al. (2018).

### 4.9. *EPIC 203710387*

EPIC 203710387 is an eclipsing pair of very low-mass stars (∼0.1 $M_\odot$) concurrently discovered and published in L15 and D16. The component stars are near the edge of the substellar boundary and thus provides an important anchor to the mass-radius relation of young stars at the very lowest masses. We re-analyzed this system using the same RVs reported in D16 but with the new EVEREST2 light curve, which has less systematic noise than previous light curves. We did not use the RVs presented in L15 due to large discrepancies between those observations and ours (Figure 13). It is also notable that the systemic velocity found by L15 ($\gamma = 0.6 \pm 1.0$ km s$^{-1}$) is incompatible with ours, and not consistent with the mean RV of Upper Sco. Our updated parameters are presented in Table 16, including results for the assumptions of equal radii and equal surface brightness. As with USco 48 and UScoCTIO 5, we ultimately adopt the solution assuming equal surface brightness since this assumption is not unreasonable for the nearly equal-mass system and leads to a more plausible ratio of radii. The overall impact of this decision on the primary conclusions of this work is negligible. Notably, the goodness-of-fit variations between the three fits are small.

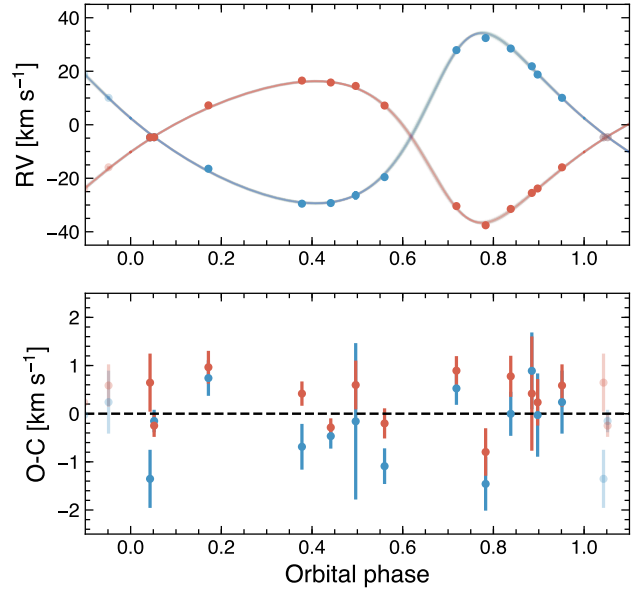

**Figure 12.** Joint fits to the radial velocity time series of the spectroscopic binary component of EPIC 203868608. The curves show fits using parameters from 100 randomly selected links in the Monte Carlo chain.

We note that the 2015 Nov 26 primary eclipse of EPIC 203710387 was observed by *Spitzer* (Program ID 11026, P.I. Werner). These observations are publicly available[14], but we do not make use of them here since a secondary eclipse would also be needed to better constrain the surface brightness ratio.

### 4.10. *Comparison with previous works*

Compared with the results of L15 for EPIC 203710387, we find masses that are $5\sigma$ (25–30%) larger and radii that are $5.5\sigma$ (11%) larger. For UScoCTIO 5, our masses are within 3% of those reported by K15, though even this small fractional difference corresponds to a $5\sigma$ difference. Our radii for UScoCTIO 5 are about 3–5% larger than the K15 values, corresponding to a 4.6–6.7$\sigma$ discrepancy. We also note the average flux ratio we measured for UScoCTIO 5 from the ratio of the CCF peak heights ($F_2/F_1 = 0.82 \pm 0.03$) varies significantly from the average of the flux ratios reported in K15 ($F_2/F_1 = 0.94 \pm 0.04$), which were also measured from HIRES data but using the ratio of the areas under the broadening function peaks. In general, the agreement with K15 is fairly good, and the fractional differences of $\lesssim 5\%$ are probably more reflective of the true uncertainties in EB parameters than the small statistical uncertainties that are often quoted in the literature.

---

[14] http://sha.ipac.caltech.edu/applications/Spitzer/SHA/

Both L15 and K15 used custom-written software to fit the EB light curves and RVs, whereas we have used JK-TEBOP. Thus, it is difficult to draw direct comparisons between the methods. For EPIC 203710387, the differences between the two studies can be understood at least in part as a result of the relatively large discrepancies in the RVs. The RVs presented here favor larger amplitudes, which at fixed orbital period yields larger masses and a larger semi-major axis. At a fixed value of the sum of the radii, a parameter typically well-determined by the light curve, the effect would be to drive down the radii, but we have found larger radii. Instead, the difference in radii we find may be due to covariance between the sum of the radii and the inclination. We find a lower inclination and larger sum of radii compared to L15.

## 5. DISCUSSION

### 5.1. *Prevalence of hierarchical multiples*

Of the nine EBs studied here, one resides in a quadruple star system (EPIC 203868608), two are definite hierarchical triples (HD 144548 and EPIC 202963882), and one more is a possible triple (EPIC 203476597). Thus, at least 30% of our sample are hierarchical multiples. This may be a consequence of observational bias. Smaller binary separations correspond to higher geometric likelihoods of eclipses and, as discussed below, the fraction of binaries with tertiary companions increases with decreasing orbital period. Tokovinin et al. (2006) found that the rate of tertiary companions to solar-type spectroscopic binaries (SBs) in the field rises steeply with decreasing SB period. For example, ∼40% of binaries with periods > 7 d have tertiary companions compared to ∼80% for $P < 7$ d or ∼96% for $P < 3$ d. The correlation between binary period and the presence of a tertiary has been interpreted as evidence for the formation of close binaries via eccentricity excitation through the Kozai-Lidov mechanism (Kozai 1962; Lidov 1962) followed by tidal circularization (Mazeh & Shaham 1979; Fabrycky & Tremaine 2007). If this interpretation is correct, the prevalence of hierarchical triples in our sample could be evidence that (1) the process is well underway for some systems by $\lesssim 10$ Myr, and/or (2) not all close binaries in hiearchical multiples form via Kozai-Lidov cycles and tidal friction. The second conclusion seems secure (though both may be true) since at least two of the close binaries studied here (USco 48 and EPIC 203710387) do not appear to presently have tertiary companions.

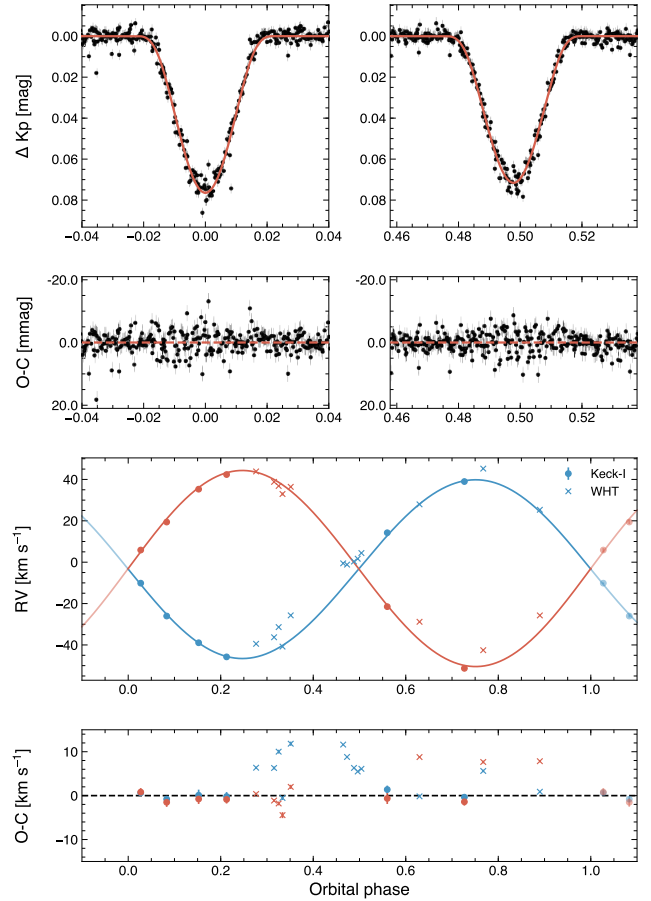

**Figure 13.** Joint fit of *K2* photometry and HIRES RVs for EPIC 203710387. In this fit the surface brightness ratio was fixed at unity. For reference, we display the RVs determined by Lodieu et al. (2015) from the ISIS spectrograph on the William Herschel Telescope (WHT), which were not included in our fit.

### 5.2. *Pre-main-sequence spin-orbit evolution*

Tidal dissipation tends to circularize binary star orbits, and synchronize the spin of each component to the orbit. In tidal equilibrium, the spin-orbit vectors of each star will also be aligned with the orbital angular momentum vector (see Ogilvie 2014, , for a review). This state can only be verified for eclipsing binaries, for which the spin-orbit angles of both stars may be directly measured. From the *K2* photometry and Keck-I/HIRES spectra, we can only comment on the eccentricity and degree of spin-orbit synchronization (by comparison of the orbital period with the rotation period as inferred from either photometric modulations, or the projected rotational velocity and measured stellar radius, or both). In Figure 14 we indicate where the binaries studied here reside in the period-eccentricity plane relative to a large catalog of spectroscopic binaries (Pourbaix et al. 2004) as well as a sampling of PMS binaries from the literature (Melo et al. 2001; Covino et al. 2001; Alencar et al. 2003; Mace et al. 2012; Torres et al. 2013; Schaefer et al. 2014; Stassun et al. 2014; Rizzuto et al. 2016).

At orbital periods less than 3 days, all of the binaries in our sample are on circular or nearly circular orbits (with small but measurable eccentricities). The binaries that fall into this category include the semi-detached EPIC 202963882 B system ($P \sim 0.6$ d, $e$=0) which has a tertiary companion, the EPIC 203476597 ($P \sim 1.4$ d, $e$=0.0) system which has an uncertain nature, the solar-type HD 144548 B binary which is in a highly compact triple system ($P \sim 1.6$ d, $e$=0.0), and finally the USco 48 ($\sim$0.74$M_\odot$+0.71$M_\odot$, $P \sim 2.9$ d, $e$ <0.02) and EPIC 203710387 ($\sim$0.1$M_\odot$+0.1$M_\odot$, $P \sim$2.8 d, $e \lesssim$0.01) systems which share very similar architectures. At periods greater than 4 days, all of the binaries in our sample are on eccentric orbits with the exception of the massive system HR 5934 discussed further below. While not conclusive, this is suggestive that for binaries with component masses $M_* \lesssim 1$ $M_\odot$, the PMS circularization period is around $\sim$4 days. Interestingly, a PMS circularization period of $\sim$4 days was already established using a sample of only 25 binaries by Mathieu (1994), although Melo et al. (2001) suggested a larger value around 7.6 days.

The theory of tidal dissipation predicts that synchronization is achieved more rapidly than circularization (Zahn 1977; Zahn & Bouchet 1989). In terms of spin-orbit synchronization, we observe one system that is clearly highly synchronized: USco 48. For that system, the orbital period and rotation period inferred from out-of-eclipse brightness modulations are indistinguishable (Figure 7). Other systems in our study are either not synchronized, or a conclusive determination is not possible from our data. For example EPIC 203710387 has a variability period that is $\sim$0.3 days shorter than the orbital period. If the difference in periods were due to surface differential rotation, the corresponding rate of differential rotation would be 0.23 rad day$^{-1}$. While rates of $\Delta\Omega \gtrsim$0.2 rad day$^{-1}$ have been observed in solar-type PMS stars (Dunstone et al. 2008; Marsden et al. 2011; Waite et al. 2011), the effect is expected to be significantly weaker for low-mass stars (e.g. Barnes et al. 2005; Collier Cameron 2007). It is possible that the difference between the rotational and orbital periods indicates the binary is not tidally synchronized. Given the youth of the system, this would not be surprising, and in fact supersynchronous rotation has been observed in PMS binaries (Melo et al. 2001; Gómez Maqueo Chew et al. 2012; Gillen et al. 2017b). The binary is also mildly eccentric, as is most evident from the light curve. Thus it is also possible that the binary is in fact in a state of pseudo-synchronous spin in which the rotation rates are synchronized to the orbital speed of the binary at periastron.

Similarly, the HD 144548 triple exhibits variability at a period about 0.095 days shorter than the orbital period of the tight binary. It is not clear whether this variability is due to the primary star, which dominates the optical flux from the system, or the secondary/tertiary. In either case, it seems likely that the period reflects that at least one star in the system is rotating supersynchronously, as would obviously be the case if the variability is due to the primary. The difference in the variability and orbital periods for the tight binary HD 144548 B corresponds to a differential rotation rate of 0.24 rad day$^{-1}$. Although a rate this high might be plausible for a hotter star, it is simply much higher than what is observed around stars of a similar temperature at field ages. The EPIC 203868608 quadruple system exhibits two variability periods, both of which are distinct from either the period of the EB or the SB2. One of the variability periods is 1.1 days. Regardless of whether this period is due to a component of the EB or SB2 it implies the responsible star is rotating super-synchronously since it is much shorter than either the EB or SB2 period. The other period detected in the light curve of EPIC 203868608 is 5.6 days, which could imply either sub-synchronous rotation if it is attributed to a component of the SB2 or super-synchronous rotation if it is due to one of the EB components.

The UScoCTIO 5 system is an interesting case with regards to synchronization and spin-orbit alignment. The binary has an orbital period of $P_{\rm orb} \sim 34.0$ days and a photometric variability period of $P_{\rm var} \sim 30.7$ days. Interestingly, the variability period of UScoCTIO 5 is one of the longest amongst Upper Sco members with *K2* observations (Fig. 15), based on the catalog of Rebull et al. (2018). Since this system is eccentric, it might seem plausible that the binary is in a pseudo-synchronous spin state where the rotational velocity of each star is commensurate with the orbital velocity at periastron passage. However, K15 found $v \sin i \sim 6.6$ km s$^{-1}$ for each component, which combined with the stellar radii and an assumption of a high inclination suggests the stars have rotation periods of ∼6 days. If this is the case, it is unusual that no such period is evident in the *K2* light curve, and the nature of the quite prominent signal at 30.7 days is entirely unclear. If the 30.7 day period is indeed due to rotation of one or both of the stars, then the implied rotational velocity is ∼1.4 km s$^{-1}$, significantly smaller than the value found by K15. Thus, the interpretation of the 30.7 day variability as arising from starspots is problematic. Two possible explanations, both of which were considered in K15, are as follows. One possibility is that the true rotation period of each star is indeed near 6 days, and the two periods are close enough to induce a beating pattern in the light curve which acts to hide the individual periods. Sustained photometric monitoring would be able to determine if this scenario is true. Another possibility is that the variability period traces the long-term evolution of a spot or spot grouping on one or both of the stars. If both components are viewed nearly pole-on and the starspots are confined to high latitudes (such that the visible hemispheres of the stars are nearly constant in brightness), then we might observe gradual changes in the spot pattern(s) while the true rotation periods might not be evident in the light curve. In this scenario, the stellar spin axes would be oriented nearly perpendicularly to the binary orbital plane. Such a scenario could be tested with Doppler tomography. An interesting, and perhaps related, observation is that the light curve for UScoCTIO 5 shows a large number of flares (Fig. 1), much more than for any other system studied here.

At the high-mass end, the orbit of HR 5934 appears to be circular with the primary rotating slightly super-synchronously. We measured $v \sin i = 25 - 30$ km s$^{-1}$ for the primary. As the inclination of the stellar rotation axis is unknown, this value represents the minimum equatorial velocity. However, for the directly measured primary radius, the spin and orbital periods become commensurate at $v_{eq} = 15.0$ km s$^{-1}$, hence our inference that the primary is rotating super-synchronously. While our measured $v \sin i$ is modest for a B-type star, its ratio with the synchronization velocity is consistent with other massive stars in close binaries (Abt et al. 2002). It is also interesting to note the existence of discrepancies in the reported $v \sin i$ for this star: 100 km s$^{-1}$ (Slettebak 1968), 14±2 (Brown & Verschueren 1997), and $5^{+9}_{-5}$ km s$^{-1}$ (Abt et al. 2002), compared with the 25–30 km s$^{-1}$ we report here. Variability over decades-long timescales in the projected rotational velocities of stars in a massive binary has previously been observed in the anomalous DI Herculis system, which has been attributed to gross misalignment of the stellar spin-orbit axes (Albrecht et al. 2009). That system shares some similarities with HR 5934, but notably has significant eccentricity. Given its brightness and young age, HR 5934 represents an intriguing system to study the early spin-orbit alignment of a massive binary through the Rossiter-McLaughlin effect (Rossiter 1924; McLaughlin 1924) or Doppler tomography.

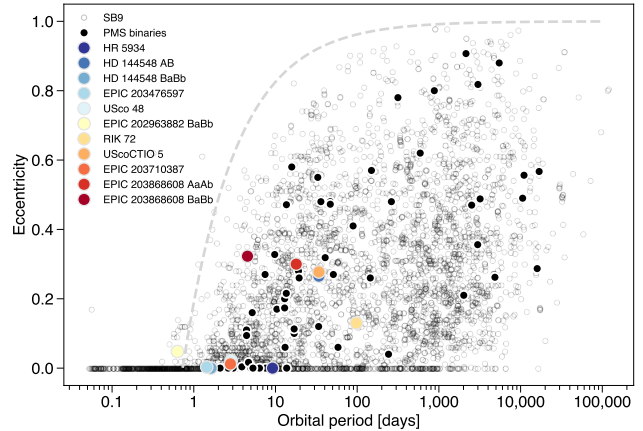

**Figure 14.** Period-eccentricity diagram for a catalog of spectroscopic binaries (Pourbaix et al. 2004) shown with open circles and EBs or SBs in Upper Scorpius (shaded circles). Other PMS binaries from the literature are indicated by filled black circles. For clarity, the eccentricities of USco 48 and UScoCTIO 5 has been offset by +0.01. Similar to Winn & Fabrycky (2015), the dashed line shows where the minimum orbital separation would be 0.02 AU for two solar-mass stars.

## 6. THE AGE OF UPPER SCORPIUS

We assess the age of Upper Scorpius from comparison of the newly characterized eclipsing binaries with theoretical predictions in the mass-radius and H-R diagrams. We make these comparisons with multiple stellar evolution models that are widely used, which are further discussed below. In Table 17 we summarize the final adopted parameters for those EBs used in the age assessment and in the evaluation of the stellar models.

### 6.1. *Summary of pre-main-sequence models*

Below, we summarize the basic properties of the PMS models considered here. We have limited our study to a manageable number of model sets, but we note that there exists a much larger number of PMS models in circulation. For each of the models considered here we adopted the solar metallicity tracks and isochrones, although different models adopt different heavy element mixtures, as discussed below.

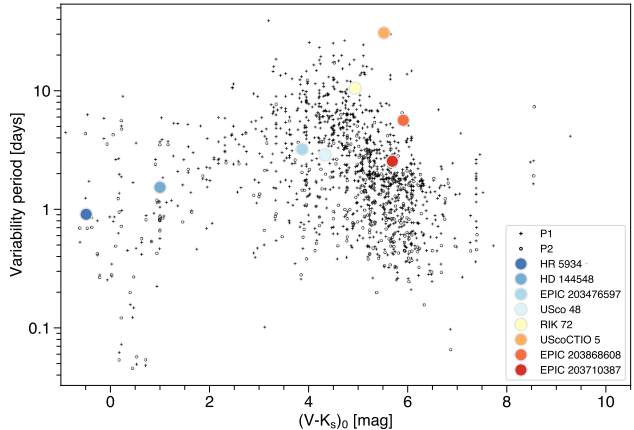

**Figure 15.** Period-color diagram. The black points indicate *K2* photometric variability periods as a function of dereddened $(V − K)$ color for known or candidate Upper Sco members, where the data originate from Rebull et al. (2018). The small crosses and open circles represent the first and second most significant periods from a periodogram analysis, respectively. The large shaded circles indicate the primary variability periods for the Upper Sco EBs studied here.

*BHAC15:* The BHAC15 evolutionary models (Baraffe et al. 2015) are an update to BCAH98 models (Baraffe et al. 1998), with the same input physics describing stellar interiors as the older model set but with new surface boundary conditions. While the BCAH98 models utilized NextGen model atmospheres (Hauschildt et al. 1999), the BHAC15 models use the updated BT-Settl models (Allard et al. 2012a,b). As with the BCAH98 models, the updated version uses the SCvH EOS (Saumon et al. 1995).

*Dartmouth:* The Dartmouth models (Dotter et al. 2008) are based on the Yale Rotating Stellar Evolution Code (Guenther et al. 1992). These models have been further developed in Feiden & Chaboyer (2012) and Feiden (2016), hereafter F16, to include the effects of magnetic fields. In short, magnetic fields inhibit convection, which in turn slows PMS contraction. Thus, as we will show, these models predict an older age in the MRD. We also note that the F16 models generally produce better agreement between the HRD and MRD.

*MIST:* The MIST models (Choi et al. 2016; Dotter 2016) are generated with the Modules and Experiments in Stellar Astrophysics (MESA) software (Paxton et al. 2011, 2013, 2015). MIST models are available with and without prescriptions for rotation. In the rotating models, solid-body rotation is commenced on the ZAMS, and the rate of rotation is gradually ramped up from zero to 40% the critical value ($\Omega/\Omega_{\rm crit} = 0.4$). Because rotation is commenced on the ZAMS in these models, there is no appreciable difference between the rotating and non-rotating models for low- and intermediate-mass stars during the pre-MS stages (Fig. 16). For more massive stars, such as the HR 5934 binary, there is a significant difference in the predicted ZAMS mass-radius relationship between the rotating and non-rotating models. Specifically, we find that the components of HR 5934 fall below the ZAMS in the mass-radius diagram when using the rotating models, but that this problem is alleviated when the non-rotating models are used instead. This is true regardless of whether our masses and radii are adopted or those reported in Maxted & Hutcheon (2018). In the analysis that follows, we will use the non-rotating models. The EOS tables used in MESA, for the solar metallicity case, are the OPAL tables (Rogers & Nayfonov 2002) with a transition to the SCvH (Saumon et al. 1995) at low temperatures and densities.

*PARSEC:* The PARSEC models (Girardi et al. 2000; Bressan et al. 2012; Chen et al. 2014) are available in three different versions (v1.0, v1.1, v1.2S), each of which makes substantially different predictions in both the MRD and HRD for PMS stars. The PARSEC v1.1 models adopt the grey atmosphere approximation as an external boundary condition, which relates the temperature, $T$, to the Rosseland mean optical depth, $\tau$. The most recent version of the models, PARSECv1.2 (Chen et al. 2014), adopt the PHOENIX BT-Settl model atmospheres as surface boundary conditions. The models are also available with an *ad hoc* shift in $T − \tau$ relations in order to reproduce the mass-radius relation of dwarf stars (PARSECv1.2S).

*SP15:* The Somers & Pinsonneault (2015) models, like the Dartmouth models, are based on the YREC evolution code. These models were specifically generated to investigate the effect of starspots on pre-main-sequence evolution. In the present work, we consider two versions of the models: one in which the stars are spot-free, and the other with spot covering fractions of 50%, representing a limiting case. Starspots impede the flow of energy near the surface, which causes the star to expand. Thus, the overall effect of starspots is similar to that of magnetic fields, and of course the two phenomena are related, in that they act to slow contraction in the PMS phase. It is notable that the SP15 models (both the spot-free and spotted versions) show more gradual contraction along the PMS than any other model set considered here. This is due to the fact that these models are commenced from the end of deuterium burning, which in turn means they bypass the rapid early phase of contraction. Furthermore, since the rate of deuterium burning is mass-dependent, this choice also affects the mass-radius slope (G. Somers, priv. communication).

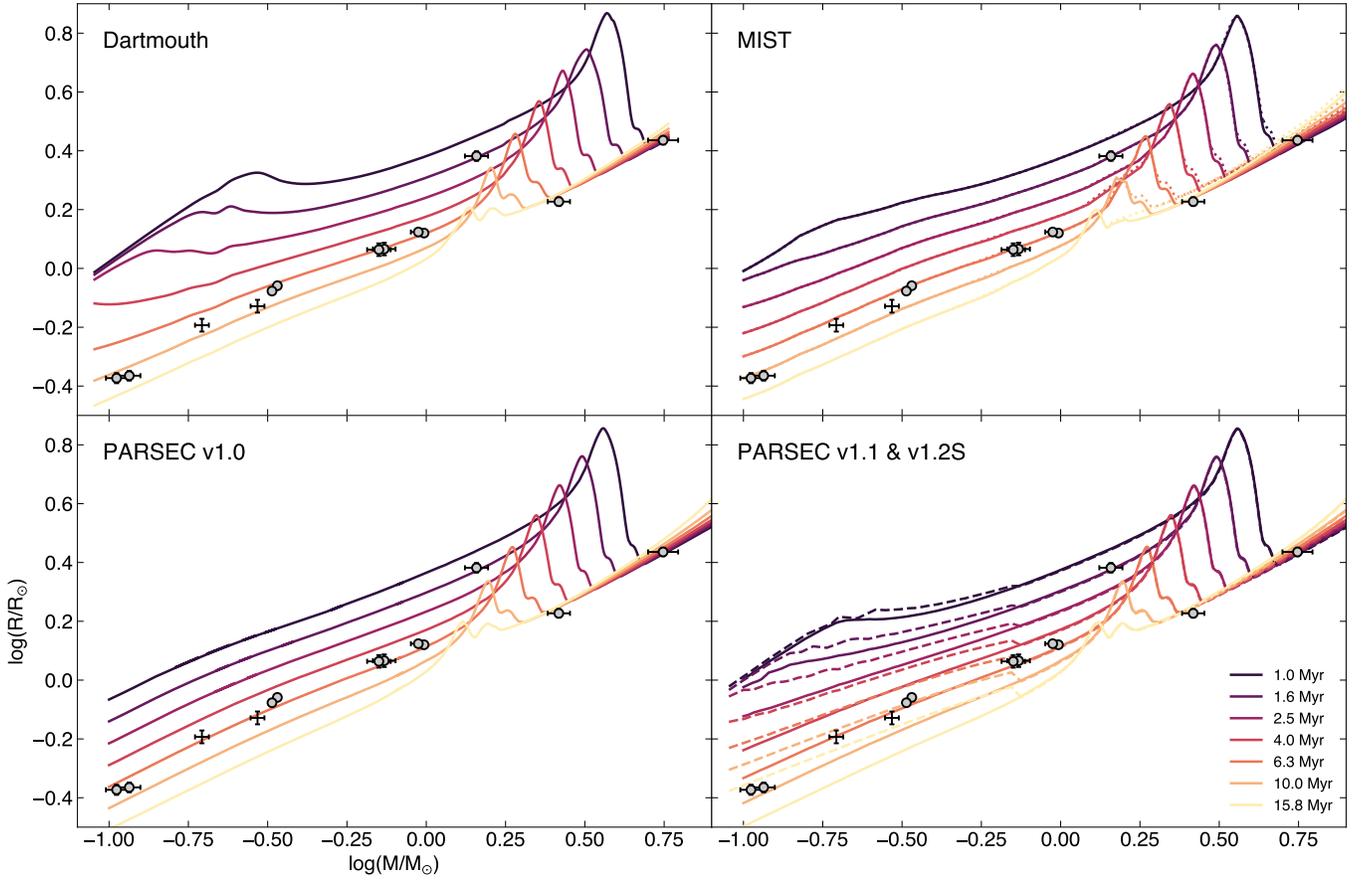

**Figure 16.** Mass-radius diagram. The shaded curves show theoretical predictions from various evolutionary models discussed in the text. The Upper Sco EBs are represented by the grey points, with the errorbars indicating the $3\sigma$ errors in mass and radius. The black crosses indicate a preliminary solution for EPIC 202963882 Bab, with 5% errors. From left to right, the EBs are EPIC 203710387 AB, EPIC 202963882 Bab, UScoCTIO AB, USco48 AB, HD 144548 Bab, HD 144548 A, HR 5934 B, and HR 5934 A. In the upper right panel, the dotted lines indicate the MIST isochrones including the effects of rotation. In the bottom right panels, the solid and dashed lines represent the PARSEC v1.1 and v1.2S models, respectively.

## 6.2. Age analysis in the mass-radius diagram

For each of the systems with fundamental mass and radius determinations presented earlier, and for each of the theoretical evolutionary models described above, we evaluated the age of Upper Sco through the following tests:

1. Using all of the individual masses and radii, we calculated $\chi^2$ over a fine grid of mass-radius isochrones for each model set. We performed this test for four cases corresponding to different mass ranges (Case 1: 0.1–6 $M_\odot$; Case 2: <1.5 $M_\odot$; Case 3: <1 $M_\odot$; and Case 4: 0.3–1.5 $M_\odot$). Each case ensured at least three EBs (or six stars) were included. Notably, HD 144548 A (the tertiary in the triply-eclipsing system) appears as an outlier in the MRD for all model sets and the 0.1+0.1 $M_\odot$ EPIC 203710387 pair appear as outliers for many, but not all, model sets. Case 3 excludes HD

144548 A from the age determination, and Case 4 excludes both that star as well as EPIC 203710387. We note some models considered here do not extend to high masses.

2. For each individual component of each EB, and for each model set, we derived an age distribution from the MRD and from the theoretical H-R diagram. This was performed using a 2D linear interpolation with the `griddata` routine in `scipy`, assuming normally distributed errors for the input parameters (mass and radius, or $T_{\rm eff}$ and $\log L_*/L_\odot$).

We present the best-fitting isochronal ages from our MRD analysis in Table 18. In Figure 18 we illustrate the variation of $\chi^2$ as a function of age for the different model sets and cases outlined above.

The mean $\chi^2_{\rm min}$ ages and standard errors across all models for Cases 1, 2, 3 and 4 are $7.3 \pm 2.0$ Myr, $7.8 \pm$

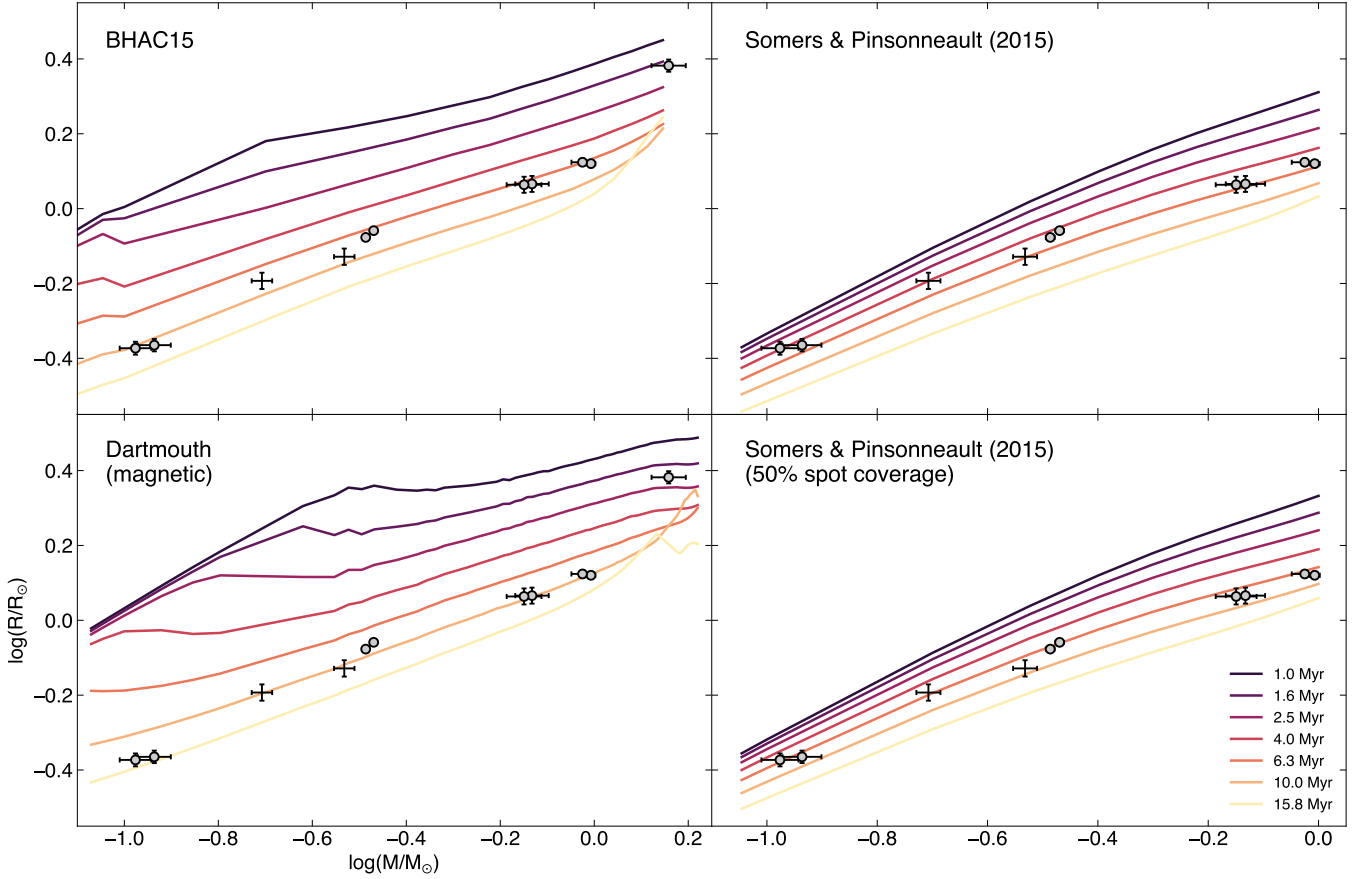

**Figure 17.** Mass-radius diagram. Same as Figure 16 but for the models that are only available at low masses.

2.1 Myr, 7.0 ± 1.4 Myr, and 6.6 ± 1.2 Myr, respectively. Considering all cases together the average age is 7 ± 2 Myr. The median of all the individual stellar ages derived from the MRD is $6^{+3}_{-2}$ Myr or $6^{+6}_{-2}$ for ages derived in the HRD, where the quoted errors correspond to the 16th and 84th percentiles.

We investigated mass-dependent trends in the inferred ages of individual stars in the MRD and present these results in Table 20 and in Figure 21. The SP15 models are best able to reproduce the mass-radius relation between 0.1–1 $M_\odot$ with a single age. The PARSEC v1.0 and v1.1 models are also able to produce a self-consistent age between 0.1–5.5 $M_\odot$, excluding the outlier HD 144548 A. However, the PARSEC v1.2S, MIST, BHAC15, and Dartmouth models (both the standard and magnetic versions) all exhibit a trend where the age of the low-mass anchor EPIC 203710387 is significantly older than the ages implied by the other systems. The effect is particularly pronounced for the PARSEC v1.2S, where there is a visible discontinuity in the mass-radius relation near 0.75 $M_\odot$ (Figure 16).

If current models of the PMS evolution of BAF stars are accurate, then the properties of HR 5934 B set a firm lower limit to age of Upper Sco. This is evident in the H-R diagram (Fig. 20) and even moreso in the MRD. HR 5934 B appears to have just completed the brief phase of radius inflation, and any age younger than 5 Myr would be strongly ruled out by the star's directly measured radius and well-determined luminosity. Thus, we argue that 5 Myr should be considered a firm lower limit to the age of HR 5934 and, by extension, Upper Sco.

### 6.3. Age analysis in the H-R diagram

We determined model-dependent masses and ages for the well-characterized EB components in the theoretical H-R diagram. For this exercise we used the radii derived earlier and summarized in Table 17, the temperatures derived from fitting BT-Settl models to the SEDs (Table 5), and bolometric luminosities calculated from the Stefan-Boltzmann law. For each component of each well-characterized EB, the mass and age was calculated by performing a 2D linear interpolation between evolutionary model grids assuming uniform distributions in $\log(T_{\rm eff})$ and $\log(L_{\rm bol})$. The individual ages are summarized in Table 21, and the offsets between model-derived

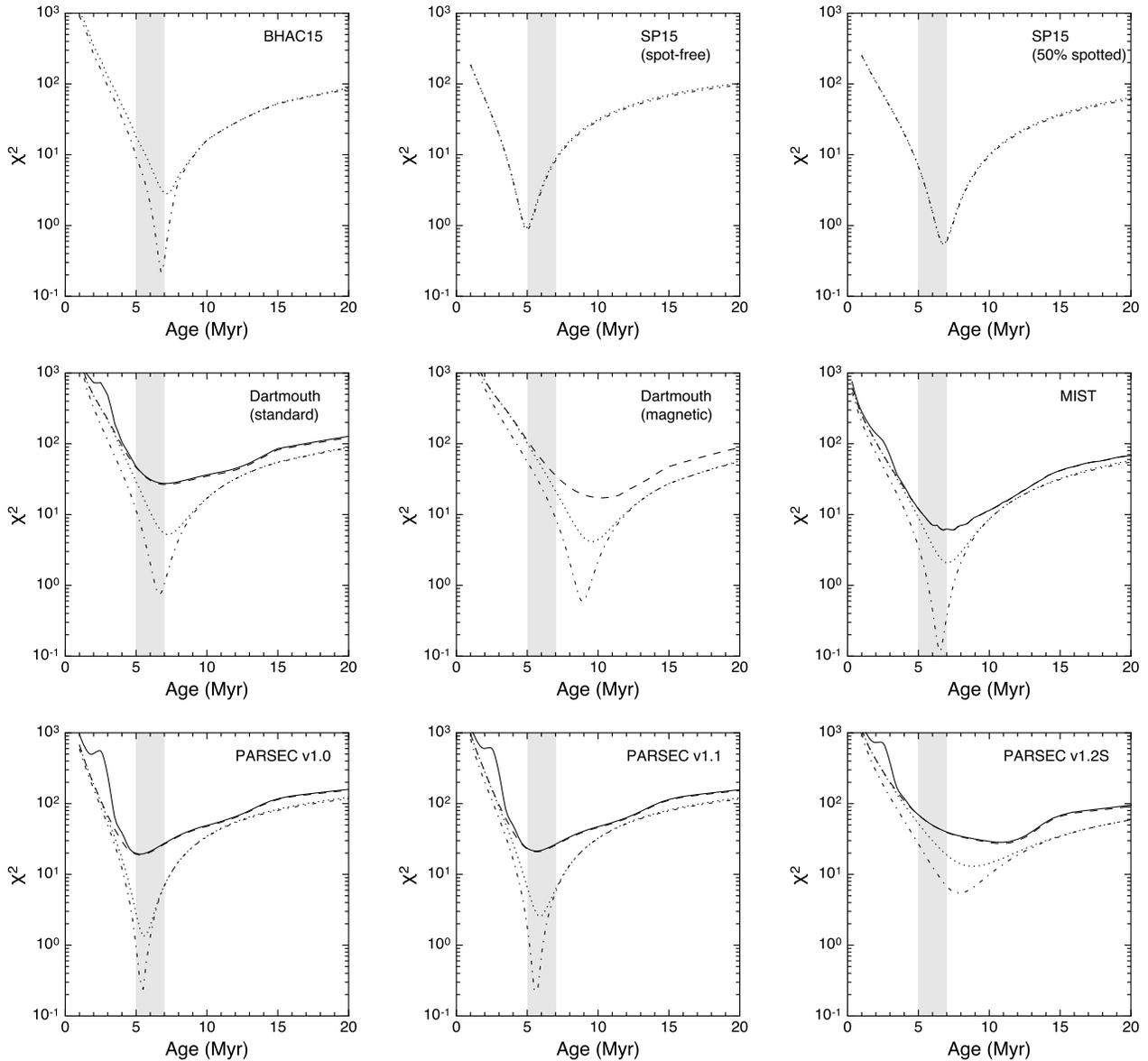

**Figure 18.** Chi-squared statistic for mass-radius isochrones from various model sets. The solid, dashed, dotted, and dash-dotted lines represent Cases 1 (all EBs), 2 ($M_* < 1.5\ M_\odot$), 3 ($M_* < 1\ M_\odot$), and 4 ($0.3\ M_\odot < M_* < 1.5\ M_\odot$), respectively. For reference, a fiducial shaded band from 5–7 Myr is shown in each panel.

masses and dynamical masses are summarized in Table 22.

In Figure 21, we demonstrate the relationship between the ages of individual EB components determined from the mass-radius (MRD) and H-R diagrams (HRD). A number of interesting features are apparent from these figures. First, there is no obvious systematic trend in the relationship between the two independent age estimates. The uncertainties in the HRD ages are significantly larger, due to the large uncertainties in $T_{\rm eff}$ that afflict young stars, while the uncertainties in the MRD ages might be underestimated due to a systematic underestimation of the masses and radii by standard EB codes. Another interesting feature of this diagram is the clustering of points with MRD ages < 2.5 Myr but HRD ages of ~7.5 Myr. These points are all due to HD 144548 A, which is apparent from Tables 20 and 21. This star, which has a radius that is significantly larger (or alternatively, a mass much smaller) than model predictions for any plausible age of Upper Sco, reliably has an HRD age of 7.5 Myr for multiple model sets. Another outlier in this figure appears to have an MRD age of ~15 Myr but an HRD age of ~7 Myr from the standard Dartmouth models. This star is HR 5934 B, which is on the ZAMS where isochrones are tightly packed. Since this system is composed of rapidly rotating B-stars, RV precision is difficult to achieve. A modest change in the mass of this star, as would be quite plausible given the differences between our parameters and those presented in Maxted & Hutcheon (2018), could thus bring the MRD age of this star into much better agreement with the HRD age, since $T_{eff}$ would not be affected and the radius is not likely to change significantly.

### 6.4. *Comparison between the mass-radius diagram and H-R diagram*

In Figure 21 we examine the mass-dependent trends in the ages determined from the MRD, as well as temperature dependent trends in ages from the HRD. Model discrepancies are most pronounced at the lowest masses and temperatures (i.e. the EPIC 203710387 system). For example, the PARSEC v1.2S and magnetic Dartmouth models predict ages for this low-mass system that are much too old, in either the MRD or HRD. Interestingly, the spotted SP15 models also produce an age for this system that is much too old in the HRD, but not in the MRD. This suggests that some PMS models may lead observers to infer HRD ages that are a factor of 2–10 too old, at least for very low mass stars. In the MRD, the age discrepancy is much smaller, but unfortunately masses and radii can only be directly determined for a very small number of systems.

It is well known that HRD and color-magnitude diagram ages in PMS associations are temperature-dependent, with cooler stars in a given association appearing younger (e.g. Naylor 2009; Bell et al. 2012, 2013; Herczeg & Hillenbrand 2015; Fang et al. 2017). In this study, we observe that the trend in HRD ages with $T_{eff}$ depends on the model set adopted. Some models do predict younger HRD ages at lower $T_{eff}$, such as PARSEC v1.0,1.1 and SP15 (spot-free) models. However, the coolest system considered here (EPIC 203710387) is an exception; all model sets predict an older age for this system relative to the stars in the 0.3–1 $M_\odot$ range. It is interesting to note that both the Dartmouth and MIST models produce an HRD age of 5–7 Myr that is self-consistent across the mass range of 0.3–5 $M_\odot$.

While both MRD and HRD ages show trends, it is clear that a more consistent association age can be obtained from the MRD compared to the HRD. The top panel of this figure shows that there is no compelling evidence that Upper Sco is younger than 5 Myr, nor is there good evidence that the association is older than 10 Myr.

The nature of the differences between ages inferred from a MRD and an HRD is unclear at this point, but may be related to the surface boundary conditions utilized by theoretical evolutionary models (see Stassun et al. 2014, for a review). It is also possible that magnetic activity impacts the measured effective temperatures and radii, and therefore the placement of stars in the MRD and HRD. Stassun et al. (2012) presented empirical relations, for quantifying the amount of temperature suppression and radius inflation (relative to theoretical models) observed for low-mass dwarf stars as a function of activity indicators. We previously showed that in the case of EPIC 203710387, for example, that the magnitude of this effect is predicted to be a ~1–4% inflation in the radius, or ~1–2% suppression in temperature (David et al. 2016a). For UScoCTIO 5, based on the Hα equivalent widths reported in K15, the empirically predicted radius inflation is $13 \pm 5\%$ and the temperature suppression is $6 \pm 2\%$. Thus, while magnetic effects are unlikely to explain the observed discrepancies at very low masses (~0.1 $M_\odot$), it may be a viable explanation at somewhat higher masses ($\gtrsim 0.3 M_\odot$).

### 6.5. *Evaluating the accuracy of PMS models*

We have so far compared the ages implied for our sample of binaries in both the MRD and the HRD. Now, using our binaries as benchmarks, we take a critical look at the accuracy and predictive power of PMS models. First, we examine masses derived from a theoretical HRD and compare them with our dynamically measured masses. For PMS stars, it has long been known that models predict masses that are 10–30% lower than dynamical measurements (Hillenbrand & White 2004). In Figure 22, we show the trends in the fractional errors in model-derived masses for various model sets. The general shapes of most of these curves are indeed quite similar to those seen in Figure 5 of (Hillenbrand & White 2004), namely: discrepancies become gradually worse as the stellar mass declines below 1 $M_\odot$ (HD 144548 Bab), with the most drastic offsets occurring near 0.3 $M_\odot$ (UScoCTIO 5), before improving substantially toward 0.1 $M_\odot$ (EPIC 203710387). The PARSEC v1.2S models and spotted SP15 models provide two clear exceptions to this trend. These models overpredict mass by 30–300% in the 0.1–0.3 $M_\odot$ range.

The models that are most accurate at predicting masses from the H-R diagram are the Feiden (2016) magnetic models, followed by the MIST models. However, both of these model sets still predict masses that are occasionally $\gtrsim 25\%$ too low or too high. The PARSEC models exhibit some of the largest systematic offsets between model-derived masses and dynamical masses. The v1.0 and v1.1 iterations of these models under-predict masses by $\gtrsim 40\%$, while the PARSEC v1.2S models over-predict the masses of EPIC 203710387 AB by ~300%. The earlier versions of these models do not extend to temperatures low enough to test the accuracy for this ultra-low-mass system.

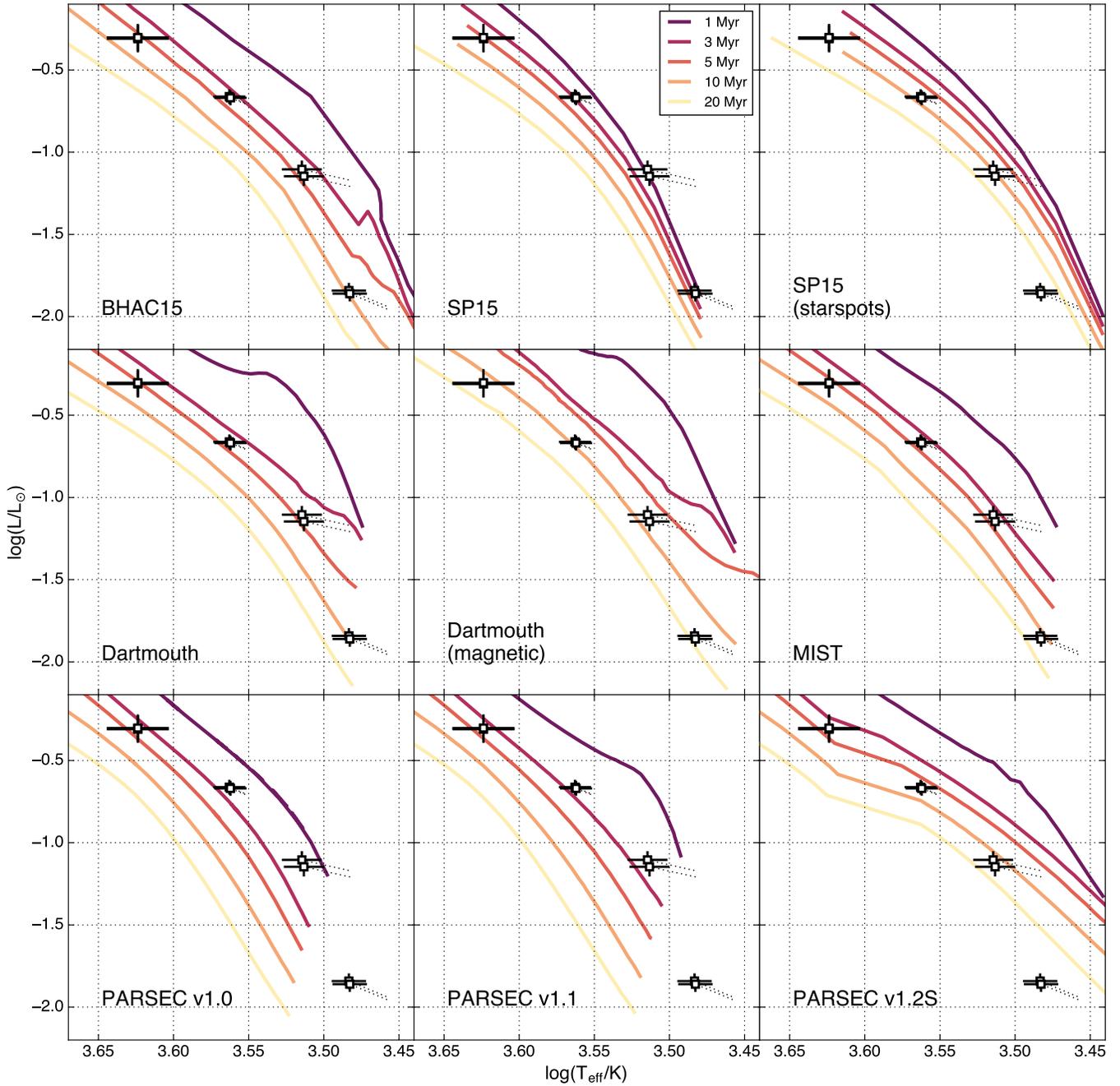

**Figure 19.** Low-mass EBs in the H-R diagram compared to isochrones from various different stellar evolution models. The temperatures and luminosities were determined from fitting the SED with BT-Settl models. The dashed lines indicate the positions of each star if PHOENIX models are used instead. From top to bottom, HD 144548 Ba and HD 144548 Bb (overlapping here), USco 48 AB, UScoCTIO 5 AB, and EPIC 203710387 AB.

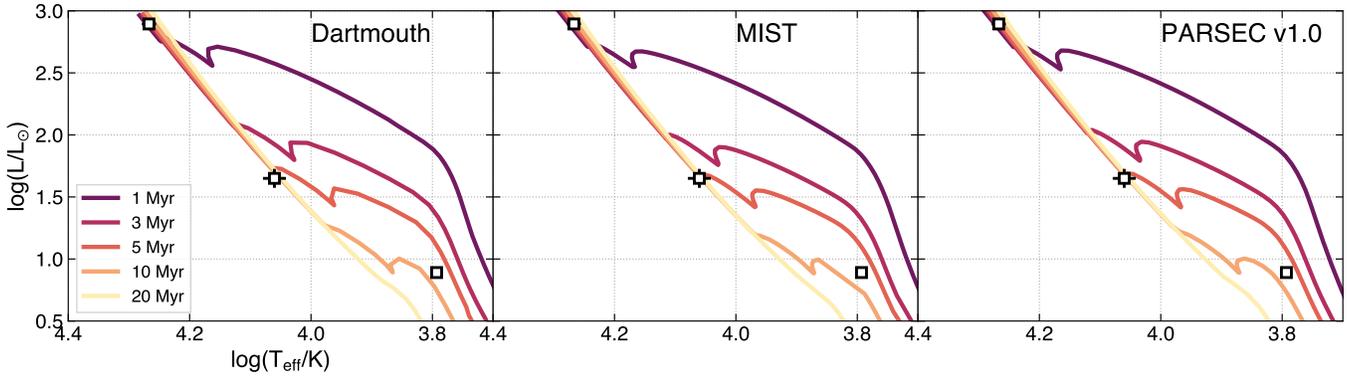

**Figure 20.** High-mass EB components in the H-R diagram compared to isochrones from different stellar evolution models. From top to bottom, HR 5934 A, HR 5934 B, and HD 144548 A.

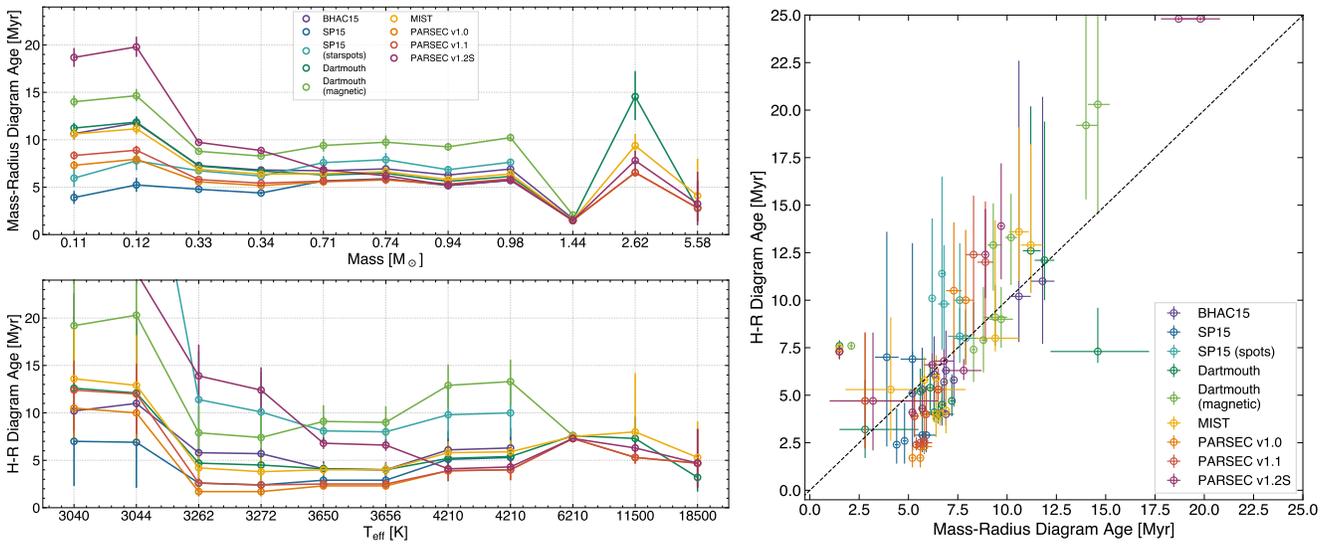

**Figure 21.** Mass-dependent and temperature-dependent trends in the ages of individual stars as determined from the mass-radius diagram (top) or the H-R diagram (bottom). Error bars indicate the 16th and 84th percentiles of the age distributions for individual stars.

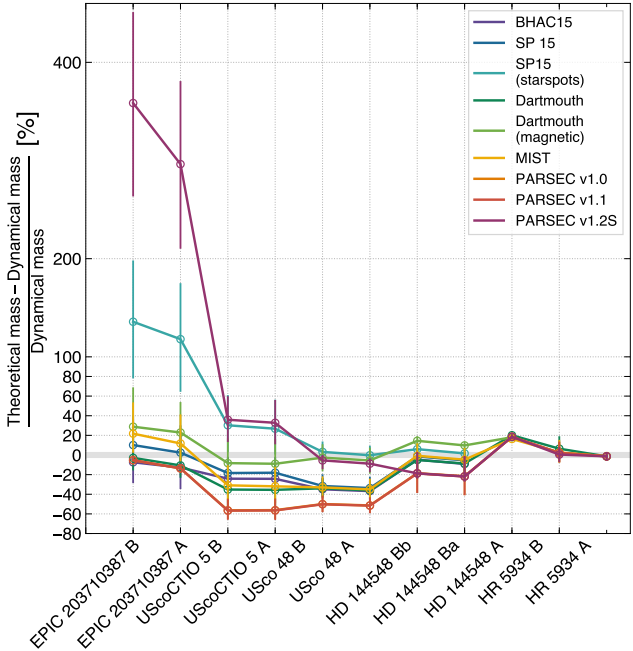

**Figure 22.** The fractional error in masses derived from the theoretical H-R diagram according to different PMS models.

Now, we turn to the ability of PMS models to predict radii. Under the assumption that the stars in Upper Scorpius share a common age without considerable dispersion we evaluated the accuracy with which stellar evolution models can predict stellar radii across a range of masses at a fixed age. For each model set considered, we calculated the expected radius for each well-characterized EB component through a 1D linear interpolation of the 5, 7, and 10 Myr theoretical mass-radius relations at the dynamical mass of each EB component. The fractional radius error as a function of mass for each isochrone and each model set is depicted in Figure 23.

Of the models considered here, the SP15 models show the least significant trend in the fractional radius error across the 0.1–1 $M_\odot$ range. For example, for an association age of 5 Myr the spot-free SP15 models successfully predict the radii to <5%. All models that extend to higher masses underestimate the radius of HD 144548 A by 30%, but we are careful to note that the analysis of this system is complicated and it is possible the radius of that star is in error. At ∼0.1 $M_\odot$, several models over-predict the radii by many tens of percent, most notably the magnetic Dartmouth models and the PARSEC v1.2S models.

To evaluate the ability of models to predict temperatures, we interpolated the mass-$T_{\rm eff}$ relations at 5 Myr and 10 Myr for each model set, and evaluated these re-lations at the dynamical masses measured for each EB component. The resulting discrepancies between the model-derived $T_{\rm eff}$ and observed $T_{\rm eff}$ are depicted in Figure 24. Most models considered here overestimate $T_{\rm eff}$ for low-mass PMS stars, sometimes in excess of 250 K (although a typical $T_{\rm eff}$ uncertainty might be 100 K). The trend somewhat mimics that observed for the model underestimation of masses, in that the temperature discrepancies are small at 1 $M_\odot$, become larger in the 0.3–0.7 $M_\odot$ range, then come back into better agreement near 0.1 $M_\odot$. Two notable exceptions to this trend are the SP15 models with starspots and the PARSEC v1.2S models, both of which underpredict $T_{\rm eff}$ for the systems in the 0.1–0.3 $M_\odot$ range.

### 6.6. Model predictions near and below the hydrogen-burning limit

To this point, we have focused on theoretical models of stars. The transiting brown dwarf RIK 72 b allows us to additionally compare observations with the predictions of substellar evolutionary models. Figure 25 compares the properties of RIK 72 b and the EPIC 203710387 binary, which is near the substellar boundary, with theoretical isochrones from Baraffe et al. (2003), hereafter COND03, and Burrows et al. (1997), hereafter B97. For comparison we also show the masses and radii of transiting brown dwarfs from a literature compilation (Csizmadia 2016) as well as more recent discoveries (Nowak et al. 2017; Gillen et al. 2017a; Cañas et al. 2018; Hodžić et al. 2018).

The fundamental properties of RIK 72 b (mass, age, radius, temperature, luminosity, and surface gravity), which notably depend on our characterization of the host star, are in remarkable agreement with these models. In particular, the B97 models can predict all of the observed parameters with a ∼50 $M_{\rm Jup}$ brown dwarf aged between 5–10 Myr. By comparison, the COND03 models predict faster evolution in the radius and luminosity between 5 and 10 Myr, suggesting the RIK 72 b parameters are more consistent with a ∼10 Myr age.

For EPIC 203710387, the COND03 models do not extend quite as far in mass to make predictions for this system, but it is clear from Figure 25 that a ∼10 Myr isochrone would reproduce the masses, radii, and luminosities of these stars fairly well. Thus, a self-consistent age is obtained from the COND03 models across the ∼55–120 $M_{\rm Jup}$ range. By comparison, the B97 models indicate the properties of these stars are better reproduced with an age of 7 Myr. These models predict slower evolution in radius and luminosity for a 55 $M_{\rm Jup}$ brown dwarf between 5 and 10 Myr, such that the properties of RIK 72 b are not so useful in discriminating its age.

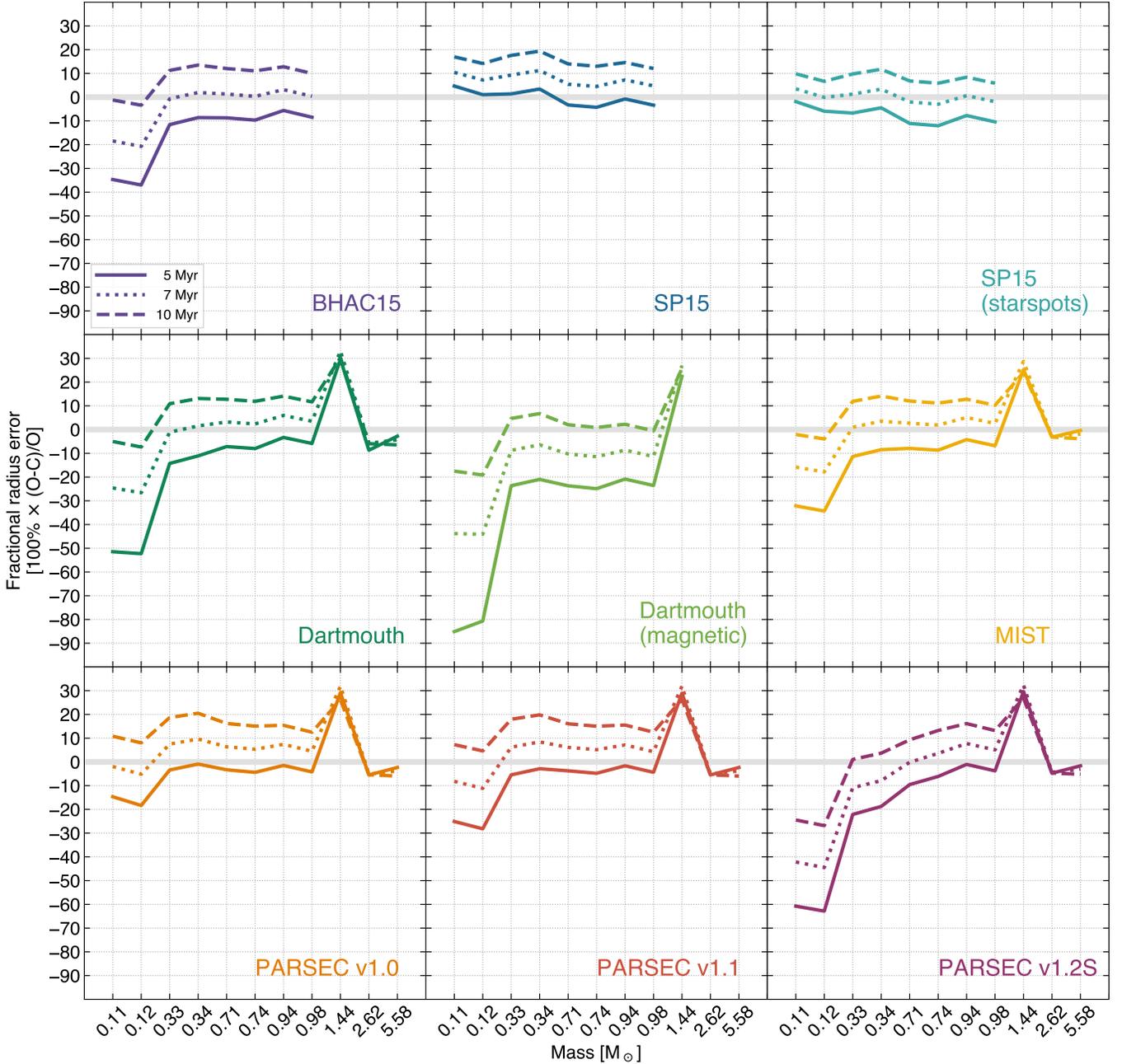

**Figure 23.** The fractional error in radius as a function of mass for different evolution models and different assumed ages.

However, a self-consistent age of 7 Myr certainly seems plausible when comparing these models to both RIK 72 b and EPIC 203710387.

Finally, we note that the COND03 models are better able to reproduce the masses and radii of the PMS eclipsing brown dwarf pair 2MASS J0535-05 in the Orion Nebula (Stassun et al. 2006), while the B97 models do not produce such large radii at ∼1 Myr.

## 7. CONCLUSIONS

We have presented a new age determination for the Upper Scorpius OB association from the locations of EBs in the mass-radius diagram and explored the systematic uncertainties resulting from differing evolutionary models. However, the degree to which any current models are accurate is not well known. Thus, while we find some models can reproduce the mass-radius relation in a self-consistent manner (i.e. requiring a population of only a single age) over a broad range of masses, it is possible that the absolute age scale of these models is biased. The discovery and characterization of a greater number of benchmarks will hopefully highlight where theoretical models are in error and lead to improvements in future iterations of said models. The *TESS* mission stands to make an important contribution in this domain as it is observing a larger number of PMS stars, including a sizable portion of the greater Scorpius-Centaurus association. For now, the primary conclusions of our study are summarized below.

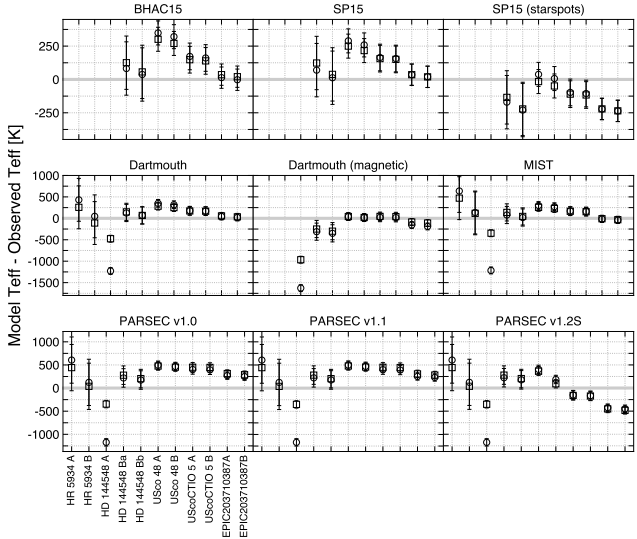

**Figure 24.** Comparison between model-derived and observed temperatures. Model temperatures are derived by evaluating the interpolated 5 Myr (circles) and 10 Myr (squares) mass-$T_{\rm eff}$ relations at the values of the EB dynamical masses.

1. **The age of Upper Sco.** Using standard PMS evolutionary models and models including starspot effects, we find an age of 5–7 Myr for Upper Sco, compared with the canonical 5 Myr and recent 11 Myr estimates based on theoretical H-R diagrams. The age we find is consistent with the MS turnoff age (7 ± 2 Myr) found by Pecaut & Mamajek (2016), which is based on a theoretical H-R diagram of the high-mass stars, where evolutionary models are considered more reliable. It is also consistent with ages found for low-mass astrometric binaries (Rizzuto et al. 2016). We also note that the region of Upper Sco observed by *K2* appears to be the youngest part of the association, according to the analysis of Pecaut & Mamajek (2016).

2. **Magnetic effects.** By invoking PMS evolutionary models with prescriptions for either magnetic fields or starspots, an older association age is inferred from both the MRD and HRD. We add support to the findings of Feiden (2016) that the magnetic Dartmouth models suggest a best-fit isochronal age of 9–10 Myr in the MRD, but we observe that our lowest-mass system appears to be older (14–15 Myr), possibly hinting at a mass-dependent systematic effect. By comparison, the Somers & Pinsonneault (2015) models that include the effect of starspots produce a consistent MRD age of 7 Myr across the mass range of 0.1–1 $M_\odot$.

3. **Coevality within binaries.** We find no compelling evidence to suggest that any of the binaries or higher-order multiples are not coeval. HD 144548, the triply-eclipsing system, exhibits the highest degree of non-coevality in the mass-radius diagram, but given the complexity of modeling this system we believe that either the current observational parameters for the tertiary are in error, PMS models are failing to accurately capture the changing stellar structure during this early period of hydrogen and carbon burning, or both. Notably, our best characterized binaries have mass ratios close to unity so it is unsurprising that they appear coeval. These results are in agreement with previous work which showed that binaries in Taurus-Auriga generally display a higher degree of coevality than randomly selected pairs of members (Kraus & Hillenbrand 2009).

4. **No appreciable radius dispersion on the PMS.** The luminosity dispersion observed for presumably coeval PMS stars has been a longstanding problem in the field (Hillenbrand 1997). Some have argued that a spread in radii could be responsible for such a luminosity dispersion, and there are claims of radius spreads in PMS (Jeffries 2007; Cottaar et al. 2014) and young MS cluster populations (Jackson et al. 2009; Jackson & Jeffries 2014). Although our sample is small, we have shown that some PMS models are able to reproduce the mass-radius relation in USco with a single age fairly well (see e.g. the PARSEC v1.0 models in Fig. 16). If there was a considerable radius dispersion, or age dispersion, in the region of Upper Sco probed by *K2*, we would not expect our sample to agree with models so well. However, one caveat is that our sample is composed of close binaries. If close binaries evolve differently from single stars or wide binaries (having e.g. different disk accretion histories), then we might not expect to see radius dispersion.

5. **Model systematics.** Few models are able to reproduce the exact slope of the mass-radius relation of Upper Sco EBs with a single age. The

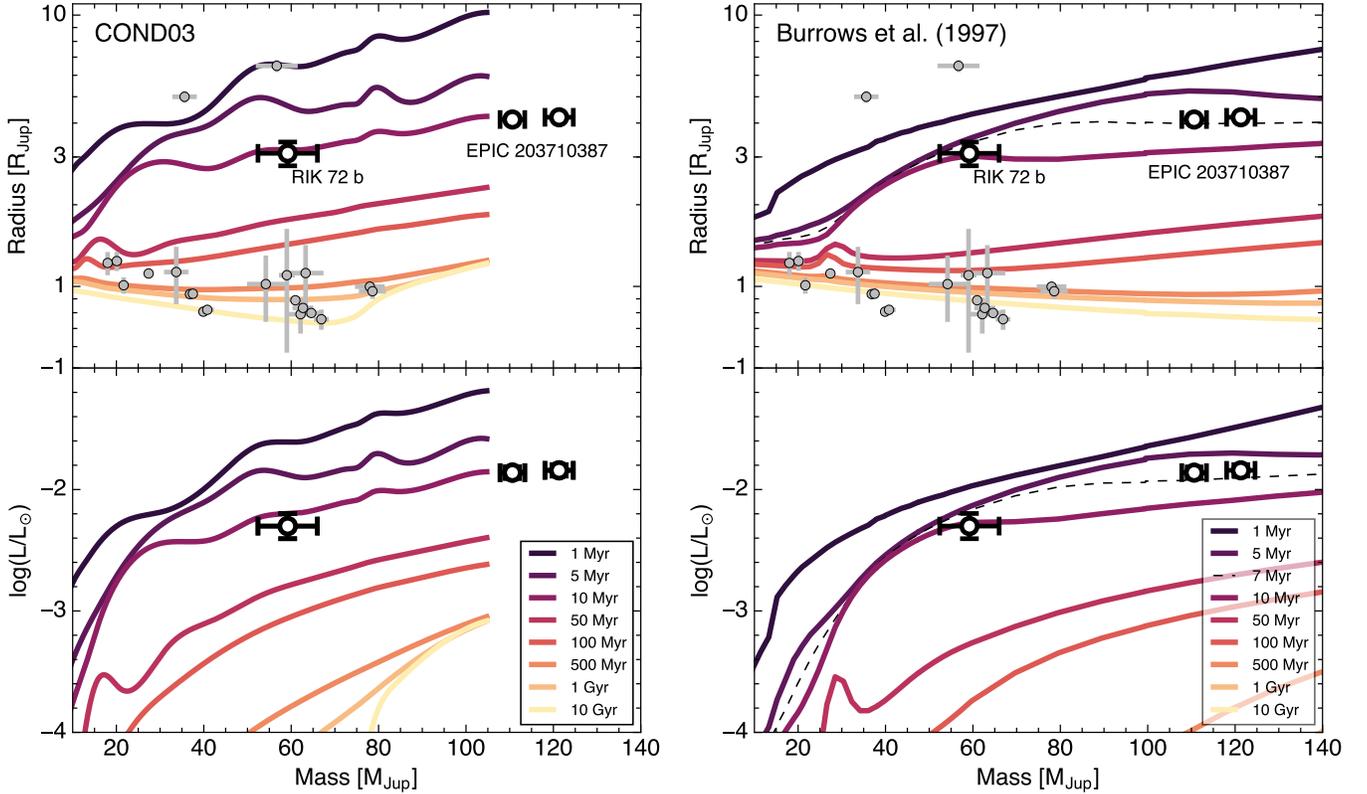

**Figure 25.** *Top left:* The masses and radii of RIK 72 b and EPIC 203710387 AB compared with transiting brown dwarfs from the literature and Baraffe et al. (2003) theoretical isochrones. The two points closest to the 1 Myr isochrone are the components of the 2M0535-05 eclipsing brown dwarf binary in the Orion Nebula (Stassun et al. 2006). *Bottom left:* Theoretical isochrones in the mass-luminosity plane. In both diagrams, RIK 72 b (denoted by the large point) is consistent with an age of ∼10 Myr. At right, the same observations now compared with the Burrows et al. (1997) evolutionary models.

PARSEC v1.2S models, for example, exhibit some of the most serious systematic offsets which are likely the result of the *ad hoc* adjustment of the surface boundary conditions for low-mass stars in those models. As such, the older age implied by the PARSEC v1.2S models is not credible. On the other hand, the SP15 models are the most successful at reproducing the data, predicting an age of 5 Myr for the spot-free case or 7 Myr for the 50% spotted case. However, we note that those models also produce unrealistically old ages for the lowest-mass system in the HRD.

6. **Agreement between mass-radius diagram and the H-R diagram.** One major pitfall of age-dating PMS populations in the H-R diagram is that self-consistent ages can not be derived across a wide range of masses, i.e. theoretical ages are mass-dependent no matter the model set adopted (e.g. Herczeg & Hillenbrand 2015; Fang et al. 2017). The EBs presented here, while they are not entirely self-consistent primarily due to one system, do exhibit a much higher degree of consistency than traditional H-R diagram analyses. In some cases, the ages inferred for an EB from the MRD are significantly different from the ages of the same stars from the HRD (e.g. EPIC 203710387 and the SP15 spotted models, or HD 144548 A and all models considered here). In other cases, e.g. the magnetic Dartmouth models, a consistent MRD and HRD age is obtained over a fairly broad range of masses, althought the lowest mass system at 0.1 $M_\odot$ remains problematic. In general, the systems studied here exhibit a much broader range of ages in the HRD than they do in the MRD.

7. **Observational and theoretical agreement at high masses.** For HR 5934 A ($M_* \sim 5.5\ M_\odot$), the dynamical mass is in excellent agreement (within 2%) with H-R diagram predictions using all of the stellar models tested here. The age of this star from an HRD analysis is generally too low ($\tau < 5$ Myr) if models including rotation are used. By contrast, if non-rotating models are considered, the age is in broad agreement with the ages of other EBs considered here, though on the younger end of our accepted range. In the case of HR 5934 B, stellar models which include rotation tend to overestimate the mass by 20%. This result may not apply generally. Since HR 5934 A is somewhat slowly rotating for a star of its mass, it is possible that tidal effects are governing the spin of the secondary and that the rotating models are indeed appropriate for more rapidly-rotating stars.

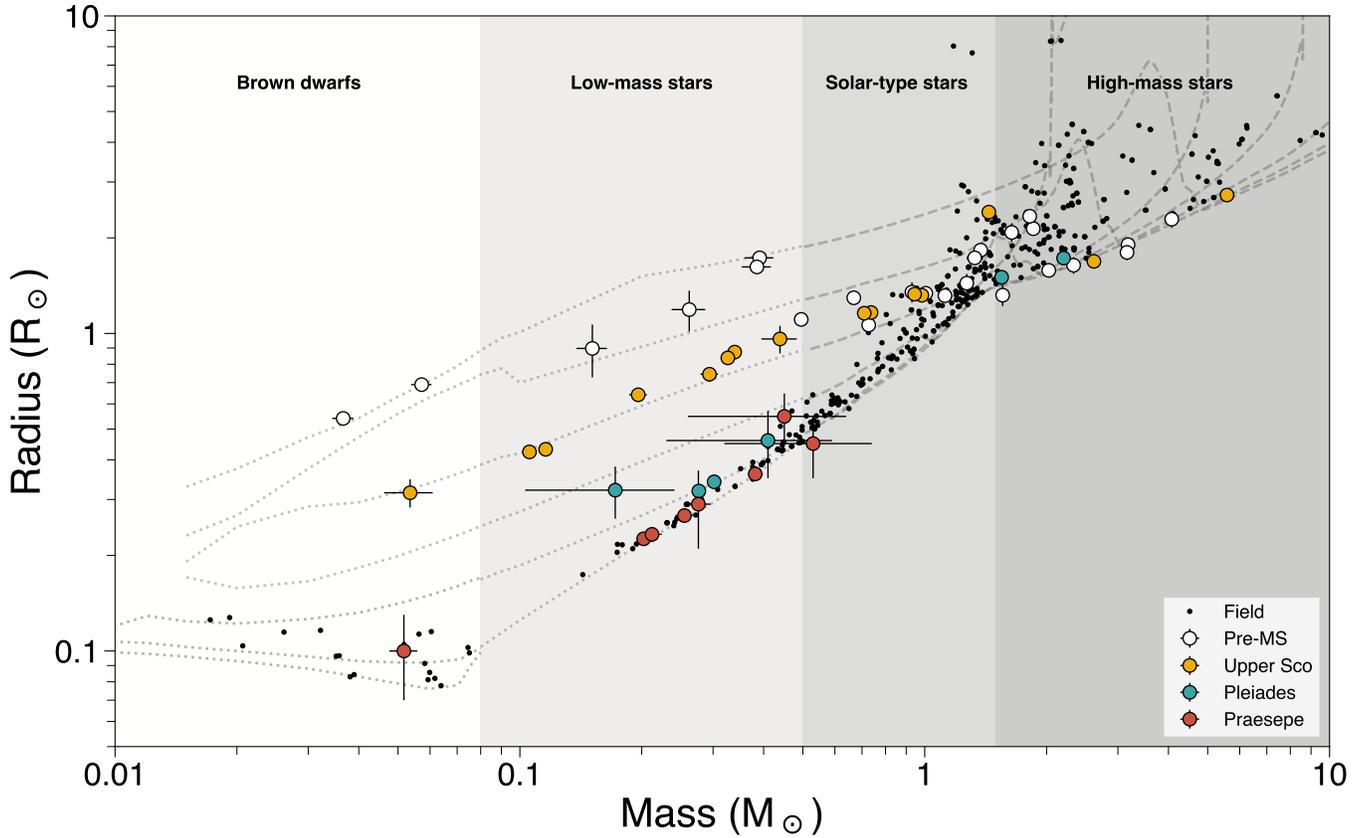

**Figure 26.** The mass-radius diagram of stars and brown dwarfs. Stars shown in this diagram belong to spectroscopically double-lined eclipsing binaries, with the exception of the single-lined RIK 72 system (the yellow circles with the largest errors). Main-sequence stars, post-main-sequence stars, and transiting brown dwarfs are shown as black points. Pre-main-sequence eclipsing binaries known prior to the *K2* mission are indicated by the white circles (compiled by Stassun et al. 2014). Eclipsing binaries, in the Upper Scorpius OB association, the Pleiades, and Praesepe open clusters are shown as yellow, teal, and red points, respectively. Nearly all of these systems were discovered from *K2* photometry. The references for systems depicted in this diagram are as summarized as follows. **Upper Scorpius:** this work, Alonso et al. (2015); Kraus et al. (2015); Lodieu et al. (2015); David et al. (2016a); Maxted & Hutcheon (2018). **Pleiades:** David et al. (2016b). **Praesepe:** Gillen et al. (2017a); Kraus et al. (2017). **Field sample:** DEBCAT[b]. **Transiting brown dwarfs:** Csizmadia (2016); Nowak et al. (2017); Gillen et al. (2017a); Cañas et al. (2018); Hodžić et al. (2018).

[a]http://www.astro.keele.ac.uk/jkt/debcat/
[b]http://www.astro.keele.ac.uk/jkt/debcat/

8. **Masses from pre-main-sequence models.** For stars less massive than the Sun, we find that many PMS models underpredict mass based on the HRD position by as much as 60%, but more typically in the 10–40% range. The magnetic Dartmouth models show the best agreement across the mass range of 0.1–1 $M_\odot$, but may overpredict mass by 20% at 0.1 $M_\odot$. In contrast to the trend described above, the spotted SP15 models and PARSEC v1.2S models overpredict mass for low-mass stars, by about 30% at 0.3 $M_\odot$ and in excess of 100% at 0.1 $M_\odot$.

9. **Choice of empirical temperature scale.** The degree of disagreement noted above depends on the empirical temperature scale adopted, which is more uncertain for PMS stars relative to field stars. In general, models of low-mass PMS stars predict temperatures that are hotter than those in-ferred from observations. Consequently, the HH15 temperature scale produces better agreement with model predictions than the PM13 scale, which is cooler at a given spectral type. Notably, we can only assess the level of agreement between models and observations here and can not comment on whether a particular empirical or theoretical temperature scale is physically accurate.

10. **Tidal circularization.** Our data suggest a circularization period of ~4 days at the age of Upper Sco. Below this orbital period, our binaries are found to be on circular or near-circular orbits. The binaries with longer orbital periods have eccentric orbits (with the exception of the high-mass HR 5934 system). This result is in good agreement with an early study on the PMS circularization period (Mathieu 1994).

11. **A low-mass pre-main-sequence EB.** We report USco 48 as a grazing EB and present the first mass and radius determinations for this system ($M_A = 0.737 \pm 0.020\ M_\odot$, $M_B = 0.709 \pm 0.020\ M_\odot$, $R_A = 1.164 \pm 0.019\ R_\odot$, $R_B = 1.158 \pm 0.019\ R_\odot$). The binary is apparently tidally synchronized, as indicated by the photometric modulations due to starspots, and is nearly circularized. The source has previously been proposed to host a debris disk (Luhman & Mamajek 2012) based on a 50% excess at 24 $\mu$m (Carpenter et al. 2009).

12. **A young transiting brown dwarf.** We report the wide eclipsing companion to RIK 72 (EPIC 205207894) as a brown dwarf. RIK 72 b is an important benchmark for brown dwarf evolutionary models which predict the size and luminosity of brown dwarfs as a function of age, and indeed current models produce remarkable agreement with our reported parameters. Further monitoring of this system with time series photometry and RVs should yield a better constraint on the orbital period and eccentricity, which will help to refine the physical parameters further.

13. **A low-mass pre-main-sequence triple.** We report EPIC 202963882 as a short-period EB in a triple system and coarsely characterize the system for the first time. Our preliminary masses and radii for the EB component are in broad agreement with the empirical mass-relation mapped by the other, well-characterized systems studied here. Better spectroscopic data and an EB code capable of modeling semi-detached systems is needed for a more physically accurate characterization.

14. **A low-mass pre-main-sequence quadruple.** We report EPIC 203868608 as a 2+2 quadruple system, with all four components having low masses. The eclipsing components are likely to have masses near or below the hydrogen-burning limit, but do not have masses as low as those previously suggested in D16. A detailed analysis of this system is presented in Wang et al. (2018). RV monitoring in the infrared, where the flux ratio between the EB and SB2 is more favorable, or AO-resolved spectroscopy will allow for the determination of the EB masses and radii. Long term astrometric monitoring could allow for a separate determination of the total masses of the EB and SB2.

15. **The mystery of EPIC 203710387.** Relative to the other EBs studied here, EPIC 203710387 is unique in that it is both underluminous for its color in a CMD and has component radii that are smaller than expectations from a best-fit mass-radius isochrone using most of the model sets considered here. It is possible that this system is genuinely older than the other EBs studied here (but still consistent with the range of ages in Sco-Cen), or that some current models overpredict the radii of ultra-low-mass pre-main-sequence stars by many tens of percent. If the former scenario is true, this may reflect populations of mixed age within the spatial and kinematic boundaries used to conventionally define Upper Sco.

As a concluding remark, we note that different authors routinely measure different masses and radii for the same EBs. In some cases, the degree of disagreement is statistically significant ($\gg 3\sigma$). This may reflect the fact that quoted parameter uncertainties in EB studies are almost always statistical, and rarely attempt to account for systematic effects. Systematic differences are likely tied to the different light curve models used (e.g., whether stellar surfaces are approximated as spheres or ellipsoids) as well as the numerous methods used to determine RVs (e.g., 1D vs. 2D cross-correlation, the broadening function, or spectral disentanglement) and perhaps the wavelength range of the spectra. For young, active stars, it may be particularly important to consider the possible effects of starspots on both light curve modeling and RV determination. Comparative studies of different RV determination (e.g. Czekala et al. 2017; Halbwachs et al. 2017) and EB modeling methods are needed to assess the amplitudes of systematic effects in different regions of binary parameter space. While author-to-author discrepancies are statistically significant for the systems studied here, the fractional uncertainties in the masses and radii are not large enough to significantly change the conclusions reached in this study on the age of Upper Sco.

*Facilities:* Keck:I (HIRES), Keck:II (NIRC2), Kepler, Gemini:South (DSSI, NESSI)

*Software:* EMCEE (Foreman-Mackey et al. 2013), GP-EBOP (Gillen et al. 2017a), JKTEBOP (Southworth 2013), SCIPY (Jones et al. 2001–)

We are grateful to anonymous referee for a thorough and thoughtful report. The authors wish to thank Mark Everett for reducing and analyzing the speckle imaging data, Andrew Howard and the California Planet Search observers for monitoring of RIK 72, and Ji Wang for collaborating on EPIC 203868608. We also thank Eric Mamajek and John Stauffer for helpful comments on an early version of this manuscript. We thank Greg Feiden, Jim Fuller, B.J. Fulton, Adam Kraus, Luisa Rebull, Garrett Somers, Erik Petigura, Marc Pinsonneault, and Josh Winn for helpful discussions.

TJD gratefully acknowledges support from the Jet Propulsion Laboratory Exoplanetary Science Initiative.

EG gratefully acknowledges support from Winton Philanthropies in the form of a Winton Exoplanet Fellowship. Some of this research was carried out at the Jet Propulsion Laboratory, California Institute of Technology, under a contract with the National Aeronautics and Space Administration. This paper includes data collected by the K2 mission. Funding for the K2 mission is provided by the NASA Science Mission directorate. Some of the data presented herein were obtained at the W. M. Keck Observatory, which is operated as a scientific partnership among the California Institute of Technology, the University of California and the National Aeronautics and Space Administration. The Observatory was made possible by the generous financial support of the W. M. Keck Foundation. The authors wish to recognize and acknowledge the very significant cultural role and reverence that the summit of Maunakea has always had within the indigenous Hawaiian community. We are most fortunate to have the opportunity to conduct observations from this mountain. 

## APPENDIX

### A. EMPIRICAL RELATIONS

Below we present empirical relations based on the derived parameters of the best-characterized EBs in the text. We stress that these relations only apply to Upper Sco and furthermore might only apply to members in a similar region to that probed by *K2*, given the proposed existence of an age gradient in the association. Furthermore, these relations might only apply to close binaries if single stars and wide binaries have evolved differently. It may be possible for these relations to be extended to stars in other associations with a similar age and metallicity to that of Upper Sco, but we urge caution.

#### A.1. *Empirical pre-main-sequence mass-radius relation for Upper Sco*

From the fundamentally-determined masses and radii of the well-characterized, double-lined EBs discussed above, we derive an empirical pre-main-sequence mass-radius relation appropriate for low-mass stars. We consider only those stars with $M_* < 1\ M_\odot$, as these stars are expected to be fully convective and lie below the prominent hump in the MRD, where the radius is expected to vary quickly with mass. As the relation is based on EBs within Upper Sco, this relation should only be used within that association or for stars that have an equivalent age and a metallicity of approximately solar.

We performed fits of two functional forms to the masses and radii of the EBs. The first fit we performed was a cubic polynomial of the form:

$$\left(\frac{R_*}{R_\odot}\right) = c_0 + c_1 \left(\frac{M_*}{M_\odot}\right) + c_2 \left(\frac{M_*}{M_\odot}\right)^2 + c_3 \left(\frac{M_*}{M_\odot}\right)^3. \qquad (A1)$$

The second fit was a power-law:

$$\left(\frac{R_*}{R_\odot}\right) = \alpha \left(\frac{M_*}{M_\odot}\right)^\beta. \qquad (A2)$$

**Table 6.** Photometry used in SED fitting.

| Band | EPIC 204432860 | EPIC 205207894 | EPIC 205030103 | EPIC 203710387 |
|---|---|---|---|---|
| PS $g_{AB}$ (mag) | $\cdots$ | $16.369 \pm 0.005$ | $16.846 \pm 0.005$ | $19.356 \pm 0.011$ |
| PS $r_{AB}$ (mag) | $\cdots$ | $14.956 \pm 0.006$ | $15.432 \pm 0.011$ | $17.934 \pm 0.005$ |
| PS $i_{AB}$ (mag) | $\cdots$ | $13.629 \pm 0.003$ | $13.820 \pm 0.003$ | $15.946 \pm 0.002$ |
| PS $z_{AB}$ (mag) | $\cdots$ | $12.979 \pm 0.100$ | $12.996 \pm 0.100$ | $14.963 \pm 0.002$ |
| PS $y_{AB}$ (mag) | $\cdots$ | $12.627 \pm 0.009$ | $12.625 \pm 0.003$ | $14.456 \pm 0.002$ |
| APASS $V$ (mag) | $13.582 \pm 0.041$ | $\cdots$ | $16.192 \pm 0.100$ | $\cdots$ |
| APASS $B$ (mag) | $15.131 \pm 0.068$ | $\cdots$ | $17.806 \pm 0.203$ | $\cdots$ |
| SDSS $g$ (mag) | $14.380 \pm 0.019$ | $16.480 \pm 0.072$ | $16.975 \pm 0.074$ | $\cdots$ |
| SDSS $r$ (mag) | $12.911 \pm 0.046$ | $14.978 \pm 0.055$ | $15.482 \pm 0.042$ | $\cdots$ |
| SDSS $i$ (mag) | $11.929 \pm 0.021$ | $13.609 \pm 0.049$ | $13.708 \pm 0.011$ | $\cdots$ |
| $Gaia$ $G$ (mag) | $12.563 \pm 0.004$ | $14.351 \pm 0.001$ | $14.552 \pm 0.001$ | $16.646 \pm 0.001$ |
| $Gaia$ $BP$ (mag) | $13.788 \pm 0.015$ | $15.910 \pm 0.003$ | $16.416 \pm 0.005$ | $18.928 \pm 0.030$ |
| $Gaia$ $RP$ (mag) | $11.465 \pm 0.009$ | $13.131 \pm 0.001$ | $13.247 \pm 0.002$ | $15.234 \pm 0.002$ |
| 2MASS $J$ (mag) | $9.824 \pm 0.021$ | $11.232 \pm 0.021$ | $11.172 \pm 0.023$ | $12.932 \pm 0.023$ |
| 2MASS $H$ (mag) | $9.101 \pm 0.023$ | $10.466 \pm 0.023$ | $10.445 \pm 0.026$ | $12.277 \pm 0.024$ |
| 2MASS $K_s$ (mag) | $8.842 \pm 0.022$ | $10.200 \pm 0.023$ | $10.170 \pm 0.021$ | $11.907 \pm 0.023$ |
| $WISE$ $W1$ (mag) | $8.752 \pm 0.022$ | $10.073 \pm 0.024$ | $10.036 \pm 0.023$ | $11.748 \pm 0.023$ |
| $WISE$ $W2$ (mag) | $8.650 \pm 0.020$ | $9.933 \pm 0.020$ | $9.838 \pm 0.020$ | $11.483 \pm 0.022$ |
| $WISE$ $W3$ (mag) | $\cdots$ | $9.822 \pm 0.056$ | $9.648 \pm 0.047$ | $\cdots$ |
| $WISE$ $W4$ (mag) | $\cdots$ | $\cdots$ | $8.777 \pm 0.467$ | $\cdots$ |

We performed initial fits using the `optimize.minimize` least-squares minimization routine within the `scipy` Python package. We then determined the statistical uncertainties on the fit parameters by sampling the following likelihood function using a Markov Chain Monte Carlo (MCMC) method:

$$\ln p(R|M, \sigma_R, c_0, c_1, c_2, c_3, f) = -\frac{1}{2} \sum_n \left[ \frac{(\text{data - model})^2}{s_n^2} + \ln\left(2\pi s_n^2\right) \right], \tag{A3}$$

where

$$s_n^2 = \sigma_{R,n}^2 + f^2(\text{model})^2, \tag{A4}$$

with $f$ being the fractional amount by which the variance is underestimated and the model is given by either equation (A1) or (A2).

To perform the MCMC sampling we used the `emcee` package with 20 walkers initialized near the parameter estimates from the least-squares fit and sampled the likelihood until the chains achieved convergence, which was diagnosed every 100 steps. The chain was considered to be converged when the length exceeded 100 times the autocorrelation length in each free parameter and when the estimates of each autocorrelation length changed by less than 1% from the previous estimate.

We present the results from the MCMC sampling in Table 23 and in Figure 27. The power-law fit to all four EBs is clearly a poor match to the data at $\sim$0.3 $M_\odot$ (UScoCTIO 5). Consequently, we investigated another power-law fit to only those EB components with masses above $\sim$0.3 $M_\odot$ (UScoCTIO 5), which does a much better job of matching the data for those six stars. The first power-law fit predicts $R_* \propto M_*^{1/2}$ between 0.1–1 $M_\odot$, while the power-law fit excluding EPIC 203710387 predicts $R_* \propto M_*^{2/5}$ between 0.3–1 $M_\odot$.

The residuals for the polynomial fit to the mass-radius relation are $\lesssim$2% in radius. The residuals for the first power-law fit are $\lesssim$10% in radius over the 0.1–1 $M_\odot$ range, or $\lesssim$2% in radius over the 0.3–1 $M_\odot$ range for the second power-law fit.

**Table 7.** Keck-I/HIRES radial velocities

| System | UT Date | BJD | $v_1$ | $\sigma_1$ | $v_2$ | $\sigma_2$ | $F_2/F_1$ | $\sigma_{F_2/F_1}$ |
|---|---|---|---|---|---|---|---|---|
| | | | km s$^{-1}$ | km s$^{-1}$ | km s$^{-1}$ | km s$^{-1}$ | | |
| HR 5934 | 2016 May 17 | 2457525.847810 | 44.6 | 4.0 | -109.9 | 2.5 | $\cdots$ | $\cdots$ |
| | 2016 May 20 | 2457528.830532 | 8.3 | 2.0 | -33.1 | 3.4 | 0.41 | 0.17 |
| | 2017 Jul 08 | 2457942.810159 | 6.2 | 2.5 | -37.5 | 1.4 | 0.46 | 0.11 |
| | 2017 Jul 09 | 2457943.807572 | -32.3 | 2.4 | 49.1 | 5.5 | $\cdots$ | $\cdots$ |
| | 2017 Jul 10 | 2457944.873724 | -62.6 | 2.6 | 126.2 | 4.7 | $\cdots$ | $\cdots$ |
| | 2017 Jul 11 | 2457945.862162 | -63.3 | 2.5 | 127.9 | 3.9 | $\cdots$ | $\cdots$ |
| USco 48 | 2015 Jun 02 | 2457175.820325 | -87.3 | 1.1 | 79.4 | 1.9 | 1.020 | 0.030 |
| | 2016 May 20 | 2457529.019142 | -61.3 | 2.0 | 56.1 | 1.5 | 0.894 | 0.073 |
| | 2017 Jul 08 | 2457942.780202 | -42.9 | 2.3 | 31.1 | 1.6 | 0.866 | 0.069 |
| | 2017 Jul 09 | 2457943.775365 | -45.8 | 2.7 | 32.1 | 2.9 | 1.019 | 0.019 |
| | 2017 Jul 10 | 2457944.786688 | 74.5 | 2.7 | -88.7 | 1.5 | 0.933 | 0.094 |
| | 2017 Jul 11 | 2457945.764455 | -57.6 | 1.8 | 50.5 | 2.1 | 0.889 | 0.068 |
| RIK 72 | 2015 Jun 02 | 2457175.872713 | -1.50 | 0.47 | $\cdots$ | $\cdots$ | $\cdots$ | $\cdots$ |
| | 2016 May 17 | 2457525.841835 | -8.73 | 0.21 | $\cdots$ | $\cdots$ | $\cdots$ | $\cdots$ |
| | 2016 May 20 | 2457528.839284 | -7.89 | 0.21 | $\cdots$ | $\cdots$ | $\cdots$ | $\cdots$ |
| | 2016 Jun 15 | 2457554.965238 | -2.46 | 0.22 | $\cdots$ | $\cdots$ | $\cdots$ | $\cdots$ |
| | 2017 Jul 08 | 2457942.790311 | -5.45 | 0.41 | $\cdots$ | $\cdots$ | $\cdots$ | $\cdots$ |
| | 2017 Jul 09 | 2457943.782008 | -4.76 | 0.41 | $\cdots$ | $\cdots$ | $\cdots$ | $\cdots$ |
| | 2017 Jul 10 | 2457944.822261 | -3.81 | 0.39 | $\cdots$ | $\cdots$ | $\cdots$ | $\cdots$ |
| | 2017 Jul 11 | 2457945.840403 | -3.33 | 0.40 | $\cdots$ | $\cdots$ | $\cdots$ | $\cdots$ |
| | 2017 Sep 03 | 2457999.770508 | -7.201 | 0.53 | $\cdots$ | $\cdots$ | $\cdots$ | $\cdots$ |
| | 2017 Sep 06 | 2458002.767394 | -7.150 | 0.92 | $\cdots$ | $\cdots$ | $\cdots$ | $\cdots$ |
| | 2017 Sep 22 | 2458018.745936 | -8.521 | 0.22 | $\cdots$ | $\cdots$ | $\cdots$ | $\cdots$ |
| | 2017 Sep 23 | 2458019.721002 | -8.419 | 0.39 | $\cdots$ | $\cdots$ | $\cdots$ | $\cdots$ |
| | 2018 May 26 | 2458264.956502 | -0.734 | 0.90 | $\cdots$ | $\cdots$ | $\cdots$ | $\cdots$ |
| EPIC 202963882 B | 2016 May 17 | 2457525.880139 | 22.2 | 10.0 | -38.9 | 10.0 | 0.976 | 0.020 |
| | 2016 May 17 | 2457526.079108 | -79.8 | 10.0 | 113.0 | 10.0 | 0.66 | 0.13 |
| | 2016 May 20 | 2457528.883113 | 71.4 | 10.0 | -122.1 | 10.0 | 0.64 | 0.12 |
| | 2016 May 20 | 2457529.067194 | 8.6 | 10.0 | -18.4 | 10.0 | $\cdots$ | $\cdots$ |
| | 2017 Jul 08 | 2457942.771142 | 44.7 | 10.0 | -95.1 | 10.0 | $\cdots$ | $\cdots$ |
| | 2017 Jul 09 | 2457943.765767 | -42.6 | 10.0 | 57.9 | 10.0 | $\cdots$ | $\cdots$ |
| | 2017 Jul 10 | 2457944.896596 | -89.9 | 10.0 | 108.4 | 10.0 | 0.821 | 0.042 |
| | 2017 Jul 11 | 2457945.803583 | 67.9 | 10.0 | -102.9 | 10.0 | $\cdots$ | $\cdots$ |
| UScoCTIO 5 | 2017 Jul 08 | 2457942.829644 | 0.27 | 0.43 | -7.12 | 0.45 | 0.830 | 0.039 |
| | 2017 Jul 09 | 2457943.819253 | 4.77 | 0.41 | -10.89 | 0.48 | 0.771 | 0.050 |
| | 2017 Jul 10 | 2457944.917664 | 9.13 | 0.41 | -14.05 | 0.46 | 0.836 | 0.051 |
| | 2017 Jul 11 | 2457945.779752 | 11.48 | 0.46 | -16.73 | 0.49 | 0.840 | 0.034 |
| EPIC 203868608 A | 2015 Jun 02 | 2457175.921333 | -4.72 | 0.60 | $\cdots$ | $\cdots$ | $\cdots$ | $\cdots$ |
| | 2015 Jul 14 | 2457217.816800 | 16.51 | 0.25 | -29.50 | 0.47 | $\cdots$ | $\cdots$ |
| | 2015 Aug 21 | 2457255.829930 | 14.51 | 0.51 | -26.39 | 1.62 | $\cdots$ | $\cdots$ |
| | 2015 Aug 28 | 2457262.799230 | -25.48 | 1.19 | 21.87 | 0.80 | $\cdots$ | $\cdots$ |
| | 2015 Aug 31 | 2457265.797000 | -4.66 | 0.23 | $\cdots$ | $\cdots$ | $\cdots$ | $\cdots$ |
| | 2015 Sep 25 | 2457290.729400 | 15.79 | 0.19 | -29.26 | 0.26 | $\cdots$ | $\cdots$ |
| | 2016 May 17 | 2457526.116009 | 7.18 | 0.31 | -19.57 | 0.37 | $\cdots$ | $\cdots$ |
| | 2016 May 20 | 2457528.959473 | -30.40 | 0.30 | 27.90 | 0.34 | $\cdots$ | $\cdots$ |
| | 2016 Jun 15 | 2457555.031623 | 7.25 | 0.34 | -16.44 | 0.37 | $\cdots$ | $\cdots$ |
| | 2017 Jul 08 | 2457942.801689 | -37.57 | 0.49 | 32.42 | 0.56 | $\cdots$ | $\cdots$ |
| | 2017 Jul 09 | 2457943.795450 | -31.46 | 0.43 | 28.49 | 0.46 | $\cdots$ | $\cdots$ |
| | 2017 Jul 10 | 2457944.858624 | -23.82 | 0.48 | 18.77 | 0.86 | $\cdots$ | $\cdots$ |
| | 2017 Jul 11 | 2457945.824092 | -15.91 | 0.44 | 10.07 | 0.65 | $\cdots$ | $\cdots$ |

**Table 8.** Literature radial velocities of HR 5934

| BJD | $v_1$ | $\sigma_1$ | $v_2$ | $\sigma_2$ | Ref. |
|---|---|---|---|---|---|
| | km s$^{-1}$ | km s$^{-1}$ | km s$^{-1}$ | km s$^{-1}$ | |
| 2442671.531650 | 4.4 | 1.2 | $\cdots$ | $\cdots$ | a |
| 2442876.891065 | -58.3 | 1.9 | 154.1[†] | 11.3 | a |
| 2442881.823965 | 48.1 | 1.3 | -129.3[†] | 8.1 | a |
| 2442179.737523 | 23.3 | 1.7 | $\cdots$ | $\cdots$ | b |
| 2442180.676523 | 44.5 | 2.3 | $\cdots$ | $\cdots$ | b |
| 2442174.817522 | -51.8[†] | 6.1 | $\cdots$ | $\cdots$ | b |
| 2442176.897523 | -61.4 | 2.6 | $\cdots$ | $\cdots$ | b |
| 2442178.643523 | -37.1 | 2.1 | $\cdots$ | $\cdots$ | b |
| 2442921.746566 | -59.5 | 0.9 | $\cdots$ | $\cdots$ | b |
| 2443297.555576 | -14.1 | 2.9 | $\cdots$ | $\cdots$ | b |
| 2443300.566576 | -41.8 | 3.7 | $\cdots$ | $\cdots$ | b |
| 2452415.300972 | -54.3 | 0.5 | $\cdots$ | $\cdots$ | c |
| 2453129.725 | 60.6 | 7.1 | -135.5 | 1.6 | d |
| 2453129.735 | 59.6 | 4.9 | -135.2 | 1.7 | d |
| 2454298.510 | 54.5 | 2.3 | -120.8 | 2.1 | d |
| 2454302.491 | -65.2 | 2.3 | 133.4 | 1.7 | d |
| 2456523.599 | 56.8 | 1.5 | -132.1 | 1.8 | d |

REFERENCES. — a: Andersen & Nordström (1983); b: Levato et al. (1987); c: Jilinski et al. (2006); d: Maxted & Hutcheon (2018)
[†]RV measurement is discrepant by >10 km s$^{-1}$ from our best-fit orbital solution and was thus excluded from our final fit.
The RVs from Maxted & Hutcheon (2018) were not included in our fit.

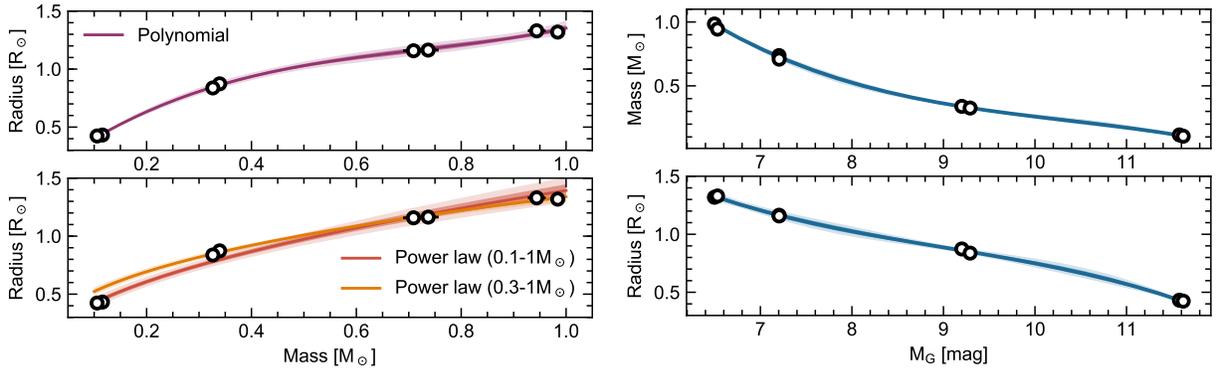

**Figure 27.** Empirical pre-main-sequence mass-radius relations (left) and relations between the absolute $G$ magnitude and mass (top right) or radius (bottom right). The points indicate the EB components used in the fits. Solid lines indicate the maximum likelihood relations, while the shaded regions indicate the $1\sigma$ and $2\sigma$ error bands as determined from the MCMC chains.

### A.2. Empirical brightness relations for Upper Sco

Four EBs studied here have mass ratios close to unity (EPIC 203710387, UScoCTIO 5, USco 48, and HD 144548 B). Only one system included in the construction of our empirical relations, HD 144548 B, is known to host a tertiary companion and, since that companion is also eclipsing, *Kepler* band luminosity ratios between all components are known (Alonso et al. 2015). Unresolved broadband photometry can thus be easily decomposed (assuming the *K2* luminosity ratios approximately reflect the $G$-band ratios) and given the *Gaia* DR2 parallaxes for these systems one can construct an empirical relationship between the absolute magnitude in a given band and mass or radius. For each of the EBs mentioned above we computed absolute $G$ magnitudes from the *Gaia* photometry and parallaxes, then decomposed the $G$-band fluxes using the luminosity ratios derived from the *K2* eclipse photometry. We then computed $M_G$ magnitudes for each individual component, and performed cubic polynomial fits following the same procedure described above. The resulting relations relate the $M_G$ magnitude to mass or radius for stars in Upper Sco, and should be valid for single stars in the mass-range of $\sim$0.1–1 $M_\odot$.

**Table 9.** System Parameters of HR 5934.

| Parameter | Symbol | Value | Units |
|---|---|---|---|
| Orbital period | $P$ | $9.199740 \pm 0.000010$ | days |
| Ephemeris timebase - 2456000 | $T_0$ | $894.35768 \pm 0.00010$ | BJD |
| Surface brightness ratio | $J$ | $0.4727 \pm 0.0013$ | |
| Sum of fractional radii | $(R_1 + R_2)/a$ | $0.11857 \pm 0.00019$ | |
| Ratio of radii | $k$ | $0.61840 \pm 0.00070$ | |
| Orbital inclination | $i$ | $88.570 \pm 0.016$ | deg |
| Primary radial velocity amplitude | $K_1$ | $65.43 \pm 0.61$ | km s$^{-1}$ |
| Secondary radial velocity amplitude | $K_2$ | $139.4 \pm 2.1$ | km s$^{-1}$ |
| Systemic radial velocity | $\gamma$ | $-2.97 \pm 0.44$ | km s$^{-1}$ |
| Mass ratio | $q$ | $0.4694 \pm 0.0081$ | |
| Orbital semi-major axis | $a$ | $37.24 \pm 0.41$ | $R_\odot$ |
| Fractional radius of primary | $R_1/a$ | $0.07326 \pm 0.00011$ | |
| Fractional radius of secondary | $R_2/a$ | $0.045307 \pm 0.000092$ | |
| Luminosity ratio | $L_2/L_1$ | $0.17147 \pm 0.00058$ | |
| Primary mass | $M_1$ | $5.58 \pm 0.20$ | $M_\odot$ |
| Secondary mass | $M_2$ | $2.618 \pm 0.070$ | $M_\odot$ |
| Primary radius | $R_1$ | $2.728 \pm 0.030$ | $R_\odot$ |
| Secondary radius | $R_2$ | $1.687 \pm 0.019$ | $R_\odot$ |
| Primary surface gravity | $\log g_1$ | $4.3124 \pm 0.0067$ | cgs |
| Secondary surface gravity | $\log g_2$ | $4.4014 \pm 0.0044$ | cgs |
| Primary mean density | $\rho_1$ | $0.2747 \pm 0.0019$ | $\rho_\odot$ |
| Secondary mean density | $\rho_2$ | $0.5453 \pm 0.0072$ | $\rho_\odot$ |
| Impact parameter of primary eclipse | $b_1$ | $0.3407 \pm 0.0035$ | |
| Impact parameter of secondary eclipse | $b_2$ | $0.3407 \pm 0.0035$ | |
| Reduced chi-squared of joint fit | $\chi^2_{\rm red}$ | $1.36$ | |
| Reduced chi-squared of light curve fit | $\chi^2_{\rm red,LC}$ | $0.92$ | |
| Residuals of light curve fit | $\rm rms_{LC}$ | $3.30$ | mmag |
| Reduced chi-squared of primary RV fit | $\chi^2_{\rm red,RV1}$ | $9.54$ | |
| Residuals of primary RV fit | $\rm rms_{RV1}$ | $4.96$ | km s$^{-1}$ |
| Reduced chi-squared of secondary RV fit | $\chi^2_{\rm red,RV2}$ | $4.77$ | |
| Residuals of primary RV fit | $\rm rms_{RV2}$ | $2.50$ | km s$^{-1}$ |
| Reduced chi-squared of light ratio fit | $\chi^2_{\rm red,LR}$ | $4.49$ | |
| Residuals of light ratio fit | $\rm rms_{LR}$ | $0.27$ | |

For the mass as a function of $M_G$ we find

$$\left(\frac{M_*}{M_\odot}\right) = 9.28337 - 2.45149 M_G + 0.228305 M_G^2 - 7.33973 \times 10^{-3} M_G^3, \tag{A5}$$

and likewise for the radius

$$\left(\frac{R_*}{R_\odot}\right) = 6.91212 - 1.74205 M_G + 0.178815 M_G^2 - 6.62350 \times 10^{-3} M_G^3. \tag{A6}$$

The standard deviation between fits randomly drawn from the MCMC sampler is about 2.6% in mass and 2.0% in radius, averaged across the entire $M_G$ range. These relations are shown in Figure 27. The fits, at the numerical precision quoted, produce residuals of $\lesssim 3\%$ in mass and $\lesssim 2\%$ in radius when evaluated for the EBs.

### B. NEWLY-IDENTIFIED PRE-MAIN-SEQUENCE SPECTROSCOPIC BINARIES

In the course of our spectroscopic follow-up program of Upper Sco and $\rho$ Oph members or candidate members with *K2* observations, we identified two new non-eclipsing, double-lined spectroscopic binaries. The two systems are HD 145655 (EPIC 204185181, spectral type G2) and [PZ99]J1609-2217 (EPIC 204447221, spectral type M0). Both systems are secure members of Upper Sco, with membership probabilities $\geq$99.7%, as assessed with the BANYAN $\Sigma$ tool (Gagné et al. 2018). Additionally, both systems are known to host debris disks (Carpenter et al. 2009; Luhman & Mamajek 2012).

**Table 10.** System Parameters of USco 48.

| Parameter | Symbol | Value | | | Units |
|---|---|---|---|---|---|
| | | $k = 1$ | $J = 1$ | $k, J$ free | |
| Orbital period | $P$ | $2.874456 \pm 0.000014$ | $2.874456 \pm 0.000014$ | $2.874456$ (fixed) | days |
| Ephemeris timebase - 2456000 | $T_0$ | $904.90027 \pm 0.00036$ | $904.90027 \pm 0.00036$ | $904.90027$ (fixed) | BJD |
| Surface brightness ratio | $J$ | $0.986 \pm 0.015$ | $1.0$ (fixed) | $0.634 \pm 0.091$ | |
| Sum of fractional radii | $(R_1 + R_2)/a$ | $0.2418 \pm 0.0036$ | $0.2415 \pm 0.0030$ | $0.2529 \pm 0.0049$ | |
| Ratio of radii | $k$ | $1.0$ (fixed) | $0.9949 \pm 0.0074$ | $1.249 \pm 0.090$ | |
| Orbital inclination | $i$ | $76.96 \pm 0.18$ | $76.98 \pm 0.17$ | $76.31 \pm 0.28$ | deg |
| Combined eccentricity, periastron longitude | $e \cos \omega$ | $-0.00299 \pm 0.00033$ | $-0.00301 \pm 0.00033$ | $-0.00300 \pm 0.00031$ | |
| Combined eccentricity, periastron longitude | $e \sin \omega$ | $0.01616 \pm 0.00089$ | $0.01645 \pm 0.00040$ | $0.0072 \pm 0.0028$ | |
| Primary radial velocity amplitude | $K_1$ | $80.9 \pm 1.2$ | $80.9 \pm 1.2$ | $81.2 \pm 1.1$ | km s$^{-1}$ |
| Secondary radial velocity amplitude | $K_2$ | $84.1 \pm 1.0$ | $84.1 \pm 1.0$ | $84.47 \pm 0.97$ | km s$^{-1}$ |
| Systemic radial velocity | $\gamma$ | $-5.49 \pm 0.68$ | $-5.48 \pm 0.67$ | $-5.38 \pm 0.67$ | km s$^{-1}$ |
| Mass ratio | $q$ | $0.962 \pm 0.019$ | $0.962 \pm 0.019$ | $0.962 \pm 0.018$ | |
| Orbital semi-major axis | $a$ | $9.62 \pm 0.38$ | $9.618 \pm 0.083$ | $9.687 \pm 0.080$ | $R_\odot$ |
| Fractional radius of primary | $R_1/a$ | $0.1209 \pm 0.0018$ | $0.1211 \pm 0.0016$ | $0.1125 \pm 0.0032$ | |
| Fractional radius of secondary | $R_2/a$ | $0.1209 \pm 0.0018$ | $0.1204 \pm 0.0016$ | $0.1404 \pm 0.0068$ | |
| Luminosity ratio | $L_2/L_1$ | $0.995 \pm 0.015$ | $0.999 \pm 0.015$ | $0.999 \pm 0.014$ | |
| Primary mass | $M_1$ | $0.738 \pm 0.035$ | $0.737 \pm 0.020$ | $0.754 \pm 0.019$ | $M_\odot$ |
| Secondary mass | $M_2$ | $0.709 \pm 0.035$ | $0.709 \pm 0.020$ | $0.725 \pm 0.020$ | $M_\odot$ |
| Primary radius | $R_1$ | $1.164 \pm 0.051$ | $1.164 \pm 0.019$ | $1.090 \pm 0.031$ | $R_\odot$ |
| Secondary radius | $R_2$ | $1.164 \pm 0.051$ | $1.158 \pm 0.019$ | $1.360 \pm 0.070$ | $R_\odot$ |
| Primary surface gravity | $\log g_1$ | $4.175 \pm 0.012$ | $4.173 \pm 0.012$ | $4.240 \pm 0.026$ | cgs |
| Secondary surface gravity | $\log g_2$ | $4.157 \pm 0.013$ | $4.161 \pm 0.013$ | $4.030 \pm 0.042$ | cgs |
| Primary mean density | $\rho_1$ | $0.469 \pm 0.019$ | $0.467 \pm 0.019$ | $0.582 \pm 0.050$ | $\rho_\odot$ |
| Secondary mean density | $\rho_2$ | $0.451 \pm 0.019$ | $0.457 \pm 0.019$ | $0.288 \pm 0.042$ | $\rho_\odot$ |
| Impact parameter of primary eclipse | $b_1$ | $1.836 \pm 0.036$ | $1.8306 \pm 0.0070$ | $2.089 \pm 0.091$ | |
| Impact parameter of secondary eclipse | $b_2$ | $1.896 \pm 0.042$ | $1.8917 \pm 0.0071$ | $2.119 \pm 0.081$ | |
| Eccentricity | $e$ | $0.01643 \pm 0.00089$ | $0.01672 \pm 0.00040$ | $0.0078 \pm 0.0025$ | |
| Periastron longitude | $\omega$ | $100.5 \pm 1.2$ | $100.3 \pm 1.1$ | $112.6 \pm 9.0$ | deg |
| Reduced chi-squared of joint fit | $\chi^2_{red}$ | $1.217$ | $1.217$ | $1.214$ | |
| Reduced chi-squared of light curve fit | $\chi^2_{red,LC}$ | $1.218$ | $1.218$ | $1.216$ | |
| Residuals of light curve fit | $rms_{LC}$ | $0.722$ | $0.722$ | $0.721$ | mmag |
| Reduced chi-squared of primary RV fit | $\chi^2_{red,RV1}$ | $0.666$ | $0.674$ | $0.512$ | |
| Residuals of primary RV fit | $rms_{RV1}$ | $1.67$ | $1.68$ | $1.51$ | km s$^{-1}$ |
| Reduced chi-squared of secondary RV fit | $\chi^2_{red,RV2}$ | $0.300$ | $0.296$ | $0.455$ | |
| Residuals of primary RV fit | $rms_{RV2}$ | $1.39$ | $1.38$ | $1.77$ | km s$^{-1}$ |
| Reduced chi-squared of light ratio fit | $\chi^2_{red,LR}$ | $1.754$ | $1.745$ | $1.745$ | |
| Residuals of light ratio fit | $rms_{LR}$ | $0.085$ | $0.087$ | $0.087$ | |

**Table 11.** Parameters of the RIK 72 eclipsing binary

| Parameter | Units | Value 1 | Value 2 |
|---|---|---|---|
| Orbital period, $P$ | days | $110.1706 \pm 0.0085$ | $97.76 \pm 0.16$ |
| Time of primary minimum, $T_0$ | BJD | $2456911.51195 \pm 0.00051$ | $2456911.51207 \pm 0.00036$ |
| Eccentricity, $e$ | | 0.0 (fixed) | $0.1079^{+0.0116}_{-0.0062}$ |
| Longitude of periastron, $\omega$ | deg | 0.0 (fixed) | $22^{+11}_{-12}$ |
| Surface brightness ratio, $J$ | | $0.1505 \pm 0.0051$ | $0.1539^{+0.0129}_{-0.0098}$ |
| Sum of fractional radii, $(R_1 + R_2)/a$ | | $0.01466 \pm 0.00016$ | $0.01702^{+0.00043}_{-0.00037}$ |
| Ratio of radii, $k$ | | $0.3268 \pm 0.0045$ | $0.3307^{+0.0036}_{-0.0028}$ |
| Orbital inclination, $i$ | deg | $89.570 \pm 0.012$ | $89.473^{+0.023}_{-0.028}$ |
| Fractional radius of primary, $R_1/a$ | | $0.011050 \pm 0.000092$ | $0.01279^{+0.00032}_{-0.00028}$ |
| Fractional radius of secondary, $R_2/a$ | | $0.003611 \pm 0.000074$ | $0.00423^{+0.00012}_{-0.00010}$ |
| *Adopted parameters* | | | |
| Primary mass, $M_1$ | $M_\odot$ | $0.439 \pm 0.044$ | |
| Primary radius, $R_1$ | $R_\odot$ | $0.961 \pm 0.096$ | |
| Primary temperature, $T_{\mathrm{eff},1}$ | K | $3349 \pm 142$ | |
| Secondary temperature, $T_{\mathrm{eff},2}$ | K | $2722 \pm 98$ | |
| *Derived parameters* | | | |
| Secondary mass, $M_1$ | $M_{\mathrm{Jup}}$ | $56.1 \pm 7.7$ | $59.2^{+6.8}_{-6.7}$ |
| Secondary radius, $R_1$ | $R_{\mathrm{Jup}}$ | $3.06 \pm 0.32$ | $3.10 \pm 0.31$ |

Values quoted for Fit 1 are best fit parameters and 1-$\sigma$ uncertainties from 5,000 Monte Carlo realizations with JKTEBOP. Values quoted for Fit 2 were calculated using the GP-EBOP model presented in Gillen et al. (2017a).

**Table 12.** MCMC Posteriors from RIK 72 RV fit.

| Parameter | Credible Interval | Maximum Likelihood | Units |
|---|---|---|---|
| $P_b$ | $97.84^{+0.3}_{-0.19}$ | 97.82 | days |
| $T\mathrm{conj}_b$ | $2456911.51195^{+0.00097}_{-0.001}$ | 2456911.512 | JD |
| $e_b$ | $0.131^{+0.1}_{-0.027}$ | 0.13 | |
| $\omega_b$ | $0.71 \pm 0.44$ | 0.71 | radians |
| $K_b$ | $4335^{+550}_{-430}$ | 4337 | m s$^{-1}$ |
| $\gamma_{\mathrm{hires}}$ | $-4504^{+380}_{-330}$ | $-4526$ | |
| $\sigma_{\mathrm{hires}}$ | $645^{+180}_{-160}$ | 520 | |

Parameters determined from 80000 links in the MCMC chain.

Priors on the sampled parameters were as follows:

$e_b$ constrained to be $< 0.99$

Bounded prior: $0.0 < \sigma_{\mathrm{hires}} < 1000.0$

Bounded prior: $70.0 < P_b < 140.0$

Gaussian prior on $T\mathrm{conj}_b$: $2456911.51195 \pm 0.001$

Gaussian prior on $\ln K_b$: $8.39 \pm 10$

Gaussian prior on $\gamma_{\mathrm{hires}}$: $-4400 \pm 1000$

Secondary eclipse prior: $2456966.59722 \pm 0.1$

**Table 13.** System Parameters of UScoCTIO 5.

| Parameter | Symbol | Value | | | Units |
|---|---|---|---|---|---|
| | | $k = 1$ | $J = 1$ | $k, J$ free | |
| Orbital period | $P$ | $34.000320 \pm 0.000057$ | $34.000259 \pm 0.000055$ | $33.999938 \pm 0.000061$ | days |
| Ephemeris timebase - 2456000 | $T_0$ | $909.253713 \pm 0.000074$ | $909.253767 \pm 0.000075$ | $909.254047 \pm 0.000078$ | BJD |
| Surface brightness ratio | $J$ | $0.9619 \pm 0.0065$ | 1.0 (fixed) | $1.254 \pm 0.023$ | |
| Sum of fractional radii | $(R_1 + R_2)/a$ | $0.044406 \pm 0.000039$ | $0.044387 \pm 0.000039$ | $0.044339 \pm 0.000039$ | |
| Ratio of radii | $k$ | 1.0 (fixed) | $0.9591 \pm 0.0035$ | $0.8578 \pm 0.0087$ | |
| Orbital inclination | $i$ | $87.9035 \pm 0.0020$ | $87.9053 \pm 0.0020$ | $87.9126 \pm 0.0021$ | deg |
| Combined eccentricity, periastron longitude | $e \cos \omega$ | $-0.2664273 \pm 0.0000066$ | $-0.2664020 \pm 0.0000057$ | $-0.266189 \pm 0.000025$ | |
| Combined eccentricity, periastron longitude | $e \sin \omega$ | $0.01977 \pm 0.00061$ | $0.02402 \pm 0.00014$ | $0.0463 \pm 0.0019$ | |
| Primary radial velocity amplitude | $K_1$ | $29.206 \pm 0.085$ | $29.178 \pm 0.082$ | $29.034 \pm 0.085$ | km s$^{-1}$ |
| Secondary radial velocity amplitude | $K_2$ | $30.368 \pm 0.079$ | $30.334 \pm 0.081$ | $30.167 \pm 0.081$ | km s$^{-1}$ |
| Systemic radial velocity | $\gamma$ | $-2.665 \pm 0.040$ | $-2.664 \pm 0.040$ | $-2.662 \pm 0.040$ | km s$^{-1}$ |
| Mass ratio | $q$ | $0.9617 \pm 0.0038$ | $0.9619 \pm 0.0038$ | $0.9624 \pm 0.0038$ | |
| Orbital semi-major axis | $a$ | $38.590 \pm 0.076$ | $38.546 \pm 0.074$ | $38.314 \pm 0.076$ | $R_\odot$ |
| Fractional radius of primary | $R_1/a$ | $0.022203 \pm 0.000019$ | $0.022656 \pm 0.000046$ | $0.02387 \pm 0.00011$ | |
| Fractional radius of secondary | $R_2/a$ | $0.022203 \pm 0.000019$ | $0.021731 \pm 0.000046$ | $0.02047 \pm 0.00011$ | |
| Luminosity ratio | $L_2/L_1$ | $0.9619 \pm 0.0065$ | $0.9199 \pm 0.0068$ | $0.9229 \pm 0.0068$ | |
| Primary mass | $M_1$ | $0.3405 \pm 0.0020$ | $0.3393 \pm 0.0020$ | $0.3331 \pm 0.0021$ | $M_\odot$ |
| Secondary mass | $M_2$ | $0.3274 \pm 0.0021$ | $0.3263 \pm 0.0020$ | $0.3206 \pm 0.0021$ | $M_\odot$ |
| Primary radius | $R_1$ | $0.8568 \pm 0.0018$ | $0.8733 \pm 0.0024$ | $0.9144 \pm 0.0042$ | $R_\odot$ |
| Secondary radius | $R_2$ | $0.8568 \pm 0.0018$ | $0.8376 \pm 0.0024$ | $0.7844 \pm 0.0050$ | $R_\odot$ |
| Primary surface gravity | $\log g_1$ | $4.1040 \pm 0.0013$ | $4.0859 \pm 0.0021$ | $4.0380 \pm 0.0045$ | cgs |
| Secondary surface gravity | $\log g_2$ | $4.0870 \pm 0.0015$ | $4.1053 \pm 0.0021$ | $4.1545 \pm 0.0048$ | cgs |
| Primary mean density | $\rho_1$ | $0.5412 \pm 0.0018$ | $0.5094 \pm 0.0033$ | $0.4357 \pm 0.0062$ | $\rho_\odot$ |
| Secondary mean density | $\rho_2$ | $0.5205 \pm 0.0017$ | $0.5553 \pm 0.0037$ | $0.664 \pm 0.011$ | $\rho_\odot$ |
| Impact parameter of primary eclipse | $b_1$ | $1.5004 \pm 0.0010$ | $1.4627 \pm 0.0028$ | $1.3521 \pm 0.0092$ | |
| Impact parameter of secondary eclipse | $b_2$ | $1.56088 \pm 0.00093$ | $1.5347 \pm 0.0027$ | $1.4835 \pm 0.0048$ | |
| Eccentricity | $e$ | $0.267160 \pm 0.000043$ | $0.267483 \pm 0.000014$ | $0.27019 \pm 0.00030$ | |
| Periastron longitude | $\omega$ | $175.76 \pm 0.13$ | $174.847 \pm 0.030$ | $170.12 \pm 0.39$ | deg |
| Reduced chi-squared of joint fit | $\chi^2_{\rm red}$ | 30.25 | 29.59 | 28.56 | |
| Reduced chi-squared of light curve fit | $\chi^2_{\rm red,LC}$ | 34.83 | 34.28 | 32.14 | |
| Residuals of light curve fit | $\rm rms_{LC}$ | 1.17 | 1.16 | 1.13 | mmag |
| Reduced chi-squared of primary RV fit | $\chi^2_{\rm red,RV1}$ | 5.88 | 5.65 | 7.66 | |
| Residuals of primary RV fit | $\rm rms_{RV1}$ | 0.487 | 0.487 | 0.64 | km s$^{-1}$ |
| Reduced chi-squared of secondary RV fit | $\chi^2_{\rm red,RV2}$ | 3.55 | 3.90 | 9.13 | |
| Residuals of primary RV fit | $\rm rms_{RV2}$ | 0.371 | 0.396 | 0.67 | km s$^{-1}$ |
| Reduced chi-squared of light ratio fit | $\chi^2_{\rm red,LR}$ | 5.49 | 2.76 | 2.75 | |
| Residuals of light ratio fit | $\rm rms_{LR}$ | 0.0915 | 0.0714 | 0.0723 | |

**Table 14.** Preliminary Parameters of EPIC 202963882.

| Parameter | Symbol | Value | Units |
|---|---|---|---|
| Orbital period | $P$ | 0.630793 | days |
| Ephemeris timebase - 2456000 | $T_0$ | 893.863021 | BJD |
| Surface brightness ratio | $J$ | 0.8051 | |
| Sum of fractional radii | $(R_1 + R_2)/a$ | 0.5678 | |
| Ratio of radii | $k$ | 0.8620 | |
| Orbital inclination | $i$ | 89.94 | deg |
| Combined eccentricity, periastron longitude | $e \cos \omega$ | -0.00452 | |
| Combined eccentricity, periastron longitude | $e \sin \omega$ | -0.04257 | |
| Primary radial velocity amplitude | $K_1$ | 78.45 | km s$^{-1}$ |
| Secondary radial velocity amplitude | $K_2$ | 117.45 | km s$^{-1}$ |
| Systemic radial velocity | $\gamma$ | -4.75 | km s$^{-1}$ |
| Mass ratio | $q$ | 0.6680 | |
| Orbital semi-major axis | $a$ | 2.44 | $R_\odot$ |
| Fractional radius of primary | $R_1/a$ | 0.3049 | |
| Fractional radius of secondary | $R_2/a$ | 0.2629 | |
| Luminosity ratio | $L_2/L_1$ | 0.5975 | |
| Primary mass | $M_1$ | 0.294 | $M_\odot$ |
| Secondary mass | $M_2$ | 0.196 | $M_\odot$ |
| Primary radius | $R_1$ | 0.744 | $R_\odot$ |
| Secondary radius | $R_2$ | 0.641 | $R_\odot$ |
| Primary surface gravity | $\log g_1$ | 4.16 | cgs |
| Secondary surface gravity | $\log g_2$ | 4.12 | cgs |
| Primary mean density | $\rho_1$ | 0.714 | $\rho_\odot$ |
| Secondary mean density | $\rho_2$ | 0.745 | $\rho_\odot$ |
| Impact parameter of primary eclipse | $b_1$ | 0.00357 | |
| Impact parameter of secondary eclipse | $b_2$ | 0.00328 | |
| Eccentricity | $e$ | 0.0428 | |
| Periastron longitude | $\omega$ | 263.9 | deg |
| Third light | $l_3$ | 0.8117 | |
| Light scale factor | $s$ | -0.00491 | |
| Photometric mass ratio | $q_{\rm phot}$ | 0.7815 | |
| Reduced chi-squared of joint fit | $\chi^2_{\rm red}$ | 1.706 | |
| Reduced chi-squared of light curve fit | $\chi^2_{\rm red,LC}$ | 1.709 | |
| Residuals of light curve fit | $\rm rms_{LC}$ | 10.38 | mmag |
| Reduced chi-squared of primary RV fit | $\chi^2_{\rm red,RV1}$ | 0.58 | |
| Residuals of primary RV fit | $\rm rms_{RV1}$ | 7.6 | km s$^{-1}$ |
| Reduced chi-squared of secondary RV fit | $\chi^2_{\rm red,RV2}$ | 0.17 | |
| Residuals of primary RV fit | $\rm rms_{RV2}$ | 4.1 | km s$^{-1}$ |
| Reduced chi-squared of light ratio fit | $\chi^2_{\rm red,LR}$ | 0.19 | |
| Residuals of light ratio fit | $\rm rms_{LR}$ | 4.77 | |

**Table 15.** Parameters of the EPIC 203868608 A spectroscopic binary

| Parameter | Units | Value |
|---|---|---|
| Orbital period, $P$ | days | $17.9420 \pm 0.0012$ |
| Epoch, $T_0$ | BJD | $2457175.182 \pm 0.031$ |
| Primary Doppler semi-amplitude, $K_1$ | km s$^{-1}$ | $26.46 \pm 0.16$ |
| Secondary Doppler semi-amplitude, $K_2$ | km s$^{-1}$ | $31.84 \pm 0.18$ |
| Systemic radial velocity, $\gamma$ | km s$^{-1}$ | $-4.436 \pm 0.072$ |
| Eccentricity, $e$ | | $0.2998 \pm 0.0041$ |
| Longitude of periastron, $\omega$ | deg | $316.36 \pm 0.93$ |
| RMS of primary RV fit | km s$^{-1}$ | $0.6$ |
| RMS of secondary RV fit | km s$^{-1}$ | $0.8$ |
| $\chi^2_{\rm red}$ of primary RV fit | | $2.2$ |
| $\chi^2_{\rm red}$ of secondary RV fit | | $2.7$ |
| | | |
| *Derived parameters* | | |
| Mass ratio, $q$ | | $0.8309 \pm 0.0062$ |
| Minimum system mass, $(M_1 + M_2) \sin^3 i$ | $M_\odot$ | $0.3685 \pm 0.0050$ |
| Orbital separation, $a$ | AU | $0.09616 \pm 0.00044$ |

Values quoted are best fit parameters and 1-$\sigma$ uncertainties from 10,000 Monte Carlo realizations with JKTEBOP.

**Table 16.** System Parameters of EPIC 203710387.

| Parameter | Symbol | Value | | | Units |
|---|---|---|---|---|---|
| | | $k = 1$ | $J = 1$ | $k, J$ free | |
| Orbital period | $P$ | $2.808851 \pm 0.000012$ | $2.808852 \pm 0.000013$ | $2.808850 \pm 0.000013$ | days |
| Ephemeris timebase - 2456000 | $T_0$ | $894.71419 \pm 0.00023$ | $894.71418 \pm 0.00022$ | $894.71421 \pm 0.00023$ | BJD |
| Surface brightness ratio | $J$ | $0.945 \pm 0.032$ | 1.0 (fixed) | $0.850 \pm 0.073$ | |
| Sum of fractional radii | $(R_1 + R_2)/a$ | $0.16881 \pm 0.00082$ | $0.16883 \pm 0.00082$ | $0.16885 \pm 0.00083$ | |
| Ratio of radii | $k$ | 1.0 (fixed) | $0.981 \pm 0.017$ | $1.062 \pm 0.048$ | |
| Orbital inclination | $i$ | $82.858 \pm 0.039$ | $82.857 \pm 0.039$ | $82.857 \pm 0.039$ | deg |
| Combined eccentricity, periastron longitude | $e \cos \omega$ | $-0.00337 \pm 0.00012$ | $-0.00337 \pm 0.00011$ | $-0.00337 \pm 0.00012$ | |
| Combined eccentricity, periastron longitude | $e \sin \omega$ | $0.0007 \pm 0.0035$ | $0.00699 \pm 0.00080$ | $-0.012 \pm 0.010$ | |
| Primary radial velocity amplitude | $K_1$ | $43.27 \pm 0.48$ | $43.20 \pm 0.50$ | $43.40 \pm 0.52$ | km s$^{-1}$ |
| Secondary radial velocity amplitude | $K_2$ | $47.49 \pm 0.55$ | $47.39 \pm 0.57$ | $47.70 \pm 0.58$ | km s$^{-1}$ |
| Systemic radial velocity | $\gamma$ | $-3.26 \pm 0.22$ | $-3.22 \pm 0.22$ | $-3.34 \pm 0.24$ | km s$^{-1}$ |
| Mass ratio | $q$ | $0.911 \pm 0.015$ | $0.912 \pm 0.015$ | $0.910 \pm 0.015$ | |
| Orbital semi-major axis | $a$ | $5.076 \pm 0.041$ | $5.066 \pm 0.042$ | $5.095 \pm 0.044$ | $R_\odot$ |
| Fractional radius of primary | $R_1/a$ | $0.08441 \pm 0.00041$ | $0.08522 \pm 0.00084$ | $0.0819 \pm 0.0019$ | |
| Fractional radius of secondary | $R_2/a$ | $0.08441 \pm 0.00041$ | $0.08361 \pm 0.00086$ | $0.0870 \pm 0.0020$ | |
| Luminosity ratio | $L_2/L_1$ | $0.945 \pm 0.032$ | $0.963 \pm 0.034$ | $0.960 \pm 0.034$ | |
| Primary mass | $M_1$ | $0.1165 \pm 0.0031$ | $0.1158 \pm 0.0031$ | $0.1179 \pm 0.0033$ | $M_\odot$ |
| Secondary mass | $M_2$ | $0.1062 \pm 0.0027$ | $0.1056 \pm 0.0027$ | $0.1073 \pm 0.0029$ | $M_\odot$ |
| Primary radius | $R_1$ | $0.4284 \pm 0.0041$ | $0.4317 \pm 0.0055$ | $0.4171 \pm 0.0093$ | $R_\odot$ |
| Secondary radius | $R_2$ | $0.4284 \pm 0.0041$ | $0.4236 \pm 0.0056$ | $0.4432 \pm 0.012$ | $R_\odot$ |
| Primary surface gravity | $\log g_1$ | $4.2404 \pm 0.0066$ | $4.231 \pm 0.010$ | $4.269 \pm 0.022$ | cgs |
| Secondary surface gravity | $\log g_2$ | $4.1999 \pm 0.0065$ | $4.207 \pm 0.010$ | $4.175 \pm 0.019$ | cgs |
| Primary mean density | $\rho_1$ | $1.482 \pm 0.024$ | $1.439 \pm 0.043$ | $1.625 \pm 0.12$ | $\rho_\odot$ |
| Secondary mean density | $\rho_2$ | $1.350 \pm 0.023$ | $1.389 \pm 0.045$ | $1.233 \pm 0.086$ | $\rho_\odot$ |
| Impact parameter of primary eclipse | $b_1$ | $1.4719 \pm 0.0051$ | $1.449 \pm 0.013$ | $1.537 \pm 0.050$ | |
| Impact parameter of secondary eclipse | $b_2$ | $1.4741 \pm 0.0054$ | $1.469 \pm 0.012$ | $1.501 \pm 0.022$ | |
| Eccentricity | $e$ | $0.0035 \pm 0.0013$ | $0.00776 \pm 0.00072$ | $0.0121 \pm 0.0083$ | |
| Periastron longitude | $\omega$ | $167.6 \pm 45.3$ | $115.8 \pm 2.7$ | $254 \pm 29$ | deg |
| Reduced chi-squared of joint fit | $\chi^2_{\rm red}$ | 1.630 | 1.632 | 1.630 | |
| Reduced chi-squared of light curve fit | $\chi^2_{\rm red, LC}$ | 1.630 | 1.630 | 1.630 | |
| Residuals of light curve fit | $\rm rms_{LC}$ | 3.28 | 3.28 | 3.28 | mmag |
| Reduced chi-squared of primary RV fit | $\chi^2_{\rm red, RV1}$ | 1.201 | 1.223 | 1.278 | |
| Residuals of primary RV fit | $\rm rms_{RV1}$ | 0.71 | 0.72 | 0.73 | km s$^{-1}$ |
| Reduced chi-squared of secondary RV fit | $\chi^2_{\rm red, RV2}$ | 0.909 | 1.197 | 0.578 | |
| Residuals of primary RV fit | $\rm rms_{RV2}$ | 0.91 | 1.06 | 0.71 | km s$^{-1}$ |
| Reduced chi-squared of light ratio fit | $\chi^2_{\rm red, LR}$ | 0.044 | 0.027 | 0.024 | |
| Residuals of light ratio fit | $\rm rms_{LR}$ | 0.017 | 0.018 | 0.016 | |

**Table 17.** Final adopted parameters of the EBs

| Star | SpT | Mass | Radius | $T_{\rm eff}$ | $\log(L_{\rm bol}/L_\odot)$ |
|------|-----|------|--------|---------------|-----------------------------|
|      |     | ($M_\odot$) | ($R_\odot$) | (K) | (dex) |
| HR 5934 A | B2.5 ± 0.5 | 5.58 ± 0.20 | 2.728 ± 0.030 | 18500 ± 500 | 2.894 ± 0.048 |
| HR 5934 B | B8.0 ± 0.5 | 2.618 ± 0.070 | 1.687 ± 0.019 | 11500 ± 500 | 1.650 ± 0.076 |
| HD 144548 A | F7.5 ± 0.5 | 1.44 ± 0.04 | 2.41 ± 0.03 | 6210 ± 80 | 0.891 ± 0.025 |
| HD 144548 Ba | K5.0 ± 0.5 | 0.984 ± 0.007 | 1.319 ± 0.010 | 4210 ± 200 | -0.310 ± 0.083 |
| HD 144548 Bb | K5.0 ± 0.5 | 0.944 ± 0.017 | 1.330 ± 0.010 | 4210 ± 200 | -0.302 ± 0.083 |
| USco 48 A | M1.0 ± 0.5 | 0.737 ± 0.020 | 1.164 ± 0.019 | 3656 ± 90 | -0.662 ± 0.045 |
| USco 48 B | M1.0 ± 0.5 | 0.709 ± 0.020 | 1.158 ± 0.019 | 3650 ± 90 | -0.669 ± 0.045 |
| UScoCTIO 5 A | M4.5 ± 0.5 | 0.3393 ± 0.0020 | 0.8733 ± 0.0024 | 3272 ± 100 | -1.105 ± 0.053 |
| UScoCTIO 5 B | M4.5 ± 0.5 | 0.3263 ± 0.0020 | 0.8376 ± 0.0024 | 3262 ± 100 | -1.146 ± 0.059 |
| EPIC 203710387 A | M4.75 ± 0.25 | 0.1158 ± 0.0031 | 0.4317 ± 0.0055 | 3044 ± 80 | -1.842 ± 0.047 |
| EPIC 203710387 B | M4.75 ± 0.25 | 0.1056 ± 0.0027 | 0.4236 ± 0.0056 | 3040 ± 80 | -1.861 ± 0.047 |

The masses and radii for HD 144548 originate from Alonso et al. (2015). All other parameters originate from this work.

**Table 18.** Best-fitting mass-radius isochrones

|  | Case 1 | Case 2 | Case 3 | Case 4 |
|------|--------|--------|--------|--------|
| Model | Age (Myr) | Age (Myr) | Age (Myr) | Age (Myr) |
| BHAC15 | ⋯ | ⋯ | 7.1 | 6.8 |
| Columbus (spot-free) | ⋯ | ⋯ | 5.0 | 5.0 |
| Columbus (50% spotted) | ⋯ | ⋯ | 6.9 | 6.9 |
| Dartmouth (standard) | 7.1 | 7.1 | 7.3 | 6.6 |
| Dartmouth (magnetic) | ⋯ | 10.2 | 9.6 | 8.9 |
| MIST | 7.4 | 7.4 | 7.1 | 6.5 |
| PARSEC v1.0 | 5.8 | 5.8 | 6.0 | 5.5 |
| PARSEC v1.1 | 5.2 | 5.2 | 5.5 | 5.5 |
| PARSEC v1.2S | 11.0 | 11.0 | 8.7 | 7.9 |

Case 1: all EB components listed in Table 17.
Case 2: EB components with $M_* < 1.5\ M_\odot$.
Case 3: EB components with $M_* < 1.0\ M_\odot$.
Case 4: EB components with $0.3\ M_\odot < M_* < 1.0\ M_\odot$.

**Table 19.** Best-fit H-R Diagram Ages.

| Model | All stars | | $M_* < 1 M_\odot$ | |
|------|-----------|-----------|-----------|-----------|
|      | Age (Myr) | $\chi^2_{\rm red}$ | Age (Myr) | $\chi^2_{\rm red}$ |
| BHAC15 | ⋯ | ⋯ | 6.3 | 9.06 |
| SP15 (spot-free) | ⋯ | ⋯ | 3.1 | 2.23 |
| SP15 (starspots) | ⋯ | ⋯ | 10.1 | 7.66 |
| Dartmouth | 6.8 | 9.28 | 4.3 | 0.41 |
| Dartmouth (magnetic) | 8.7 | 1.59 | 9.1 | 1.42 |
| MIST | 6.8 | 9.04 | 4.3 | 0.67 |
| PARSEC v1.0 | 5.2 | 39.0 | 2.2 | 2.34 |
| PARSEC v1.1 | 5.2 | 36.2 | 2.5 | 1.02 |
| PARSEC v1.2S | 5.2 | 22.0 | 5.2 | 11.5 |

**Table 20.** Individual ages from the mass-radius diagram

| Star | BHAC15 | SP15 (spot-free) | SP15 (spotted) | Dartmouth (standard) | Dartmouth (magnetic) | MIST | PARSEC (v1.0) | PARSEC (v1.1) | PARSEC (v1.2S) |
|---|---|---|---|---|---|---|---|---|---|
| HR 5934 A | $\cdots$ | $\cdots$ | $\cdots$ | $2.8^{+2.7}_{-1.3}$ | $\cdots$ | $4.1^{+3.8}_{-2.3}$ | $2.8^{+2.8}_{-1.3}$ | $2.8^{+2.8}_{-1.3}$ | $3.2^{+3.3}_{-2.2}$ |
| HR 5934 B | $\cdots$ | $\cdots$ | $\cdots$ | $14.6^{+2.6}_{-2.4}$ | $\cdots$ | $9.4^{+1.2}_{-1.0}$ | $6.5^{+0.4}_{-0.3}$ | $6.5^{+0.3}_{-0.3}$ | $7.8^{+0.9}_{-0.9}$ |
| HD 144548 A | $1.7^{+0.1}_{-0.1}$ | $\cdots$ | $\cdots$ | $1.5^{+0.1}_{-0.1}$ | $2.1^{+0.1}_{-0.1}$ | $1.5^{+0.1}_{-0.1}$ | $1.5^{+0.1}_{-0.1}$ | $1.5^{+0.1}_{-0.1}$ | $1.5^{+0.1}_{-0.1}$ |
| HD 144548 Ba | $6.9^{+0.2}_{-0.2}$ | $5.7^{+0.2}_{-0.2}$ | $7.6^{+0.3}_{-0.3}$ | $6.1^{+0.2}_{-0.2}$ | $10.2^{+0.3}_{-0.3}$ | $6.4^{+0.2}_{-0.2}$ | $5.8^{+0.2}_{-0.2}$ | $5.9^{+0.2}_{-0.2}$ | $5.7^{+0.2}_{-0.2}$ |
| HD 144548 Bb | $6.3^{+0.3}_{-0.3}$ | $5.2^{+0.2}_{-0.2}$ | $6.8^{+0.3}_{-0.3}$ | $5.6^{+0.2}_{-0.2}$ | $9.3^{+0.4}_{-0.4}$ | $5.8^{+0.2}_{-0.2}$ | $5.3^{+0.2}_{-0.2}$ | $5.3^{+0.2}_{-0.2}$ | $5.2^{+0.2}_{-0.2}$ |
| USco 48 A | $6.9^{+0.4}_{-0.4}$ | $5.9^{+0.5}_{-0.4}$ | $7.9^{+0.6}_{-0.6}$ | $6.5^{+0.4}_{-0.4}$ | $9.7^{+0.6}_{-0.6}$ | $6.6^{+0.5}_{-0.4}$ | $5.8^{+0.4}_{-0.4}$ | $5.8^{+0.4}_{-0.4}$ | $6.2^{+0.5}_{-0.5}$ |
| USco 48 B | $6.7^{+0.4}_{-0.4}$ | $5.7^{+0.4}_{-0.4}$ | $7.6^{+0.6}_{-0.6}$ | $6.3^{+0.4}_{-0.4}$ | $9.4^{+0.6}_{-0.6}$ | $6.4^{+0.4}_{-0.4}$ | $5.6^{+0.4}_{-0.3}$ | $5.6^{+0.4}_{-0.3}$ | $6.8^{+0.5}_{-0.5}$ |
| UScoCTIO 5 A | $6.8^{+0.1}_{-0.1}$ | $4.4^{+0.1}_{-0.1}$ | $6.2^{+0.1}_{-0.1}$ | $6.7^{+0.1}_{-0.1}$ | $8.3^{+0.1}_{-0.1}$ | $6.4^{+0.1}_{-0.1}$ | $5.2^{+0.1}_{-0.1}$ | $5.4^{+0.1}_{-0.1}$ | $8.9^{+0.1}_{-0.1}$ |
| UScoCTIO 5 B | $7.3^{+0.1}_{-0.1}$ | $4.8^{+0.1}_{-0.1}$ | $6.7^{+0.1}_{-0.1}$ | $7.2^{+0.1}_{-0.1}$ | $8.8^{+0.1}_{-0.1}$ | $6.9^{+0.1}_{-0.1}$ | $5.6^{+0.1}_{-0.1}$ | $5.8^{+0.1}_{-0.1}$ | $9.7^{+0.1}_{-0.1}$ |
| EPIC 203710387 A | $11.8^{+0.6}_{-0.6}$ | $5.2^{+0.7}_{-0.6}$ | $7.8^{+1.0}_{-0.9}$ | $11.9^{+0.5}_{-0.5}$ | $14.6^{+0.6}_{-0.5}$ | $11.2^{+0.6}_{-0.5}$ | $7.9^{+0.4}_{-0.4}$ | $8.9^{+0.4}_{-0.4}$ | $19.8^{+1.0}_{-1.0}$ |
| EPIC 203710387 B | $10.6^{+0.6}_{-0.5}$ | $3.9^{+0.6}_{-0.6}$ | $6.0^{+0.9}_{-0.8}$ | $11.2^{+0.5}_{-0.4}$ | $14.0^{+0.6}_{-0.5}$ | $10.6^{+0.5}_{-0.5}$ | $7.3^{+0.4}_{-0.4}$ | $8.3^{+0.4}_{-0.4}$ | $18.7^{+0.9}_{-0.9}$ |

Ages are in Myr.

**Table 21.** Individual ages from the H-R diagram

| Star | BHAC15 | SP15 (spot-free) | SP15 (spotted) | Dartmouth (standard) | Dartmouth (magnetic) | MIST | PARSEC (v1.0) | PARSEC (v1.1) | PARSEC (v1.2S) |
|---|---|---|---|---|---|---|---|---|---|
| HR 5934 A | $\cdots$ | $\cdots$ | $\cdots$ | $3.2^{+2.6}_{-1.5}$ | $\cdots$ | $5.3^{+3.8}_{-2.7}$ | $4.7^{+3.6}_{-2.5}$ | $4.7^{+3.6}_{-2.5}$ | $4.7^{+3.6}_{-2.6}$ |
| HR 5934 B | $\cdots$ | $\cdots$ | $\cdots$ | $7.3^{+2.3}_{-0.6}$ | $\cdots$ | $8.0^{+6.2}_{-0.7}$ | $5.3^{+0.9}_{-0.7}$ | $5.3^{+1.1}_{-0.6}$ | $6.3^{+0.6}_{-0.5}$ |
| HD 144548 A | $\cdots$ | $\cdots$ | $\cdots$ | $7.6^{+0.3}_{-0.4}$ | $7.6^{+0.2}_{-0.2}$ | $7.5^{+0.3}_{-0.4}$ | $7.3^{+0.3}_{-0.3}$ | $7.3^{+0.3}_{-0.3}$ | $7.3^{+0.3}_{-0.3}$ |
| HD 144548 Ba | $6.3^{+2.1}_{-1.8}$ | $5.3^{+2.2}_{-1.4}$ | $10.0^{+3.0}_{-2.5}$ | $5.4^{+1.3}_{-1.2}$ | $13.3^{+2.3}_{-2.5}$ | $5.9^{+1.2}_{-1.4}$ | $4.0^{+1.5}_{-1.1}$ | $4.0^{+1.5}_{-1.1}$ | $4.3^{+1.2}_{-0.3}$ |
| HD 144548 Bb | $6.1^{+2.0}_{-1.7}$ | $5.1^{+2.2}_{-1.4}$ | $9.8^{+3.1}_{-2.5}$ | $5.2^{+1.2}_{-1.2}$ | $12.9^{+2.2}_{-2.4}$ | $5.8^{+1.2}_{-1.3}$ | $3.9^{+1.4}_{-1.1}$ | $3.9^{+1.4}_{-1.1}$ | $4.1^{+1.2}_{-0.3}$ |
| USco 48 A | $4.0^{+0.8}_{-0.6}$ | $2.9^{+1.0}_{-0.9}$ | $8.0^{+1.5}_{-1.4}$ | $4.0^{+0.6}_{-0.5}$ | $9.0^{+1.7}_{-1.5}$ | $4.0^{+0.7}_{-0.6}$ | $2.3^{+0.4}_{-0.4}$ | $2.3^{+0.4}_{-0.4}$ | $6.6^{+0.6}_{-0.6}$ |
| USco 48 B | $4.1^{+0.8}_{-0.7}$ | $2.9^{+1.1}_{-0.9}$ | $8.1^{+1.5}_{-1.4}$ | $4.1^{+0.6}_{-0.5}$ | $9.1^{+1.7}_{-1.5}$ | $4.0^{+0.7}_{-0.6}$ | $2.3^{+0.4}_{-0.4}$ | $2.5^{+0.4}_{-0.4}$ | $6.8^{+0.6}_{-0.6}$ |
| UScoCTIO 5 A | $5.7^{+0.6}_{-1.3}$ | $2.4^{+1.9}_{-1.0}$ | $10.1^{+4.2}_{-3.4}$ | $4.5^{+1.1}_{-0.9}$ | $7.4^{+2.5}_{-1.7}$ | $3.8^{+1.5}_{-1.0}$ | $1.7^{+0.6}_{-0.5}$ | $2.4^{+0.4}_{-0.3}$ | $12.4^{+2.4}_{-2.3}$ |
| UScoCTIO 5 B | $5.8^{+2.3}_{-1.4}$ | $2.6^{+2.0}_{-1.2}$ | $11.4^{+5.1}_{-4.0}$ | $4.7^{+1.4}_{-0.8}$ | $7.9^{+2.8}_{-1.7}$ | $4.2^{+1.7}_{-1.2}$ | $1.7^{+0.7}_{-0.5}$ | $1.7^{+0.7}_{-0.5}$ | $13.9^{+3.3}_{-2.8}$ |
| EPIC 203710387 A | $11.0^{+9.7}_{-3.3}$ | $6.9^{+6.1}_{-4.8}$ | $46.8^{+20.1}_{-19.6}$ | $12.1^{+7.3}_{-2.1}$ | $20.3^{+7.3}_{-5.8}$ | $12.9^{+5.3}_{-2.5}$ | $10.0^{+3.7}_{-3.4}$ | $12.0^{+3.2}_{-4.2}$ | $24.8^{+0.1}_{-0.1}$ |
| EPIC 203710387 B | $10.2^{+12.4}_{-2.4}$ | $7.0^{+6.6}_{-4.7}$ | $48.7^{+20.4}_{-20.2}$ | $12.6^{+7.6}_{-2.2}$ | $19.2^{+7.0}_{-3.9}$ | $13.6^{+5.5}_{-2.6}$ | $10.5^{+3.6}_{-3.4}$ | $12.4^{+3.1}_{-4.3}$ | $24.8^{+0.1}_{-0.1}$ |

Ages are in Myr.

**Table 22.** Offsets between dynamical and model-derived masses from the H-R diagram

| Star | BHAC15 | SP15 (spot-free) | SP15 (spotted) | Dartmouth (standard) | Dartmouth (magnetic) | MIST | PARSEC (v1.0) | PARSEC (v1.1) | PARSEC (v1.2S) |
|---|---|---|---|---|---|---|---|---|---|
| USco 48 A | $-36^{+6}_{-6}$% | $-34^{+7}_{-6}$% | $0^{+6}_{-6}$% | $-36^{+5}_{-5}$% | $-5^{+8}_{-5}$% | $-35^{+6}_{-5}$% | $-51^{+4}_{-4}$% | $-51^{+4}_{-4}$% | $-8^{+4}_{-5}$% |
| | ($-13.6\sigma$) | ($-12.6\sigma$) | ($-0.1\sigma$) | ($-13.3\sigma$) | ($-2.1\sigma$) | ($-12.9\sigma$) | ($-19.0\sigma$) | ($-19.1\sigma$) | ($-3.3\sigma$) |
| USco 48 B | $-34^{+6}_{-6}$% | $-32^{+7}_{-6}$% | $3^{+7}_{-6}$% | $-33^{+5}_{-5}$% | $-2^{+8}_{-6}$% | $-33^{+6}_{-6}$% | $-50^{+5}_{-4}$% | $-50^{+5}_{-4}$% | $-5^{+5}_{-5}$% |
| | ($-12.4\sigma$) | ($-11.4\sigma$) | ($1.1\sigma$) | ($-12.0\sigma$) | ($-1.0\sigma$) | ($-11.7\sigma$) | ($-17.8\sigma$) | ($-17.8\sigma$) | ($-1.9\sigma$) |
| UScoCTIO 5 A | $-26^{+11}_{-8}$% | $-23^{+4}_{-5}$% | $25^{+18}_{-16}$% | $-36^{+11}_{-10}$% | $-9^{+14}_{-12}$% | $-32^{+10}_{-10}$% | $-57^{+7}_{-6}$% | $-57^{+7}_{-6}$% | $32^{+14}_{-13}$% |
| | ($-45.8\sigma$) | ($-39.3\sigma$) | ($43.5\sigma$) | ($-61.3\sigma$) | ($-16.4\sigma$) | ($-55.2\sigma$) | ($-98.4\sigma$) | ($-98.4\sigma$) | ($54.4\sigma$) |
| UScoCTIO 5 B | $-24^{+12}_{-9}$% | $-22^{+5}_{-6}$% | $29^{+19}_{-17}$% | $-35^{+12}_{-10}$% | $-8^{+15}_{-12}$% | $-31^{+12}_{-11}$% | $-58^{+7}_{-6}$% | $-58^{+7}_{-6}$% | $35^{+14}_{-13}$% |
| | ($-39.5\,\sigma$) | ($-37.4\sigma$) | ($48.9\sigma$) | ($-57.7\sigma$) | ($-13.6\sigma$) | ($-50.7\sigma$) | ($-95.0\sigma$) | ($-95.0\sigma$) | ($58.4\sigma$) |
| EPIC 203710387 A | $-14^{+25}_{-12}$% | $-4^{+13}_{-11}$% | $115^{+35}_{-36}$% | $-15^{+13}_{-6}$% | $21^{+20}_{-19}$% | $5^{+15}_{-13}$% | $\cdots$ | $\cdots$ | $291^{+51}_{-52}$% |
| | ($-5.3\,\sigma$) | ($-1.8\sigma$) | ($43.2\sigma$) | ($-5.9\,\sigma$) | ($8.1\sigma$) | ($2.1\sigma$) | | | ($109.0\sigma$) |
| EPIC 203710387 B | $-8^{+29}_{-14}$% | $2^{+15}_{-12}$% | $133^{+39}_{-40}$% | $-8^{+14}_{-6}$% | $27^{+29}_{-15}$% | $15^{+17}_{-14}$% | $\cdots$ | $\cdots$ | $354^{+58}_{-59}$% |
| | ($-3.3\sigma$) | ($1.1\sigma$) | ($52.3\,\sigma$) | ($-3.2\sigma$) | ($10.8\,\sigma$) | ($6.1\sigma$) | | | ($138.7\sigma$) |

Mass offsets are calculated as (model-dynamical)/dynamical, such that negative values correspond to under-predictions by the models. In parentheses, the mass offset is given in units of $\sigma$.

**Table 23.** Empirical mass-radius relations.

| Parameter | Value | Prior |
|---|---|---|
| *Fit 1* | | |
| $c_0$ | $0.1054^{+0.0230}_{-0.0304}$ | $\mathcal{U}(0,1)$ |
| $c_1$ | $3.306^{+0.347}_{-0.224}$ | $\mathcal{U}(0,10)$ |
| $c_2$ | $-3.731^{+0.526}_{-0.913}$ | $\mathcal{U}(-10,0)$ |
| $c_3$ | $1.665^{+0.620}_{-0.339}$ | $\mathcal{U}(0,10)$ |
| *Fit 2* | | |
| $\alpha$ | $1.393^{+0.051}_{-0.049}$ | $\mathcal{U}(0,10)$ |
| $\beta$ | $0.5166^{+0.0291}_{-0.0291}$ | $\mathcal{U}(0,10)$ |
| *Fit 3* | | |
| $\alpha$ | $1.341^{+0.014}_{-0.018}$ | $\mathcal{U}(0,10)$ |
| $\beta$ | $0.4100^{+0.0139}_{-0.0168}$ | $\mathcal{U}(0,10)$ |

Fits 1, 2, and 3 are as follows: polynomial ($0.1 \leq M_*/M_\odot \leq 1$), power law ($0.1 \leq M_*/M_\odot \leq 1$), and power law ($0.3 \leq M_*/M_\odot \leq 1$), respectively. For a given parameter, the quoted value is the maximum likelihood value from the MCMC chain after removal of burn-in (twice the average autocorrelation length). The associated uncertainties are given by the 16th and 84th percentiles. These relations are only valid for stars in the mass ranges indicated with ages equivalent to that of Upper Sco.